%% file: 2DMirror.tex
\documentclass[11pt]{article}

\usepackage{xr}

\makeatletter
\newcommand*{\addFileDependency}[1]{% argument=file name and extension
  \typeout{(#1)}
  \@addtofilelist{#1}
  \IfFileExists{#1}{}{\typeout{No file #1.}}
}
\makeatother

\newcommand*{\myexternaldocument}[1]{%
    \externaldocument{#1}%
    \addFileDependency{#1.tex}%
    \addFileDependency{#1.aux}%
}

\myexternaldocument{2DMirrorSupp_aux}

\newcommand{\blind}{1}
\usepackage[utf8]{inputenc}
\usepackage{amsmath}
\usepackage{bm}
\usepackage[figuresright]{rotating}
\usepackage{amsthm}
\usepackage{comment}
\usepackage{amssymb}
\usepackage{enumerate}
\usepackage{epsfig}
\usepackage{graphics}
\usepackage[letterpaper]{geometry}
\usepackage{algpseudocode}
\usepackage{algorithm}
\usepackage{color,xcolor}
\usepackage{mathrsfs}
\usepackage{enumitem}
\usepackage[authoryear]{natbib}
\usepackage{multirow}
\usepackage{booktabs} 
\usepackage {setspace}
\newcommand{\RNum}[1]{\uppercase\expandafter{\romannumeral #1\relax}}
\newcommand{\rNum}[1]{\lowercase\expandafter{\romannumeral #1\relax}}
\usepackage{makecell}
\usepackage{multirow}
\usepackage{subfigure}
\usepackage{caption}
\usepackage{tikz}
\usetikzlibrary{shapes,arrows}
\tikzstyle{startstop} = [rectangle,rounded corners, minimum width=3cm, text width = 3cm,minimum height=1cm,text centered, draw=black]
\tikzstyle{io} = [trapezium, trapezium left angle = 70,trapezium right angle=110,minimum width=4cm,minimum height=1cm, text width = 3.5cm,text centered,draw=black]
\tikzstyle{process} = [rectangle,minimum width=6cm, minimum height=1cm,text centered,text width =5cm,draw=black]
\tikzstyle{decision} = [diamond,aspect = 3,text centered,draw=black]
\tikzstyle{arrow} = [thick,->,>=stealth]
\tikzstyle{arrowdash} = [dashed,->,>=stealth]
\tikzstyle{arrowsum} = [dotted, line width=1.5pt,->,>=stealth]
\tikzstyle{arrowbig} = [thick, line width=2pt,->,>=stealth]

\tikzstyle{item} = [rectangle,rounded corners, minimum width=1.25cm, text width = 1.25cm, minimum height=1cm,text centered, draw=black]
\tikzstyle{itemlong} = [rectangle,rounded corners, minimum width=5.75cm, text width = 5.75cm, minimum height=1cm,text centered, draw=black]

\newtheorem{remark}{Remark}

\newtheorem{theorem}{Theorem}
\newtheorem{defn}{Definition}
\newtheorem{ex}{Example}

\newcounter{subtheorem}[theorem]

\makeatletter
\renewcommand{\p@subtheorem}{\thetheorem}% Counter prefix.
\makeatother

\newcounter{subassumption}[assumption]

\makeatletter
\renewcommand{\p@subassumption}{\theassumption}% Counter prefix.
\makeatother

\newcounter{sublemma}[lemma]
\renewcommand{\thesublemma}{(\textit{\alph{sublemma}})}
\makeatletter
\renewcommand{\p@sublemma}{\thelemma}% Counter prefix.
\makeatother

\newcounter{subsublemma}[sublemma]

\makeatletter
\renewcommand{\p@subsublemma}{\thelemma\thesublemma}% Counter prefix.
\makeatother

\input{Terminology}
\usepackage[CJKbookmarks=true,
bookmarksnumbered=true,
bookmarksopen=true,
colorlinks=true,
citecolor=blue,
linkcolor=blue,
anchorcolor=blue,
urlcolor=blue]{hyperref}
\usepackage{enumitem}

\begin{document}

\if1\blind
{
  \title{\bf Joint Mirror Procedure: Controlling False Discovery Rate for Identifying Simultaneous Signals}
  \author{Linsui Deng$^{1}$,
  Kejun He$^{1}$\thanks{Correspondence: \href{mailto:kejunhe@ruc.edu.cn}{kejunhe@ruc.edu.cn} and \href{mailto:zhangxiany@stat.tamu.edu}{zhangxiany@stat.tamu.edu}.}~,
  Xianyang Zhang$^{2}$\footnotemark[1] %\thanks{Zhang acknowledges support from NSF DMS-2113359, NIH 1R01GM144351-01 and NIH 1R21HG011662.}
  %\\
%^{1}$The Center for Applied Statistics, Institute of Statistics and Big Data, \\ Renmin University of China, Beijing, China \\
%$^{2}$Department of Statistics, Texas A\&M University, College Station, USA\\
}
  \footnotetext[1]{Institute of Statistics and Big Data, Renmin University of China, Beijing, China}
  \footnotetext[2]{Department of Statistics, Texas A\&M University, College Station, USA}
  \date{\today}  
  \maketitle
} \fi

\if0\blind
{
  \bigskip
  \bigskip
  \bigskip
  \begin{center}
    {\LARGE\bf Joint Mirror Procedure: Controlling False Discovery Rate for Identifying Simultaneous Signals}
\end{center}
  \medskip
} \fi

%\title{Borrowing Neighboring Information Improves Detection Power in Large-Scale Spatial Multiple Testing}
 % \author[1]{Linsui Deng}
  %\author[1]{Kejun He}
  %\author[2]{Xianyang Zhang}
  %\affil[1]{Renmin University of China}
  %\affil[2]{Texas A\&M University}
%\date{\normalsize This version: \today}
%\maketitle
\begin{abstract}
\small
In many applications, the process of identifying a specific feature of interest often involves testing multiple hypotheses for their joint statistical significance. Examples include mediation analysis which simultaneously examines the existence of the exposure-mediator and the mediator-outcome effects, and replicability analysis aiming to identify simultaneous signals that exhibit statistical significance across multiple independent experiments. In this study, we present a new approach called joint mirror (JM) procedure that effectively detects such features while maintaining false discovery rate (FDR) control in finite samples. The JM procedure employs an iterative method that gradually shrinks the rejection region based on progressively revealed information until a conservative estimate of the false discovery proportion (FDP) is below the target FDR level. Additionally, we introduce a more stringent error measure, known as the modified FDR (mFDR), which assigns weights to each false discovery based on its number of null components. We demonstrate that, under appropriate assumptions, the JM procedure controls the mFDR in finite samples.
To implement the JM procedure, we propose an efficient algorithm that can incorporate partial ordering information. Through extensive simulations, we demonstrate that our procedure effectively controls the mFDR and enhances statistical power across various scenarios.
Finally, we showcase the utility of our method by applying it to real-world mediation and replicability analyses.
\\
\noindent%
\end{abstract}
%Simultaneous signals.
%\vfill
%\newpage
%\spacingset{1.71} % DON'T change the spacing!
%\section{Background}
%Spatial statistical analysis 

\section{Introduction}
Consider a set of features, where for each feature, one can compute multiple p-values (associated with multiple hypothesis tests) to examine the statistical significance of the feature. Identifying a specific feature of interest requires the joint statistical significance of all hypotheses concerning that feature. Two notable examples of such a setup are testing high-dimensional mediation hypotheses and detecting replicable signals across multiple studies. Below we provide a brief introduction to these two problems that motivate our research.

Mediation analysis entails dissecting a mechanistic relationship between an exposure (denoted by $X$, e.g., smoking), a set of  candidate mediators (denoted by $M_i$ for $1\leq i\leq m$, e.g., DNA methylation), and an outcome (denoted by $Y$, e.g., lung cancer progression). Under the assumption of no unmeasured confounding, if $M_i$ mediates the effect of $X$ on $Y$, then the exposure $X$ should have a causal effect on the mediator $M_i$, and meanwhile, the mediator $M_i$ has a causal effect on the outcome $Y$. Suppose the causal effect $X\rightarrow M_i$ is parameterized by $\alpha_i$, and the causal effect $M_i\rightarrow Y$ is summarized by $\beta_i$. 
%{\color{red}\sout{1. Then identifying the $M_i$'s that have the mediation effect constitutes of $m$ joint significance tests by testing the composite null hypotheses $H_{0,i}: \alpha_i=0 \text{ or } \beta_i=0$ for $1\leq i\leq m.$} 
Then identifying the $M_i$s that have a mediation effect consists of conducting $m$ joint significance tests to test the composite null hypotheses $\Hcal_{0i}: \alpha_i=0 \text{ or } \beta_i=0$ for $1\leq i\leq m$.
This problem has been extensively studied in the literature. For example, 
%{\color{red} \cite{Sobel1982} \sout{2. studied the product of two normally distributed statistics for testing exposure-mediator and mediator-outcome effects and proposed the Wald-type Sobel’s test.} 
\cite{Sobel1982} proposed a Wald-type test to study the product of two normally distributed statistics for testing indirect effects, which includes testing exposure-mediator and mediator-outcome effects as a special case.
%}\linsuin{Sobel test is not limited to the scope of mediation analysis. }  
\cite{MacKinnon2002} and \cite{Huang2018} recommended the conventional joint significance (JS) test that delivers power improvement over Sobel’s test. Observing that the composite null hypothesis consists of three null types, the two conventional methods are conservative because of ignoring the composite nature of the hypotheses. To alleviate the conservatism, \cite{Huang2019} proposed to adjust the variances of the test statistics without directly estimating the proportions of different null types. An alternative approach is to estimate the null proportions and incorporate them into large-scale inference. For example, \cite{Dai2020} enhanced the conventional JS test by accounting for the mixture null distribution. \cite{Liu2021} constructed a composite p-value by aggregating the p-values weighted by the relative proportions. However, these methods depend on the estimation accuracy of the variances or the null proportions and hence are not guaranteed to control the FDR in finite samples.

Another application that falls into our framework is replicability analysis which has become a cornerstone of modern scientific research \citep{Moonesinghe2007}. Consistent results obtained from different studies with different data provide stronger scientific evidence and are a crucial component of reproducible research. For example, in genomics, replication can significantly improve the credibility of identified genotype-phenotype associations \citep{kraft2009replication}. The partial conjunction null hypothesis (PCH) test, introduced by \cite{friston2005conjunction}, provides a framework for evaluating the consistency of scientific discoveries across multiple studies. \cite{Benjamini2008} further investigated the problem of the PCH test in large-scale inference, facing challenges of composition and multiplicity, and suggested combining p-values from different single tests. \cite{Wang2022} proposed to filter the features unlikely to be significant to alleviate the multiplicity burden and improve power, where the combined p-values are still ``valid'' conditioning on filtering because of the composite structure. Adopting the idea of filtering, \cite{Dickhaus2021} constructed new p-values by re-scaling the combined p-values that survive after filtering. Other related works that concern the PCH framework or its variants include \cite{Heller2014,Zhao2020,Liang2022} to name a few. However, these methods either fail to provide FDR control in finite samples or are too conservative when the number of studies is large.

Motivated by the applications in mediation and replicability analyses, we formulate the following multiple joint significance testing problem. Consider a set of $m$ features, where for the $i$th feature, we observe a sequence of $K$ p-values, denoted by $\{p_{ki}\}^{K}_{k=1}$ for testing the $K$ hypotheses $\{H_{ki}\}^{K}_{k=1}$ concerning the feature $i$, $i=1,\dots,m$. A feature is declared to be statistically significant if all the hypotheses $H_{ki}$ for $1\leq k\leq K$ associated with that feature are found to be significant simultaneously. See Table \ref{tab:setup} for a summary of the setup and notation. In mediation analysis, we have $K=2$ tests for each feature corresponding to testing the exposure-marker and marker-outcome effects. For replicability analysis, we are given $K\geq 2$ p-values for each feature, resulting from $K$ independent experiments. 

\begin{figure}[!h]
     \centering
    \subfigure[$K=2$. \label{fig:2d}]{
    \includegraphics[width=0.45\textwidth]{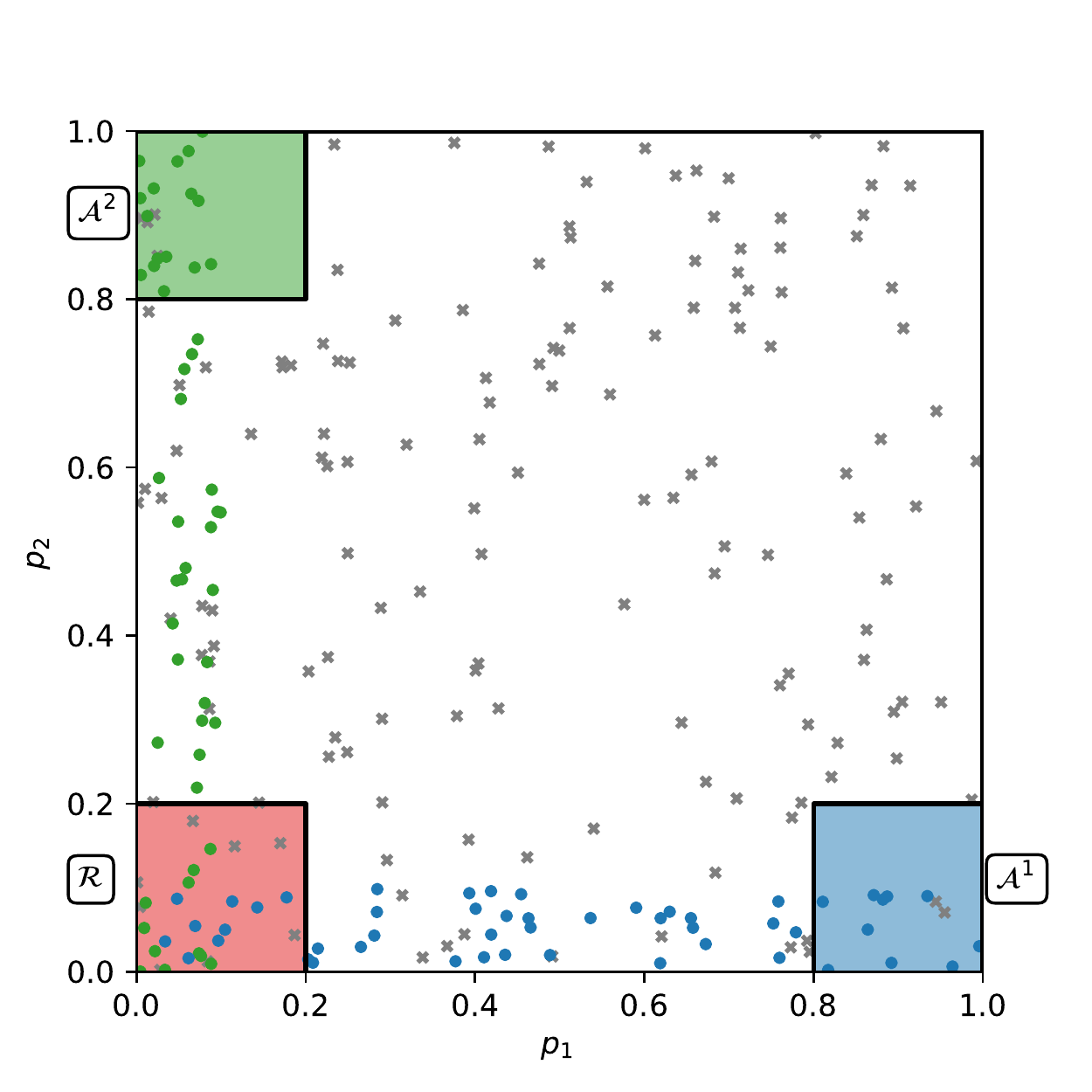}}
    \subfigure[$K=3$.\label{fig:3d}]{
    \includegraphics[width=0.45\textwidth]{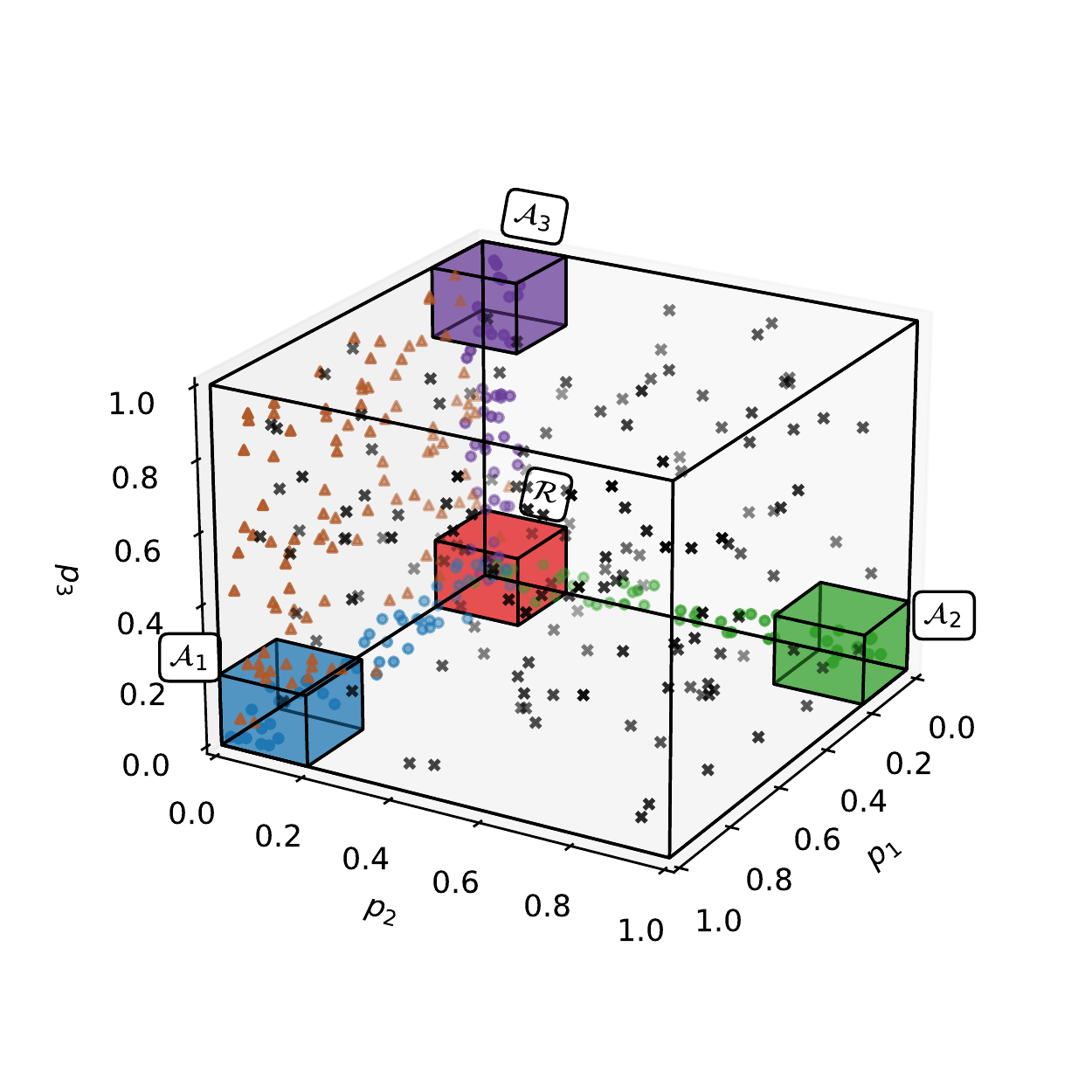}}
        \caption{Illustration of the rejection region (the red square) and the $K$ mirror regions (the blue, green, and purple squares). In Figure~\ref{fig:2d}, $\Acal^1$ estimates the number of false discoveries under the states $\btheta=(0,0)$ (gray $\times$) and $(0,1)$ (blue $\bullet$). In Figure~\ref{fig:3d}, $\Acal^1$ estimates the number of false discoveries under the states $\btheta=(0,0,0)$ (gray $\times$), $(0,1,0)$ (brown $\blacktriangle$), $(0,0,1)$ (omitted) and $(0,1,1)$ (blue $\bullet$). }
        \label{fig:illu2d3d}
\end{figure}

%\subsection{Main contributions}
In this study, we introduce a novel procedure called the joint mirror (JM) procedure, which effectively controls the false discovery rate (FDR) while detecting and identifying signals where all corresponding hypothesis tests are non-null. Our procedure involves two main components (i) a new false discovery proportion (FDP) estimator built on counting the numbers of p-value vectors located at a rejection region and a control region consisting of $K$ mirror regions; (ii) a sequential rule of shrinking the rejection region. The JM procedure sequentially updates the rejection region subject to the shrinkage and partial masking principles (see Section \ref{sec:sequnmaskrule}) until the FDP estimator is below the target FDR level. Figure~\ref{fig:illu2d3d} depicts the rejection region as well as the $K$ mirror regions for the cases of $K=2$ and $3$. We briefly summarize our contribution below.
\begin{itemize}
    \item \textit{Methodological innovation.} Our work is related to the line of research on constructing knockoff copies and finding negative controls in multiple hypothesis testing  \citep{Barber2015, Lei2016, Candes2018}. Our setting is challenging and different from most existing works in that (i) each feature has a vector of p-values lying in the $K$-dimensional unit cube, where no natural ordering of these p-value vectors is available when $K\geq 2$. (ii) the p-value vectors have different numbers of null components and unknown alternative distributions, leading to the composite null problem; (iii) the $K$ p-values from the same feature are potentially dependent, adding another layer of complication. The JM procedure is designed to tackle these challenges and it enjoys several nice features: (i) it constructs $K$ mirror regions to estimate the number of false discoveries for each experiment. Such a construction is conceptually simple and computationally efficient for large $K$; (ii) the FDP estimate in the JM procedure was shown to be less conservative than the conventional JS test (see Example~\ref{ex:sharpest}) and hence often led to higher power in our simulation studies; (iii) practitioners have the flexibility to choose the sequence of rejection regions and can incorporate partial ordering information to improve the interpretability of the findings.

    \item \textit{Finite sample FDR control.} We establish the finite sample FDR control for the JM procedure under the PCH, including the composite null in the joint significance testing problem as a special case.
    
    \item \textit{Modified FDR.}
    We introduce a more stringent error measure, known as the modified FDR
    (mFDR), which assigns weights to each false discovery based on its number of null components. Such an error measure aims to distinguish the null features based on its number of null components. It is particularly desirable in a situation where falsely rejecting a null hypothesis with
    more null components is more harmful than rejecting one with fewer null components.
    We demonstrate that, under appropriate assumptions, the JM procedure controls the mFDR in finite samples. 
    %Numerical results suggest that our procedure can control a modified FDR, a more stringent error measure than the conventional FDR, that weights each null hypothesis based on its number of null components. 
    %This new measure provides an explanation for the conservatism of the JM procedure, which is not easily quantified in other methods \citep{Wang2022,Dickhaus2021}.
    
    \item \textit{Power improvement in mediation and replicability analyses.} %The JM procedure is anticipated to enhance detection power due to the sharper FDP estimator compared to the conventional JS method. Numerically, 
    The JM procedure was shown to outperform the state-of-the-art methods in mediation \citep{Dai2020,Liu2021} and replicability analyses \citep{Zhao2020,Wang2022,Dickhaus2021} with more reliable FDR and mFDR control and higher power.

\end{itemize}

\begin{table}[]
    \centering
\resizebox{\linewidth}{!}{
\fontsize{12}{14}\selectfont
\begin{tabular}{cccccccc}
   \toprule
 \multicolumn{6}{c}{Hypothesis}& The Num.   &  The Estimated Num. 
 \\
 \cmidrule{1-6}
   \multirow{2}*{Experiment} & \multicolumn{4}{c}{Features}& \multirow{2}*{Null Set} &  of & of \\
 \cmidrule(r){2-5}
   &1st  & 2nd &$\cdots$ &  $m$th&  & False Rejection  &False Rejection  \\
   \midrule
   1st & $H_{11}$&$H_{12}$&$\cdots$&$H_{1m}$ & $\Hcal_0^1$ & $V^1=\sum_{i\in\Hcal_0^1}\boldsymbol{1}\{\pf_i\in\Rcal\}$ & $A^1=\sum_{i=1}^m\boldsymbol{1}\{\pf_i\in\Acal^1\}$ \\
   2nd & $H_{21}$&$H_{22}$&$\cdots$&$H_{2m}$ & $\Hcal_0^2$ & $V^2=\sum_{i\in\Hcal_0^2}\boldsymbol{1}\{\pf_i\in\Rcal\}$ & $A^2=\sum_{i=1}^m\boldsymbol{1}\{\pf_i\in\Acal^2\}$ \\
    \makecell[c]{$\vdots$}& 
    \makecell[c]{$\vdots$}& 
    \makecell[c]{$\vdots$}& 
    \makecell[c]{$\ddots$}& 
    \makecell[c]{$\vdots$}& 
    \makecell[c]{$\vdots$}  &
    \makecell[c]{$\vdots$}  &
    \makecell[c]{$\vdots$} \\
   $K$th  & $H_{K1}$&$H_{K2}$&$\cdots$&$H_{Km}$& $\Hcal_0^K$ & $V^K=\sum_{i\in\Hcal_0^K}\boldsymbol{1}\{\pf_i\in\Rcal\}$ & $A^K=\sum_{i=1}^m\boldsymbol{1}\{\pf_i\in\Acal^K\}$ \\
   \midrule
   Composite & $H_{1}$&$H_{2}$&$\cdots$&$H_{m}$ & $\Hcal_0$ & \multirow{2}*{$V=\sum_{i\in\Hcal_0}\boldsymbol{1}\{\pf_i\in\Rcal\}$ } & \multirow{2}*{$A=\sum_{k=1}^K\sum_{i=1}^m\boldsymbol{1}\{\pf_i\in\Acal^k\}$} \\
   Hypothesis & $\cup_{k=1}^KH_{k1}$&$\cup_{k=1}^KH_{k2}$&$\cdots$&$\cup_{k=1}^KH_{km}$ & $\cup_{k=1}^K \Hcal_0^k$ &  & \\
   \bottomrule
\end{tabular}}
    \caption{Basic setup and notation for the JM procedure.}
    \label{tab:setup}
\end{table}

The rest of the article is organized as follows. In Section~\ref{sec:methodology}, we present a detailed description of the JM procedure and investigate its theoretical properties.
Section~\ref{sec:sequnmaskrule} covers implementation specifics of the proposed procedure, including the oracle/feasible unmasking rule and the utilization of partial ordering information. To demonstrate the superior performance of the JM procedure, Sections~\ref{sec:simu} and \ref{sec:app} provide illustrative examples using both simulated datasets and real-world data from mediation and replicability analyses.
Section~\ref{sec:discussion} concludes and outlines several future research directions. The supplement includes additional implementation details, the proofs of the main theoretical results, two extensions of the procedure, and numerical results.

\section{Methodology}\label{sec:methodology}

\subsection{Basic setups}
In the following discussions, we should adopt the terminology from replicability analysis for clarity.
Consider a sequence of p-values arising from $K$ experiments $\pf_i=(p_{1i},\cdots,p_{Ki})$ for $i\in[m]:=\{1,\ldots,m\}$, where $p_{ki}$ is the p-value for testing $H_{ki}$, the hypothesis associated with the $k$th experiment and the $i$th feature. Let $\btheta_i=(\theta_{1i},\dots,\theta_{Ki})\in\{0,1\}^K$ be the indicator of the underlying truth of the hypotheses $(H_{1i},\dots,H_{Ki})$, where $\theta_{ki}=0$ 
($\theta_{ki}=1$) if $H_{ki}$ is under the null (the alternative). 
Consider a class of probability measures $\Pb_{\btheta}$ parameterized by $\btheta\in \Theta$. 
We assume that the p-values $\{\pf_i\}^{m}_{i=1}$ are independently generated from 
\begin{equation}\label{equ:model}
    \pf_i=(p_{1i},\cdots,p_{Ki})\sim\Pb_{\btheta_i}
    \qquad
    \text{for}
    \qquad
    \btheta_{i}\in\{0,1\}^K.
\end{equation}
The null set for experiment $k$ is defined as $\Hcal_0^k=\{1\leq i\leq m:\theta_{ki}=0\}$ and the alternative set is given by $\Hcal_1^k=\{1\leq i\leq m:\theta_{ki}=1\}$. We define the composite null set as $\Hcal_0=\cup_{k=1}^K\Hcal_0^k$. Our goal is to select a set $\whScal\subseteq [m]$ with the largest cardinality such that the FDR, defined as 
\begin{equation}\label{equ:FDRH0}
    \FDR(\whScal)=\mathbb{E}\left\{\frac{|\whScal\cap \Hcal_0|}{|\whScal|\vee 1}\right\},
\end{equation}
is controlled at a target FDR level $q\in(0,1)$, where $|\cdot|$ denotes the cardinality of a set.

\subsection{Conditional mirror conservatism}
%Mirror  conservatism has been widely adopted in multiple testing for the simple null hypothesis \citep{Lei2016, Chao2021, Yun2020}. 
For $\mathbf{p}=(p_1,\dots,p_K) \sim \Pb_{\btheta}$ with $\btheta=(\theta_1,\dots,\theta_K)$, let $\pf_{-k}=(p_1,\dots,p_{k-1},p_{k+1},\dots,p_K)$.
Denote the conditional distribution of $p_k$ given $\pf_{-k}$ and the marginal distribution of $\pf_{-k}$ by $\Pb_{\btheta,k\mid-k}$ and $\Pb_{\btheta,-k }$, respectively, for any $k=1,\ldots,K$. We introduce the following concept that generalizes which notion of mirror conservatism in \cite{Lei2016}.

\begin{defn}[Conditional mirror  conservatism]
Let $\mathcal{S}_0=\{1\leq k\leq K: \theta_k=0\}$. For $k\in \mathcal{S}_0$, $p_{k}$ is said to be conditionally mirror conservative if
\begin{equation}\label{equ:cond_mirror}
    \Pb_{\btheta,k\mid-k}(p_{k}\in[a_1,a_2]\mid \pf_{-k}=\tf)
    \leq 
    \Pb_{\btheta,k\mid-k}(p_{k}\in[1-a_2,1-a_1]\mid \pf_{-k}=\tf)
\end{equation}
for all $0\leq a_1\leq a_2\leq 1/2$ and $\tf\in[0,1]^{K-1}$. We say that $\pf$ is conditionally mirror conservative if all its components in $\mathcal{S}_0$ are conditionally mirror conservative. 
\end{defn}

It is worth noting that the components of $\pf$ are allowed to be dependent in the above definition. When the p-values are conditionally mirror conservative and for a given rejection region $\Rcal\subseteq [0,1/2)^K$, we propose to estimate the number of false discoveries using the number of p-value vectors located at a control region that consists of $K$ separate mirror regions of the form
\begin{equation}\label{equ:control_reg}
    \Acal^k=\Acal^k(\Rcal)=\{(t_1,\cdots,t_{k-1},1-t_{k},t_{k+1},\cdots,t_m):\tf\in\Rcal\}
\end{equation}
for $k=1,\ldots,K$. We now explain the intuition behind the construction of the control region. If $i\in\Hcal_0$, then at least one component of $\pf_i$ is conditionally mirror conservative, i.e., \eqref{equ:cond_mirror} stands for some $k=1,\ldots,K$. %{\color{red}We denote as $i\in\Hcal^k$ when there is no ambiguity.} 
We note that 
\begin{equation}\label{equ:cntrIdeal}
    \sum_{i\in\mathcal{H}_0^k}\boldsymbol{1}(\pf_{i}\in\Rcal)\approx \sum_{i\in \mathcal{H}_0^k}\Pb_{\btheta_i}(\pf_i\in\Rcal)\leq \sum_{i\in \mathcal{H}_0^k}\Pb_{\btheta_i}(\pf_i\in\Acal^k)\approx \sum_{i\in\mathcal{H}_0^k}\boldsymbol{1}(\pf_{i}\in\Acal^k),
%\boldsymbol{1}(\pf_{i}\in\Rcal)\approx \Pb_{\btheta_i}(\pf_i\in\Rcal)\leq \Pb_{\btheta_i}(\pf_i\in\Acal^k)\approx \boldsymbol{1}(\pf_{i}\in\Acal^k),
\end{equation}
where the approximation is due to the law of large numbers, and the inequality is because of mirror conservatism. Thus we see that $\Acal^k$ is designed to estimate the number of false discoveries for experiment $k$. By the definition of $\mathcal{H}_0$ and using (\ref{equ:cntrIdeal}), we have
\begin{align}\label{equ:cons.est}
\sum_{i\in\Hcal_0} 
\boldsymbol{1}(\pf_{i}\in\Rcal)
\leq &
\sum_{k=1}^K 
\sum_{i\in\Hcal_0^k} 
\boldsymbol{1}(\pf_{i}\in\Rcal)
\lesssim 
\sum_{k=1}^K 
\sum_{i\in\Hcal_0^k} 
\boldsymbol{1}(\pf_{i}\in \Acal^k)\leq \sum_{k=1}^K 
\sum_{i=1}^m 
\boldsymbol{1}(\pf_{i}\in \Acal^k),
\end{align}
where $\lesssim$ means ``asymptotically less or equal to''. %{\color{red}Compared to approximating the expected number of false discoveries, the above formula brilliantly bounds the number of false discoveries with an easily calculable quantity, which is the summation of the estimated number of false discoveries for each $\Hcal_0^k$. Scrutinizing $K$ hypotheses separately is non-trivial when adopting the BH-type approaches that require $p_{ki}$ to be super-uniform if $i\in\Hcal_0^k$. These approaches just can conclude $\Pb_{\btheta_i}(p_i\in\Rcal)\leq \sup_{\tf\in\Rcal}t_k$ with $\theta_{ki}=0$ and hence the induced bound is quite loose and impracticable. Therefore, it is no exaggeration to say that our approach considers $\Hcal_0^k$ separately and then merges the results for the first time. It does not just answer the question ``How many discoveries are under the $\Hcal_0$", but also ``How many discoveries are under $\Hcal_0^k$".
%Our mirror regions not only allow the procedure  approach can we. Though this idea is easy and intuitive at first glance, simply bounding the left hand of $\lesssim$ with its expectation is intractable. However, this idea shows its true strength before. has never been introduced before because it shows its true strength.} {\color{blue} Try to avoid self-boasting. Avoid using terms like `` brilliantly'' and ``no exaggeration to say''. Also it is a bit hard to follow the discussions. What you want to say is that the bound we derive is better than $\sup_{\tf\in\Rcal}t_k$. But there is no evidence showing that this is the case here. I don't see the point of adding the discussions here. It seems content-less to me.}
Subsequently, we obtain an approximate upper bound for the FDP using $\Rcal$ as the rejection region:
\begin{equation}\label{equ:FDP.est}
\FDP(\Rcal)
=
\frac{\sum_{i\in\Hcal_0} 
\boldsymbol{1}(\pf_{i}\in\Rcal)}{1\vee \sum_{i=1}^m
\boldsymbol{1}(\pf_{i}\in\Rcal)}
\lesssim
\frac{
\sum_{k=1}^K 
\sum_{i=1}^m
\boldsymbol{1}(\pf_{i}\in\Acal^k)+1}{1\vee \sum_{i=1}^m 
\boldsymbol{1}(\pf_{i}\in\Rcal)}
:=
\widehat{\FDP}(\Rcal).
\end{equation}
We add a constant one in the numerator, which is crucial for achieving the finite sample FDR and mFDR control; see Theorems \ref{thm:fdr} and \ref{thm:mfdr} in Section \ref{sec:JM} and their proofs in Section \ref{sec:proof} of the supplement. The upper bound in \eqref{equ:cons.est} appears to be quite conservative as (i) each $i\in\mathcal{H}_0$ could belong to multiple $\mathcal{H}_0^k$ and (ii) bounding $\sum_{i\in\Hcal_0^k} 
\boldsymbol{1}(\pf_{i}\in\Rcal)$ by $\sum_{i=1}^m 
\boldsymbol{1}(\pf_{i}\in \Acal^k)$ for each $k$ may be loose.  
%{\color{red}The above formula indicates that the estimated number of false discoveries is the summation of the estimated number of false discoveries for each $\Hcal_k$, which is easy and intuitive at first glance. However, this idea show its true strength before. has never been introduced before because it show its true strength.}

Interestingly, we argue that the proposed FDP estimate is not as conservative as it initially appears, and it exhibits improvement over the FDP estimate obtained through a conventional JS test (MacKinnon, 2002; Huang et al., 2018) when the rejection region concentrates around the origin. To elucidate this point, let $\Scal_{0i}=\{1\leq k\leq K:\theta_{ki}=0\}$. Define $\Hcal^{(\kappa)}=\{i\in [m]:|\Scal_{0i}|=\kappa\}$ for $\kappa=0,1,\ldots,K$.
In the context of replicability analysis, $\Hcal^{(\kappa)}$ represents the set of features that are null in $\kappa$ out of the $K$ experiments. It is evident that $\Hcal_0=\cup_{\kappa=1}^K\Hcal^{(\kappa)}$ and $\Hcal_1=\Hcal^{(0)}$ (hence $[m]=\cup_{\kappa=0}^K\Hcal^{(\kappa)}$). In Example~\ref{ex:sharpest} provided below, we compare the number of false discoveries estimated by \eqref{equ:FDP.est} and the conventional JS test within a general setting. Furthermore, Example~\ref{ex:ex1} offers a specific numerical example to demonstrate this phenomenon.
\begin{ex}\label{ex:sharpest}
    In this example, we assume that $\{p_{ki}: \theta_{ki}=0, 1\leq k\leq K, 1\leq i\leq m\}$ are uniformly distributed and mutually independent. 
Fixing $\Rcal=[0,t]^K$ as the rejection region, the expectation of the number of false discoveries can be calculated as
\begin{equation}\label{equ:exp_fd}
\Eb\left(\sum_{i\in\Hcal_0}\boldsymbol{1}\{\pf_{i}\in \Rcal\}\right) =  \sum_{\kappa=1}^K\sum_{i\in\Hcal^{(\kappa)}} \left\{t^{\kappa}\prod_{k\not\in\Scal_{0i}}F_{ki}(t)\right\}
\end{equation}
where $F_{ki}:\Rb\rightarrow[0,1]$ is the distribution function of $p_{ki}$ under the alternative. The FDR-controlling procedure with the conventional JS test would bound \eqref{equ:exp_fd} from above by $(m-m_0)t$, where $m_\kappa$ denotes the cardinality of $\Hcal^{(\kappa)}$. As a comparison, the expectation of the RHS of \eqref{equ:cons.est} is
\begin{equation}\label{equ:exp_control}
\begin{aligned}
    & \Eb\left(\sum_{k=1}^K\sum_{i=1}^m\boldsymbol{1}\{\pf_{i}\in \Acal^k\}\right) = \sum_{\kappa=0}^K\sum_{i\in\Hcal^{(\kappa)}} 
    \left[\kappa t^{\kappa}\prod_{l\not\in\Scal_{0i}}F_{li}(t)
    +\sum_{k\not\in\Scal_{0i}}t^{\kappa} \left\{1-F_{ki}(1-t)\right\}\prod_{l\not\in\Scal_{0i}\cup\{k\}}F_{li}(t) \right],
\end{aligned}
\end{equation}
which bounds \eqref{equ:exp_fd} from above. Let us examine each summand in the RHS of \eqref{equ:exp_control} when $t$ is small. 
Since the p-values under the alternative concentrate around zero, $F_{ki}(1-t)$ is expected to be close to one, and thus, the summand is dominated by its first term $\kappa t^{\kappa}\prod_{l\not\in\Scal_{0i}}F_{li}(t)$. 
We further note that 
$$
\sum_{\kappa=0}^K\sum_{i\in\Hcal^{(\kappa)}} 
\kappa t^{\kappa}\prod_{l\not\in\Scal_{0i}}F_{li}(t)\leq 
\sum_{\kappa=0}^K\sum_{i\in\Hcal^{(\kappa)}} 
\kappa t^\kappa 
= 
m_1 t +2 m_2 t^2 +\cdots + K m_K t^K, 
$$
which can be much smaller than $(m-m_0)t= (m_1+\cdots+m_K)t$ provided that $\kappa m_\kappa t^\kappa \ll m_\kappa t$ for $\kappa\geq 2$, i.e., $t\ll \min_{\kappa=2}^K \kappa^{1/(1-\kappa)}=1/2$ and $m_\kappa>0$ for some $\kappa\geq 2$. In this way, our proposed estimator for the number of false discoveries reduces the gap to the true expected number of false discoveries from $\sum^{K}_{\kappa=2} m_\kappa (t-t^\kappa)$ to $\sum^{K}_{\kappa=2}(\kappa-1)m_\kappa t^\kappa$. This improvement can be quite significant for small $t$ as our estimate captures the correct power of $t$ at the expense of a larger coefficient.
\end{ex}

\begin{figure}[!h]
     \centering
    \includegraphics[width=0.8\textwidth]{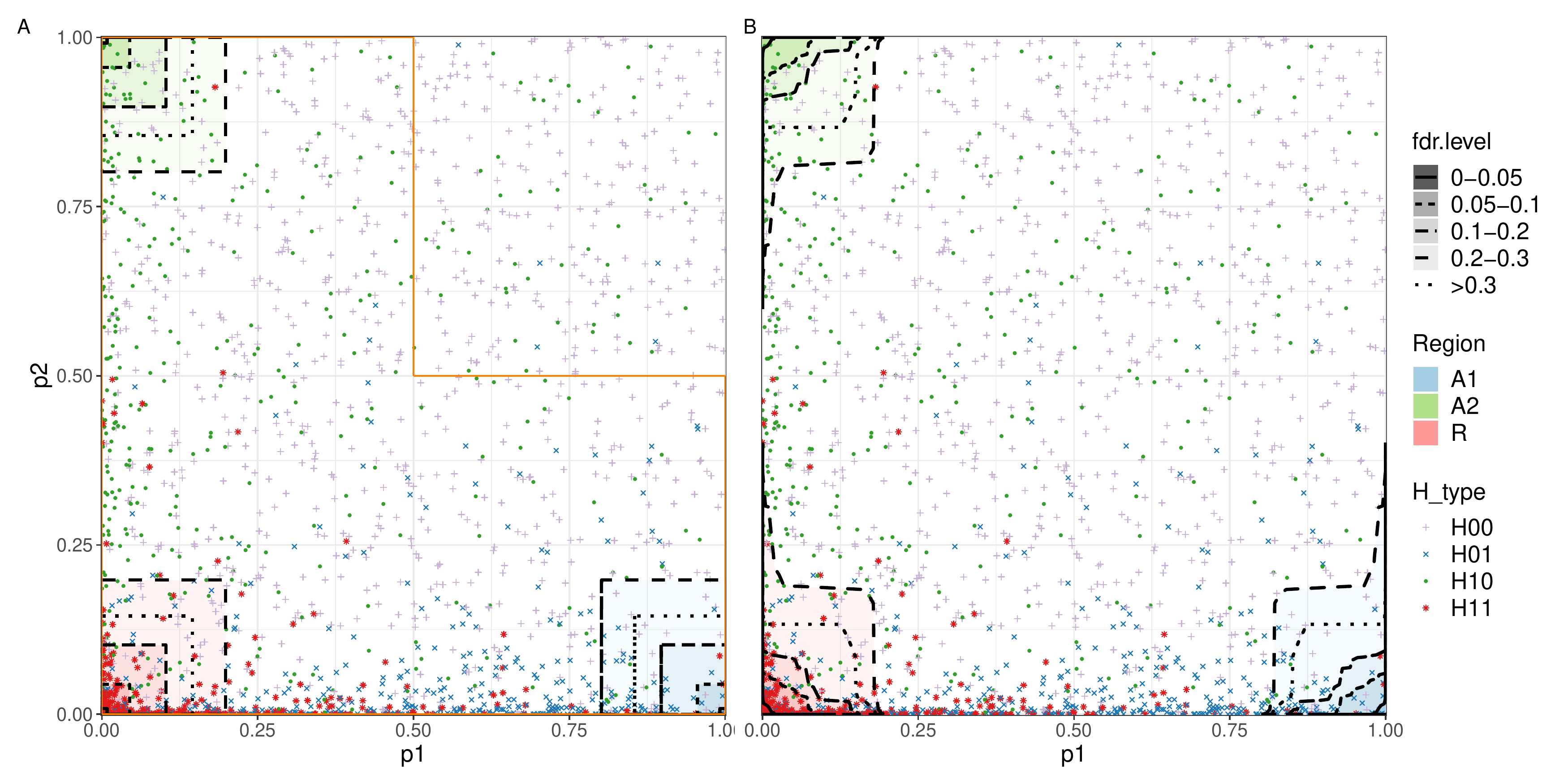}
        \caption{Illustration of different rejection/control regions.
        Panel A corresponds to the square rejection region, while Panel B corresponds to the rejection region associated with the product order.}
        \label{fig:illustration}
\end{figure}

\begin{figure}[!h]
    \centering
    \includegraphics[width=0.8\textwidth]{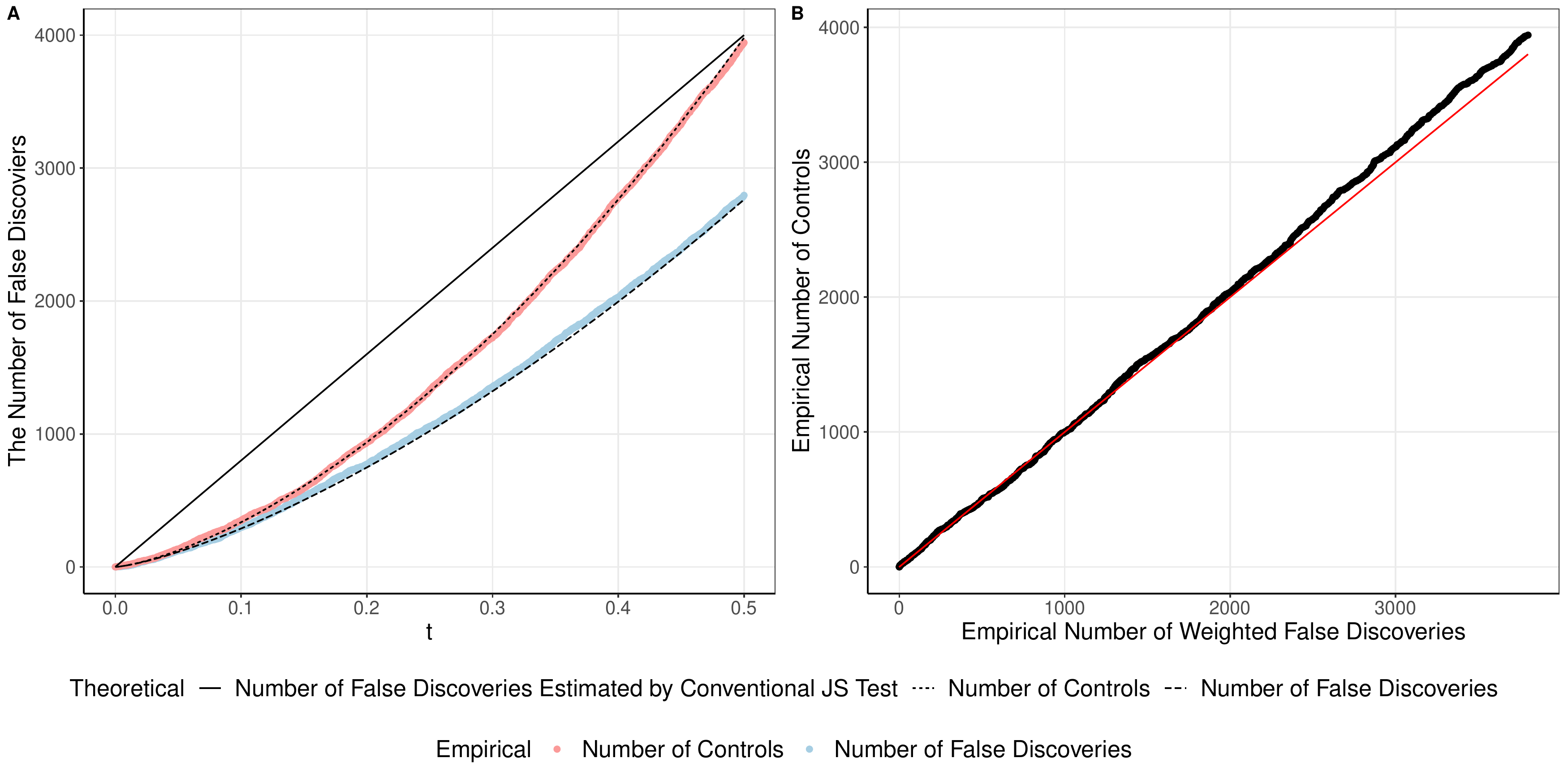}
    \caption{Panel A presents (\rNum{1}) the theoretical numbers of false discoveries and controls, the theoretical numbers of false discoveries estimated by the conventional joint significance test, and (\rNum{2}) the empirical numbers of false discoveries and controls, using the rejection region $\Rcal=[0,t]^2$. Panel B compares the empirical numbers of controls with the weighted false discoveries, i.e., the enumerator of the mFDR defined in Definition~\ref{def:mfdr}.
    }
    \label{fig:appEff}
\end{figure}

\begin{ex} \label{ex:ex1}
Consider the case of $K=2$. We generate $\btheta_i=(\theta_{1i},\theta_{2i})$ for $1\leq i\leq m$ from a mixture distribution of point masses: 
$$
\btheta_i \sim 
\pi_{(0,0)}\delta_{(0,0)}
+ \pi_{(0,1)}\delta_{(0,1)}
+ \pi_{(1,0)}\delta_{(1,0)}
+  \pi_{(1,1)}\delta_{(1,1)},
$$
where $\delta_{(a,b)}$ denotes a point mass at $(a,b)$, $\pi_{(0,0)}=0.4$ and $\pi_{(1,0)}=\pi_{(0,1)}=\pi_{(1,1)}=0.2$. Given $\btheta_i$ we generate two test statistics $X_{1i}\sim N(\mu_{i1},\sigma_1^2)$ and $X_{2i}\sim N(\mu_{i2},\sigma_2^2)$ independently. Specifically, we take $\sigma_1^2=\sigma_2^2=1$, $(\mu_{i1},\mu_{i2})=(0,0)$ for $\btheta_i=(0,0)$, $(\mu_{i1},\mu_{i2})=(0,2.5)$ for $\btheta_i=(0,1)$, $(\mu_{i1},\mu_{i2})=(1.5,0)$ for $\btheta_i=(1,0)$, and $(\mu_{i1},\mu_{i2})=(2,3)$ for $\btheta_i=(1,1)$. Figure~\ref{fig:illustration} presents the corresponding two-sided p-values for testing $(\mu_{i1},\mu_{i2})=(0,0)$ and the rejection and control regions produced by the JM procedure with two different unmasking rules (see Section~\ref{sec:poosequnmaskrule} for the details), various target FDR levels and $m=2,000$. 
%\kejun{Which two methods? JS VS JM or two unmasking rules of JM?}
With the rejection region being $\Rcal=[0,t]^2$ and the sample size $m=10,000$, Panel A of Figure~\ref{fig:appEff} shows the number of controls, compared to the number of false discoveries estimated by the conventional JS test, serving         
       as a better estimator of the number of false rejections, especially when $t<0.1$. 
It can also be seen that the empirical numbers of controls and false discoveries (red and blue points, respectively) are quite close to their theoretical values (dotted and dashed lines, respectively). 
\end{ex}

\subsection{Joint mirror procedure}\label{sec:JM}
In this section, we propose an FDR-controlling procedure named the JM procedure for testing the composite nulls $\{H_i\}_{i=1}^m$.  
It is a sequential procedure that starts with an initial rejection region $\Rcal_0\subseteq[0,1/2)^K$. At each step $t=0,1,\cdots$, we calculate the FDP estimate $\widehat{\FDP}(\Rcal_t)$ according to \eqref{equ:FDP.est}. If $\widehat{\FDP}(\Rcal_t)\leq q$, the procedure terminates and returns the set of rejections $\whScal = \{1\leq i\leq m:\pf_i\in\Rcal_t\}$. If $\widehat{\FDP}(\Rcal_t)>q$, we propose a new rejection region $\Rcal_{t+1}$ that satisfies (\rNum{1}) the shrinkage principle and (\rNum{2}) the partial masking principle. 
The shrinkage principle requires that $\Rcal_{t+1}\subset\Rcal_{t}$. The partial masking principle loosely refers to that only part of the information is available for determining the next rejection region.  
Specifically, the information available at step $t$ include the number of controls, $A_t=\sum_{k=1}^K\sum_{i=1}^m\boldsymbol{1}\{\pf_i\in\Acal_t^k\}$, the number of discoveries, $R_t=\sum_{i=1}^m\boldsymbol{1}\{\pf_i\in\Rcal_t\}$ with $\Acal_t^k=\Acal^k(\Rcal_t)$, and the masked p-value vectors $\{\tpf_{t,i}\}_{i=1}^m$, defined as 
\begin{align*}
\tpf_{t,i}=\begin{cases}
\Proj(\pf_i) \, , &\text{ if } i\in\Mcal_t\\
\pf_i \, , &\text{ if } i\in\Ucal_t
\end{cases}
\end{align*}
with the projection function $\Proj(\tf)=\left(\min\left\{1-t_1,t_1\right\},\cdots,\min\left\{1-t_K,t_K\right\}\right)$
for $\tf\in[0,1]^K$. In the above, we let $\Mcal_t=\{1\leq i\leq m:\pf_i\in\cup_{k=1}^K\Acal^k_t\cup\Rcal_t\}$ be the masked set and $\Ucal_t=[m]\backslash\Mcal_t$ be the unmasked set. When $i\in\Mcal_t$, the value of $\tpf_{t,i}$ is not informative about the region to which $\pf_i$ belongs and hence $\tpf_{t,i}$ is called the masked p-value vector. 
Mathematically, the information available at step $t$ is summarized by the sigma field $\Fcal_t=\sigma\left\{(\tpf_{t,i})_{i=1}^m,A_t,R_t,\Mcal_t,\Ucal_t\right\}
$. The partial masking principle can be formally stated as that 
$$\text{$\Rcal_{t+1}$ is measurable with respect to $\Fcal_t$, i.e., $\Rcal_{t+1}\in\Fcal_t$}.$$
For the ease of presentation, we let $\Rcal_{-1}=[0,1/2)^K$ and generate the corresponding $\Fcal_{-1}$ so that the pre-specified rejection region satisfies that $\Rcal_{0}\subset\Rcal_{-1}$ and $\Rcal_0\in\Fcal_{-1}$. 
To distinguish $(\Rcal_{-1}, \{\Acal_{-1}^k\}^{K}_{k=1})$ from $(\Rcal_{t}, \{\Acal_{t}^k\}^{K}_{k=1})$ for $t \geq 0$, we call $\Rcal_{-1}$ the rejection side and $\{\Acal_{-1}^k\}^{K}_{k=1}$ the control side (consisting of $K$ mirror sides). Obviously, the two principles ensure that $(\Fcal_t)_{t=-1,0,1,\cdots}$ is a filtration as the information at each step increases with $t$, and the trajectory of updating $A_t$, $R_t$, $\Mcal_t$ and $\Ucal_t$ can be inferred from the initial state. Note that at the initial state, $\pf_i$ with two of its components greater than $1/2$ are in the unmasked set and thus are excluded from estimating the FDP. Equivalently, only the p-values located at the ``L"-Shape region in Panel A of Figure~\ref{fig:illustration} are used in estimating the FDP. 

\begin{algorithm}[!h]
	\small
		\caption{Joint Mirror Procedure. \label{alg:JM}} %算法的名字
		\hspace*{0.02in} {\bf Input:} %算法的输入， \hspace*{0.02in}用来控制位置，同时利用 \\ 进行换行
	 P-values $\{\pf_i\}_{i=1}^m$, target FDR level $q\in(0,1)$;
		
		\hspace*{0.02in} {\bf Initialization:} 
		\begin{algorithmic}[1]
			\State Set $\Rcal_{-1}=[0,1/2)^K$ and initialize $\Fcal_{-1}$;
			\State Set $t=0$ and $\Rcal_0$ such that $\Rcal_0\in\Fcal_{-1}$ and $\Rcal_0\subset\Rcal_{-1}$;
            \State Update $\Fcal_0$ and calculate $A_0$, $R_0$ and $\widehat{\FDP}(\Rcal_0)=(1+A_0)/(R_0\vee 1)$;
		\end{algorithmic}
		\hspace*{0.02in} {\bf Search Step:}
		\begin{algorithmic}[1]
			\While{$\widehat{\FDP}(\Rcal_t)> q$ and $R_t>0$}
   \State Choose the next rejection region satisfying that $\Rcal_{t+1}\in\Fcal_t$ and $\Rcal_{t+1}\subset\Rcal_t$ ;
   \State Update $\Fcal_{t+1}$ and calculate $A_{t+1}$, $R_{t+1}$ and $\widehat{\FDP}(\Rcal_{t+1})=(1+A_{t+1})/(R_{t+1}\vee 1)$;
			\State Update $t=t+1$;
			\EndWhile
		\end{algorithmic}
	\hspace*{0.02in} {\bf Output:} %算法的结果输出
 $\whScal=\{i:\pf_i\in\Rcal_t\}$.
	%Rejection region $\Rcal_{t}$.
\end{algorithm}
%Conditional on the event of the p-values locating at a similar ``L"-Shaped region (called the filtering region), \cite{Wang2022} developed an FDP estimator that is much tighter than the conventional method. 
In a related study, \cite{Wang2022} developed an FDP estimator that is tighter than conventional methods. The core idea behind their approach is to estimate the number of falsely rejected null hypotheses using the expected number of null hypotheses located in a similar ``L"-shaped region (called the filtering region). The latter value is approximated by counting the p-values (including both the null and alternative p-values) in the ``L"-shaped region. In contrast, in the case of $K=2$, our proposed FDP estimator improves the detection power by partitioning the ``L"-shaped region into three sections: a rejection side (the bottom-left corner) and two mirror sides (the top-left and bottom-right corners). By using the p-values in the mirror sides to estimate the number of false discoveries, our method is less affected by the p-values under the alternative hypothesis, which are more likely to be on the rejection side than on the mirror sides.

We summarize the JM procedure in Algorithm~\ref{alg:JM}. Similar to \cite{Barber2015} and \cite{Lei2016}, the idea of proving the finite sample FDR control relies on the observation that
$$\begin{aligned}
\Eb\{\FDP(\Rcal_\tau)\}
&=
\Eb\left\{\widehat{\FDP}(\Rcal_\tau)
\frac{\sum_{i\in\Hcal_0} 
\boldsymbol{1}(\pf_{i}\in\Rcal _\tau)}{
\sum_{k=1}^K 
\sum_{i=1}^m
\boldsymbol{1}(\pf_{i}\in\Acal_\tau^k)+1}\right\} \\
& \leq
q
\Eb\left\{\frac{\sum_{i\in\Hcal_0} 
\boldsymbol{1}(\pf_{i}\in\Rcal _\tau)}{
\sum_{i\in\Hcal_0}\sum_{k=1}
^K\boldsymbol{1}(\pf_{i}\in\Acal_\tau^k)+1}\right\},
\end{aligned}$$
where $\tau:=\inf\{t:\widehat{\FDP}(\Rcal_t)\leq q\}$ is a stopping time for a backwards super-martingale. 
We expect the expectation of the ratio at the RHS to be less or equal to one because of \eqref{equ:cons.est}. In fact, we are able to obtain a more sophisticated inequality by considering the problem of testing the PCH. The PCH concerns about whether or not at least $\kappa$ out of the $K$ experiments are under the alternative. In other words, the set of null hypotheses is given by $\Hcal^{\kappa,partial}_0=\cup_{k=\kappa}^K\Hcal^{(\kappa)}$ for $\kappa=1,\ldots,K$. Fixing $i\in\Hcal^{\kappa,partial}_0$, we obtain that 
\begin{equation}\label{equ:controlPCH}
\kappa\Pb_{\btheta_i}\left(\pf_i\in\Rcal\right)\leq 
\sum_{k\in\Scal_{0i}}\Pb_{\btheta_i}\left(\pf_i\in\Acal^k\right)
=\Pb_{\btheta_i}\left(\pf_i\in\cup_{k\in\Scal_{0i}}\Acal^k\right)
\leq 
\Pb_{\btheta_i}\left(\pf_i\in\cup_{k=1}^K\Acal^k\right),
\end{equation}
where the first inequality is because of the mirror conservatism and $|\Scal_{0i}|\geq \kappa$, and the equality holds as $\{\Acal^k\}_{k=1}^K$ are mutually disjoint. 
Using the optional stopping time theorem, we show that 
\begin{align*}
\Eb\left\{\frac{\sum_{i\in\Hcal^{\kappa,partial}_0} 
\boldsymbol{1}(\pf_{i}\in\Rcal _\tau)}{
\sum_{i\in\Hcal^{\kappa,partial}_0}\sum_{k=1}
^K\boldsymbol{1}(\pf_{i}\in\Acal_\tau^k)+1}\right\}\leq\frac{1}{\kappa},    
\end{align*}
which implies the finite sample FDR control for the JM procedure. The detailed arguments are given in the supplement. 

\begin{theorem}[Finite sample FDR control]\label{thm:fdr}
Consider the problem of testing $\Hcal^{\kappa,partial}_0$ for any $\kappa=1,\ldots,K$, where the corresponding FDR
is defined in (\ref{equ:FDRH0}) with $\mathcal{H}_0$ being replaced by $\Hcal^{\kappa,partial}_0$. 
Suppose that the null p-values $\{\pf_{i}\}_{i\in\Hcal^{\kappa,partial}_0}$ are independent of each other and of the non-null
p-values $\{\pf_{i}\}_{i\in\Hcal^{\kappa,partial}_1}$ with $\Hcal^{\kappa,partial}_1=[m]\setminus \Hcal^{\kappa,partial}_0$. Suppose $\pf_{i}$ is conditionally mirror conservative for all $i\in\Hcal^{\kappa,partial}_0$. Then the JM procedure controls the
$\FDR$ at level $q/\kappa$, or equivalently controls $\kappa \FDR$ at level $q$. As a consequence, when $\kappa=1$ (i.e., under the composite null $\mathcal{H}_0$), the JM procedure controls the FDR at level $q$.
\end{theorem}

Theorem~\ref{thm:fdr} shows that the JM procedure controls the FDR in finite samples for testing the PCH. Researchers have the flexibility to choose the sequence of rejection regions that adheres to the shrinkage
and the partial masking principles.
Section~\ref{sec:sequnmaskrule} offers a feasible procedure that is interpretable and computationally efficient.

As seen from Theorem \ref{thm:fdr}, the FDR control of the JM procedure becomes more stringent as the number of null components $\kappa$ gets larger. To understand this finding, let us assume that the null p-values are all symmetric about 0.5. Recall that the JM procedure estimates the number of false discoveries by $\sum_{k=1}^K
\sum_{i=1}^m 
\boldsymbol{1}(\pf_{i}\in \Acal^k)$ which is lower bounded by
\begin{align*}%\label{equ:weigh_fd}
\sum_{k=1}^K 
\sum_{i\in \Hcal_0}
\boldsymbol{1}(\pf_{i}\in \Acal^k)
\approx \sum^{K}_{\kappa=1}\kappa\sum_{i\in \Hcal^{(\kappa)}}
\boldsymbol{1}(\pf_{i}\in \Rcal),
\end{align*}
where the RHS weights the number of false discoveries by the corresponding number of null components. Put simply, the JM procedure inherently considers the heterogeneity present among the null hypotheses by assigning weights to them based on the number of null components. This characteristic becomes especially advantageous when the consequences of erroneously rejecting a null hypothesis with more null components are more severe compared to one with fewer null components. This observation inspires us to introduce the following error measure, which can be seen as a modified version of the weighted false discovery rate (wFDR, \citealp{Benjamini1997}), tailored to the specific context at hand.

\begin{defn}[Modified FDR, mFDR] \label{def:mfdr}\rm
        The modified FDR is defined as
$$
\mFDR(\whScal)
=
\Eb\left\{\mFDP(\whScal)\right\}
=
\Eb\left\{
\frac{
\sum_{\kappa=1}^K \kappa|\whScal\cap\Hcal^{(\kappa)}|
}{|\whScal|\vee 1}\right\},
$$
where the numerator is a weighted sum of the number of false discoveries (also called the number of weighted false discoveries). 
\end{defn}
The mFDR differs from both the traditional FDR and the wFDR in several ways. Specifically, the mFDR is a more stringent error measure than the conventional FDR because $\mFDR(\whScal)\geq \FDR(\whScal)$. In spatial signal detection, the wFDR calculates the numerator and denominator as weighted sums of the number of false and total discoveries, respectively. The weights $w_i$ in the wFDR are related to the cluster sizes, reflecting the fact that discovering a cluster with a larger size would account for either a larger error if it is under the null or more meaningful discoveries if it is indeed a signal \citep{Benjamini2007,Basu2018}. In contrast, the weights in the mFDR are based on the numbers of null components and only affect the false discoveries. Falsely rejecting a hypothesis with a larger number of null components would account for a larger error in mFDR.
%{\color{red} One of the most significant points is rejecting a hypothesis corresponding to a large cluster reflects discovering more interested locations.} %increases the weighted sum of the total discoveries more, which is desirable if it contains a signal because a large cluster contains more individual hypotheses 
%In Basu et al. 2008, the authors commented that ``the weighted false discovery rate was used to reflect that a false positive cluster with larger size would account for a larger error.'' I think this is easier to understand than what you described.} 
The mFDR approach offers a fresh viewpoint on joint significance testing, as it penalizes more for rejecting a hypothesis that contains more null components. Panel B of Figure~\ref{fig:appEff} demonstrates that the empirical number of controls is very close to the empirical number of weighted false discoveries. Using the leave-one-out technique introduced in \cite{Barber2020}, we establish the finite sample mFDR control for the JM procedure when the rejection regions are only determined by the masked p-value vectors.

\begin{theorem}[Finite Sample mFDR control]\label{thm:mfdr}
    Suppose that the null p-values $\{\pf_{i}\}_{i\in\Hcal_0}$ are independent of each other and of the non-null
p-values $\{\pf_{i}\}_{i\in\Hcal_1}$. Suppose $\pf_{i}$ is conditionally mirror conservative for all $i\in\Hcal_0$. If we additionally require that $\Rcal_{t}\in\sigma\left\{(\tpf_{i})_{i=1}^m\right\}$ with $\tpf_i=\mathrm{Proj}(\pf_i)$, then the JM procedure controls the $\mFDR$ at level $q$.
\end{theorem}

JM.Max (a variant of the JM procedure introduced in Section \ref{sec:poosequnmaskrule}) fulfills the assumption that $\Rcal_{t}\in\sigma\left\{(\tpf_{i})_{i=1}^m\right\}$, and hence is guaranteed to control the mFDR in finite samples.
Our simulation studies in Section~\ref{sec:simu} demonstrate that different variants of the JM procedure can empirically control the mFDR in all the numerical examples.

\subsection{JM procedure with unmasking rule}
We note that the method used for updating the rejection region does not affect the FDR control as long as $\Rcal_{t+1}\in\Fcal_t$. 
Here we provide a way of updating the rejection region, which is useful for implementing the JM procedure in practice. 
At each step, we aim to construct a new rejection region $\Rcal_{t+1}$ using the information in $\mathcal{F}_t$. As $\Rcal_{t+1}\subset\Rcal_{t}$, the control region is also shrinking over time, i.e., $\cup_{k=1}^K \Acal_{t+1}^k\subset \cup_{k=1}^K \Acal_{t}^k$. Our idea is to remove one hypothesis from the masked set at a time using a sequential unmasking rule (to be described in the next section). The p-value vector associated with this particular hypothesis is then revealed. It belongs to either the current rejection region $\Rcal_t$ or one of the $\Acal_t^k$s. We remove this p-value vector from the corresponding rejection or control region and update $A_t$ (the conservative estimate of the number of false rejections) and $R_t$ (the number of rejections) accordingly. The search step in Algorithm~\ref{alg:JM} can be described as follows:
\begin{enumerate}%[topsep=0pt,itemsep=0pt,parsep=0pt]
    \item Choose $i\in\Mcal_t$ using the information in $\Fcal_t$, and set $\Mcal_{t+1}=\Mcal_{t}\backslash\{i\}$ and $\Ucal_{t+1}=\Ucal_{t}\cup\{i\}$;
    \item Update $\Fcal_{t+1}$. Set $A_{t+1}=A_{t}$ and $R_{t+1}=R_{t}-1$ if $\pf_i\in\Rcal_{-1}$, and set $A_{t+1}=A_{t}-1$ and $R_{t+1}=R_t$ otherwise;
    \item Calculate $\widehat{\FDP}_{t+1}=(1+A_{t+1})/(R_{t+1}\vee 1)$.
\end{enumerate}

In this way, updating the rejection region amounts to finding an appropriate unmasking rule, i.e., determining the next p-value vector to be revealed from the masked set. % when there is no tie in $\{\tpf_{0,i}\}_{i\in\Mcal_{0}}$. %Specifically, the rejection region at step $t$ can be selected as {\color{red} a point set $\Rcal_t=\{\tpf_{t,i}\}_{i\in\Mcal_t}$.} 
%We will also provide a careful discussion about the connection between the partial order-assisted unmasking rule and its corresponding rejection region in Section \ref{sec:poosequnmaskrule}. 
%Particularly,} Remark~\ref{rmk:poarr} provides the construction of a partial order-assisted rejection region {\color{red} when considering unmasking rule using partial order} , which is intuitive and naturally extends the {\color{red} concept of rejection region in one-dimensional case. Is there any reason to mention this remark a couple of times? It shouldn't be just a remark if it is so important.}  For convenience, we say updating rejecting region and employing the unmasking rule interchangeably {\color{red} There is a difference between the words ``updating'' and ``employing''}.

\section{Sequential Unmasking Rule}\label{sec:sequnmaskrule}

In Section \ref{sec:or}, we develop an oracle sequential unmasking rule by adopting a Bayesian viewpoint. The oracle procedure enjoys certain optimality as shown in Theorem \ref{thm:pow}. 
We describe a feasible sequential unmasking rule in Section \ref{sec:f_sur}. In Section \ref{sec:poosequnmaskrule}, we discuss how to incorporate external partial ordering information to improve the interpretability of the findings.

\subsection{Oracle procedure}\label{sec:or}
Consider a mixture model with $2^K$ components for the p-values: 
\begin{equation}
 \label{equ:twoNmodel} 
 f(\tf)=\sum_{\btheta\in\{0,1\}^K}\pi_{\btheta}f_{\btheta}(\tf)%=\sum_{\btheta\in\{0,1\}^K}\Pb(\bTheta=\btheta)f_{\btheta}(\tf),
\end{equation}
with $\sum_{\btheta\in\{0,1\}^K}\pi_{\btheta}=1$ and $\pi_{\btheta} \geq 0$, where $\pi_{\btheta}$ is the probability of the underlying state being $\btheta$ and $f_{\btheta}$ is the density of $\pf$ given the state $\btheta$. We shall first consider $\pf\in\cup_{k=1}^K\Acal^k_{-1}\cup\Rcal_{-1}$ and denote by $\tpf=\Proj(\pf)$ the corresponding masked p-value vector. The density of $\tpf$ is given by
\begin{equation}\label{eq-unmask}
g(\ttf)
\propto\sum_{\tf:\Proj(\tf)=\ttf} f(\tf)\boldsymbol{1}\{\tf\in \cup_{k=1}^K\Acal^k_{-1}\cup\Rcal_{-1}\}.\end{equation}
At step $t$, the sequential unmasking rule selects the hypothesis $i_t^*$ satisfying that
\begin{equation}\label{equ:SeqUnmaskR}
i_t^{*}=\arg\min_{i\in\Mcal_t} q_{i}\, ,
\quad\text{where}\quad
q_{i}= \Pb(\pf=\tpf_i\mid\tpf=\tpf_i)=f(\tpf_i)/g(\tpf_i)
\end{equation}
represents the conditional probability of a p-value vector falling into the rejection region given its masked p-value vector.

Roughly speaking, we prefer to reveal the hypothesis whose p-values are more likely to be in the control side.
%{\color{red}\sout{6. region} side}. 
Subsequently, the FDP estimate calculated by \eqref{equ:FDP.est} would get smaller, and the procedure may stop earlier, leading to a higher power. Such an idea has been implemented by \cite{Ren2020,Chao2021}
to determine the next statistic or p-value to be revealed from the masked set using side information. 
We show that the unmasking rule defined in (\ref{eq-unmask}) is optimal 
in the sense of delivering the most discoveries among all possible unmasking procedures. 
Analogous to \cite{Chao2021}, although this rule is updated stepwise and appears to be only stepwise optimal, it is indeed globally optimal once the initial state has been determined.

\begin{theorem}\label{thm:pow}
 Suppose $\{\pf_i\}_{i=1}^m$ are independently generated from model \eqref{equ:twoNmodel}. The sequence of the masked sets induced by \eqref{equ:SeqUnmaskR} is the most powerful among all possible sequences in the sense of maximizing $\Pb(R_{\tau}\geq r)$ for every $r=1,\cdots,|\Mcal_{0}|$. 
\end{theorem}

Another interesting observation is that the conditional probability in \eqref{equ:SeqUnmaskR} is closely related to a generalized version of the local false discovery rate (lfdr, \citealp{Efron2001}) under some conditions.
\begin{theorem}\label{thm:lmfdr}
Suppose $\{\pf_i\}_{i=1}^m$ are independently generated from model \eqref{equ:twoNmodel} that additionally satisfies: For $\pf \sim f_{\btheta}$, (a) $p_{k}< 1/2$ if $\theta_k=1$; (b) the distribution of $p_k$ is symmetric about $1/2$ if $\theta_k=0$ conditional on $\pf_{-k}$. Setting $\kappa_{\btheta} = \sum_{k=1}^K(1-\theta_k)$, we have 
$$f(\ttf)/g(\ttf)\propto \left\{1+\frac{\sum_{\btheta\not=(1,\cdots,1)}\kappa_{\btheta}\pi_{\btheta}f_{\btheta}(\ttf)}{f(\ttf)}\right\}^{-1} ~ \text{ for }~ \ttf\in\Rcal_{-1}.$$
%    for $\ttf\in\Rcal_{-1}$. 
\end{theorem}
When $K=1$, $\sum_{\btheta\not=(1,\cdots,1)}\kappa_{\btheta}\pi_{\btheta}f_{\btheta}(\ttf)/f(\ttf)$ becomes  $\pi_{0}f_0(\tt)/f(\tt)$, the usual lfdr, where $\pi_0$ is the null probability, $f_0(\cdot)$ is the probability density of the p-value under the null, and $f(\cdot)$ is the marginal density of the p-value. When $K>1$, it can be viewed as a ``modified local false discovery rate" (mlfdr) because $\pi_{\btheta} f_{\btheta}(\tf)/f(\tf)=\Pb(\bTheta=\btheta\mid \pf=\tf)$, the posterior probability of $\bTheta=\btheta$ given $\pf=\tf$, and $\kappa_{\btheta}$ plays the same role as $\kappa$ in the mFDR. Theorem~\ref{thm:lmfdr} suggests that the hypothesis $i^*_t$ in \eqref{equ:SeqUnmaskR} is also the one with the maximum mlfdr value in $\Mcal_t$. %{\color{red}Consider a special case for $K=2$ such that $\pi_{(0,0)}=\pi_{(1,0)}=1/2$, $\pi_{(0,1)}=\pi_{(1,1)}=0$ and p-values are uniformly distributed on $[0,1]^2/[0,0.25]\times[0,1]$ for $\btheta=(0,0)/(1,0)$. Consider two masked p-values $\tpf_1\in[0,0.25]\times[0,0.5]$ (corresponding to the hypothesis with one and two null components) and $\tpf_2\in[0.25,0.5]\times[0,0.5]$ (corresponding to the hypothesis with two null components). In this case, mlfdr is $5/6$ at $\tpf_1$ and two at $\tpf_2$, and therefore $\tpf_2$ is revealed before $\tpf_1$.} 
%{\color{red} You're making up an example. But is this what really happens in a numerical example?}
%This indicates that the oracle procedure tends to rule out null hypotheses with more null components first, which is the main factor causing conservatism. 

\subsection{Feasible sequential unmasking rule}\label{sec:f_sur}

In practice, the conditional probability $q_i$ in \eqref{equ:SeqUnmaskR} is unknown and we need to estimate it from the data. 
As long as the estimation procedure uses only the information contained in $\Fcal_t$ at step $t$, our procedure controls the FDR regardless of the estimation accuracy. However, the estimation method affects the sequence of the masked sets and hence has a critical impact on the power. In this work, we consider a kernel-based estimator for $q_i$ at step $t$ defined as 
\begin{equation}\label{equ:est_q}
\hq_{t,i}
= 
\frac{{\sum_{i^\prime\in\Ucal_t\cap\Mcal_{-1}}\boldsymbol{1}\{\pf_{i^\prime}\in\Rcal_{-1}\}v_{H}(\tpf_{i},\tpf_{i^\prime})}}{{\sum_{i^\prime\in\Ucal_t\cap\Mcal_{-1}}v_{H}(\tpf_{i},\tpf_{i^\prime})}}:=\frac{\hq^n_{t,i}}{\hq^d_{t,i}}, ~ i =1, \dots, m,
\end{equation}
where $\tpf_{i}=\Proj(\pf_i)$, $v_{H}(\mathbf{x},\mathbf{x}^\prime)={\Kcal_H(\xf-\xf^\prime)}/{\Kcal_H(\mathbf{0})}$, $\Kcal_H(\mathbf{t})=\det(H)^{-1/2}\Kcal(H^{-1/2}\mathbf{t})$, $H\in\Rb^{K\times K}$ is a positive definite bandwidth matrix, and $\Kcal:\mathbb{R}^{K}\rightarrow\mathbb{R}$ is a positive, bounded and symmetric kernel function. 
In \eqref{equ:est_q}, $\hq^n_{t,i}$ and $\hq^d_{t,i}$ can be viewed as the kernel density estimators for $f(\tpf_i)$ and $g(\tpf_i)$ up to a multiplicative %{\color{red} multiplicative?} 
constant, respectively, based on the information in $\Fcal_t$. 
An interesting fact is that the numerator and the denominator of \eqref{equ:est_q} can be updated separately, which allows us to speed up the computation using the result from the previous step. Specifically, given $\hi_{t+1}^*$ the index of the next p-value vector to be unmasked, we only need to add $\boldsymbol{1}\{\pf_{\hi^*_{t+1}}\in\Rcal_{-1}\}v_{H}(\tpf_{i},\tpf_{\hi^*_{t+1}})$ and $v_{H}(\tpf_{i},\tpf_{\hi^*_{t+1}})$ to the numerator and the denominator of \eqref{equ:est_q} respectively to get the value of $\hat{q}_{t+1,i}$. 
As for the choice of the bandwidth matrix $H$, we adopt the rule of thumb bandwidth matrix 
$\{4/m(K+2)\}^{2/(K+4)}\widehat{\var}(\tpf)$
proposed in \cite{Silverman1986}. 
In our numerical experiments, we found that this rule works reasonably well. 
There are other ways to select the bandwidth matrix, including the plug-in method \citep{duong2003plug,Chacon2010} and the cross-validation method \citep{Duong2005}. 

\subsection{Sequential unmasking rule with partial ordering information}\label{sec:poosequnmaskrule}
The feasible sequential unmasking rule has two issues. Firstly, the accuracy of the estimator in \eqref{equ:est_q} relies on the information available within the unmasked set. This raises concerns when revealing a hypothesis whose masked p-value vector is isolated from the masked p-value vectors of those hypotheses in the unmasked set. An illustrative example using the real data from Section \ref{sec:dna} demonstrates this issue. In that example, the SNP with the strongest mediation effect (depicted as a pink solid triangle) between smoking and lung disease was not rejected by JM.EmptyPoset (the JM procedure without utilizing partial ordering information), potentially due to the isolated nature of its masked p-value vector.

The second concern can be demonstrated by considering the scenario where $K=1$. Let us assume that the p-values $\{p_i\}_{i=1}^m$ are independently generated from a two-group mixture model defined as:
$f(x) = \pi_0 f_0(x)+(1-\pi_0)f_1(x)$, 
where $\pi_0=0.9$ represents the probability of a hypothesis being under the null, and $f_0(x)=2x$ and $f_1(x)=2(1-x)$ represent the densities of the p-values under the null and alternative, respectively.
In this case, the conditional probability $\Pb(p=x|\tp=x)$, where $p$ denotes the original p-value and $\tp$ represents the masked p-value, can be calculated as $f(x)/g(x) = 0.1 + 0.8x$ for $0\leq x\leq 0.5$.
Based on the sequential unmasking rule defined in \eqref{equ:SeqUnmaskR}, hypotheses with smaller masked p-values $\tp$ are revealed earlier. However, this approach tends to unveil a substantial proportion of alternative p-values too early in the process, thus diminishing the detection power.

To address these two concerns, we suggest imposing some additional restrictions on the unmasking/revealing orders. Specifically, we introduce a general framework based on the notion of partially ordered sets that
generalizes the intuitive concept of an ordering, sequencing, or arrangement of the elements of a set.

\begin{defn}[Partially Ordered Set, \citealp{wallis2011beginner}]
 A strict partial order on a set $\Pcal$ is a relation $\prec$ on $\Pcal$ that is irreflexive, asymmetric, and transitive.
 \begin{enumerate}%[(a)]%[topsep=0pt,itemsep=0pt,parsep=0pt]
     \item[(a)] Irreflexivity: for any $a\in\Pcal$, $a\prec a$ is not true;
     \item[(b)] Asymmetry: for any $a,b\in\Pcal$ such that $a\prec b$, $b\prec a$ is not true;
     \item[(c)] Transitivity: for any $a,b,c\in\Pcal$ such that $a\prec b$ and $b\prec c$, then $a\prec c$ is always true.
 \end{enumerate} We call the pair $(\Pcal,\prec)$ a partially ordered set (or a poset for short). If $a\prec b$, we say $a$ is less than $b$ or $b$ is larger than $a$.
\end{defn}

\begin{defn}[Maximal Set, \citealp{wallis2011beginner}]
Consider a poset $(\Pcal,\prec)$. We call $a\in P$ a maximal element of $(\Pcal,\prec)$ if there is no other element $b\in P$ such that $a\prec b$. The maximal set $\Pcal^{maximal}$ is the collection of all maximal elements of $(\Pcal,\prec)$.
\end{defn}

Consider a partial order $\prec$ on $\{\tpf_i\}_{i=1}^m$ satisfying that a small value favors the alternative hypothesis. For example, $\prec$ is defined as the usual ``less than sign $<$'' when $K=1$. % {\color{red} How to construct the partial order? A small value of what?} %Then, $(\tPcal,\prec):=(\{\tpf_i\}_{i=1}^m,\prec)$ is a poset and we will omit $\prec$ for short. 
The candidates for the next masked p-value vector to be revealed are in the maximal set of $\tPcal_t:=\{\tpf_i\}_{i\in\Mcal_t}$, denoted by
$\tPcal^{maximal}_t$. For convenience, we define a partial order $\prec$ on $\Mcal_t$ satisfying that $i\prec j$ in $\Mcal_t$ iff $\pf_i\prec\pf_j$ in $\tPcal_t$, and write $\Mcal_t^{maximal}$ as the corresponding maximal set. %Combine itThe JM procedure with partial order oriented sequential unmasking rule is described in Algorithm~\ref{alg:alg1}. 
Algorithm~\ref{alg:JMposet} combines the kernel method in \eqref{equ:est_q} with the sequential unmasking rule using partial order. %The maximal set $\Mcal_t^{maximal}$ changes during the JM procedure. 
If more than one hypothesis attain the minimum value of $\hq_{t,i}$ in Line 2 of the search step of Algorithm \ref{alg:JMposet}, we randomly select one of them. This ensures that only one hypothesis is revealed at each step.
Throughout the JM procedure, we continuously update the maximal set $\Mcal_t^{maximal}$ by removing the revealed hypothesis and incorporating new elements into $\Mcal_t^{add,maximal}$ at each step. The specifics of this implementation are provided in Section~\ref{sec:discussionAlg2} of the supplement.

\begin{algorithm}[!h]
	\small
		\caption{JM Procedure with the Sequential Unmasking Rule using Partial Order. \label{alg:JMposet}} %算法的名字
		\hspace*{0.02in} {\bf Input:} %算法的输入， \hspace*{0.02in}用来控制位置，同时利用 \\ 进行换行
	 P-values $\{\pf_i\}_{i=1}^m$, partial order $\prec$, target FDR level $q\in(0,1)$;
		
		\hspace*{0.02in} {\bf Initialization:} 
		\begin{algorithmic}[1]
			\State Set $\Rcal_{-1}=[0,1/2)^K$ and initialize $\Fcal_{-1}$;
			\State Set $t=0$ and $\Rcal_0$ such that $\Rcal_0\in\Fcal_{-1}$ and $\Rcal_0\subset\Rcal_{-1}$;
            \State Update $\Fcal_0$ and calculate $A_0$, $R_0$ and $\widehat{\FDP}_0:=\widehat{\FDP}(\Rcal_0)=(1+A_0)/(R_0\vee 1)$;
			\State Obtain $\Mcal_0$, $\Ucal_0$, $\Mcal_0^{maximal}$, and calculate $\hq^n_{t,i}$ and $\hq^d_{it}$ according to \eqref{equ:est_q} for $i\in\Mcal_0^{maximal}$;
		\end{algorithmic}
		\hspace*{0.02in} {\bf Search Step:}
		\begin{algorithmic}[1]
			\While{$\widehat{\FDP}_t> q$}
			\State Calculate $\hq_{t,i}=\hq^n_{t,i}/\hq^d_{t,i}$ for $i\in\Mcal_t^{maximal}$ and find $\hi^*_t=\arg\min_{i\in\Mcal_t^{maximal}}\hq_{t,i}$;
   \State Update the mask and unmask sets $\Mcal_{t+1}=\Mcal_t\backslash\{\hi^*_t\}$ and $\Ucal_{t+1}=\Ucal_t\cup\{\hi^*_t\}$;
   \State Find new hypotheses to be added to the maximal set, denoted as $\Mcal_{t}^{add,maximal}$; %{\color{red} Can you explain this step?}
   \State Update $\Mcal_{t+1}^{maximal}=\Mcal_{t}^{maximal}\cup\Mcal_{t+1}^{add,maximal}\backslash\{\hi_t^*\}$;
   \State For $i\in\Mcal_{t}^{add,maximal}$, calculate $\hq^n_{t
   +1,i}$ and $\hq^d_{t+1,i}$ according to \eqref{equ:est_q};
   \State For $i\in\Mcal_{t}^{maximal}\backslash\{\hi^*_t\}$, update $\hq^d_{t+1,i}=\hq^d_{t+1,i}+v_{H}(\tpf_{i},\tpf_{\hi^*_t})$;
			\If{$\pf_{\hi^*_t}\in\Rcal_{-1}$}
			\State $R_{t+1}=R_t-1$;
   \State For $i\in\Mcal_{t}^{maximal}\backslash\{\hi^*_t\}$, update $\hq^n_{t+1,i}=\hq^n_{t+1,i}+v_{H}(\tpf_{i},\tpf_{\hi^*_t})$;
   \Else
			\State $A_{t+1}=A_t-1$;
			\EndIf
   \State Update $\widehat{\FDP}_{t+1}=(1+A_{t+1})/(R_{t+1}\vee 1)$;
			\State Update $t=t+1$;
			\EndWhile
		\end{algorithmic}
	\hspace*{0.02in} {\bf Output:} %算法的结果输出
	 $\whScal=\{i:\pf_i\in\Rcal_{-1}\}\cap\Mcal_{t}$. 
\end{algorithm}

%We now introduce some possible choices of partial order and discuss their influences.

The JM procedure encompasses various variants that rely on different choices of partial orders. Below we introduce three specific partial orders, each exhibiting a distinct influence on the determination of the subsequent hypothesis to be unveiled.
\begin{enumerate}%[topsep=0pt,itemsep=0pt,parsep=0pt]
    \item The JM.Max procedure employs the infinity norm to order $K$-dimensional vectors: $\tf\prec \tf^\prime$ iff $|\tf|_{\infty}<|\tf^\prime|_{\infty}$ for $\tf,\tf^\prime\in\Rb^K$, where $|\tf|_{\infty}=\max_{k} t_k$. It should be noted that JM.Max requires the rejection region to take the form of a cube, specifically $\{[0,s]^K:s\in[0,1]\}$. In cases where there are no ties in the set $\{|\tpf_i|_{\infty}\}_{i=1}^m$, the maximal set $\Mcal_t^{maximal}$ contains only a single element at each step.

    \item The JM.Product procedure utilizes the product order: $\tf\prec \tf^\prime$ iff $t_k\leq t^\prime_k$ for $k=1,\ldots,K$ and $\tf\not=\tf^\prime\in\Rb^K$. In this case, the maximal set $\Mcal_t^{maximal}$ often contains multiple elements at each step. Therefore, both the partial order and the kernel method contribute to determining the next hypothesis to be revealed.
    
    \item The JM.EmptyPoset procedure assumes that the set of partial order relations $\prec$ is empty, meaning that any two vectors are non-comparable. In this case, the procedure simplifies to the JM approach solely utilizing the feasible sequential unmasking rule.
\end{enumerate}

The different performances of the three JM procedures reflect the impact of the choice of partial ordering. Given a partial order and setting the target FDR level to zero, the full unmasking order can be determined. The later a hypothesis is revealed, the more likely it is under the alternative.  Algorithm~\ref{alg:JMposet} induces a topological order $\prec_t$ of $(\{i:\pf_i\in\Rcal_{0}\},\prec)$ such that $a\prec b$ implies $a\prec_t b$, where $\prec_t$ is a strict total order that is irreflexive, asymmetric, transitive, and additionally connected: 
\begin{enumerate}%[topsep=0pt,itemsep=0pt,parsep=0pt]
    \item[(d)] Connectivity: for any $a\not= b\in\Pcal$, either $a\prec_t b$ or $b\prec_t a$.
\end{enumerate}
% that satisfies  \citep{Kahn1962}. {\color{red} What are topological order and total order?} 
The induced order, derived from the JM procedure, can be used as the prior information in subsequent research by ranking $i$ before $j$ if $i\prec_t j$ \citep{GSell2016,Li2017,Ang2018}. The real data example in Section \ref{sec:crohn} shows that the ordering generated by JM.Product is empirically superior to the other procedures; see Figure~\ref{fig:CrohnGWAS:b}. 
%Finally, the utilization of partial order enables us to explicitly specify the rejection region that corresponds to the JM procedure with the sequential unmasking rule.

\begin{remark}[Partial-order-assisted rejection region]\label{rmk:poarr}
\rm Algorithm~\ref{alg:JMposet} can be seen as a specific implementation of Algorithm~\ref{alg:JM}, where the rejection region $\Rcal_t$ at step $t$ is chosen as follows: $\Rcal_t=\{\tf\in[0,1]^K:\tf\prec\tpf_i \text{ for some } i\in\Mcal_t\}\cup\tPcal_t$. In other words, the rejection region comprises the masked p-value vectors in the masked set $\tPcal_t$ and all points in $[0,1]^K$ that are considered ``less than'' at least one element of $\tPcal_t$ based on the partial ordering. This insight enables us to define the rejection region assisted by the partial order.
The bottom-left corners in Figure~\ref{fig:illustration} correspond to the rejection regions induced by JM.Max and JM.Product, where nominal FDR levels are selected as $\{0.05,0.1,0.2,0.3\}$.
\end{remark}

\section{Simulation Studies}\label{sec:simu}
We evaluate the performance of the JM procedure in the mediation analysis and the reproducible study. As described in Section~\ref{sec:poosequnmaskrule}, three variants of the JM procedure, namely JM.Max, JM.Product and JM.EmptyPoset, are implemented in the simulation studies below. Throughout, we consider the following three criteria, i.e., 
$$
\FDP = \frac{|\whScal\cap \Hcal_0|}{|\whScal|\vee 1},\quad 
\mFDP =
\frac{
\sum_{\kappa=1}^K \kappa|\whScal\cap\Hcal^{(\kappa)}|
}{|\whScal|\vee 1},\quad
\text{and}\quad
\mathrm{power} = \frac{|\whScal\cap \Hcal_1|}{|\Hcal_1|\vee 1},
$$
to evaluate the performance of different procedures.
%We consider two target fdr level $q$: $0.05$ or $0.2$.
\subsection{Mediation analysis}\label{sec:simuMed}

In this section, we consider four popular competitors for mediation analysis: the divide-aggregate composite-null test (DACT, \citealp{Liu2021}), the joint significance test with the asymptotic mixture null distribution (JS.Mix.Asy,  \citealp{Dai2020}) and with the finite mixture null distribution (JS.Mix.Finite, \citealp{Dai2020}), and the mediation test with the composite null hypothesis (MT.Comp, \citealp{Huang2019}). 
We follow the settings in \cite{Dai2020} and compare the abilities of the aforementioned tests in assessing the mediation effect of an exposure on quantitative outcomes by independent molecular markers. Specifically, we simulate $n=250$ subjects from the following model:
$$X\sim \text{Bernoulli}(0.2),\quad M_i = \alpha_iX+\epsilon_i,\quad Y_i = \beta_iM_i+\beta_0 X+e_i, \quad i=1,\ldots,5000,$$
where for each $i$, both $\epsilon_i$ and $e_i$ are independently drawn from a standard normal distribution, $\beta_0$ is the common effect of the exposure on the outcomes, $\alpha_i$ is the effect of the exposure on the $i$th molecular marker, and $\beta_i$ corresponds to the effect of the $i$th marker on the $i$th outcome. 
Our goal is to identify the markers possessing exposure-marker and marker-outcome effects simultaneously. Equivalently, we are interested in testing $\Hcal_{i0}:\alpha_i=0$ or $\beta_i=0$. 
For each marker, there are four different cases $\Hcal_{00}$, $\Hcal_{10}$, $\Hcal_{01}$, and $\Hcal_{11}$: Under $\Hcal_{00}$, $\alpha_i=0$ and $\beta_i=0$; under $\Hcal_{10}$, $\alpha_i=0.25$ and $\beta_i=0$; under $\Hcal_{01}$, $\alpha_i=0$ and $\beta_i=0.375$; under $\Hcal_{11}$, $\alpha_i=0.25$ and $\beta_i=0.375$. For the common effect, we take $\beta_0=0.3$. Let $(\pi_{00},\pi_{10},\pi_{01},\pi_{11})$ be the proportions for each case. We consider two different values for $\pi_{00}$, namely $\pi_{00}=0.4$ or 0.88. Fixing $\pi_{00}$, we set $\pi_{11}=\tilde{\pi}_{1}(1-\pi_{00})$ with $\tilde{\pi}_{1}$ varying from zero to one, and $\pi_{01}=\pi_{10}=(1-\pi_{00}-\pi_{11})/2$. We obtain 5,000 pairs of p-values for testing $\alpha_i=0$ and $\beta_i=0$
(by regressing $M_i$ on $X$, and $Y_i$ on $(M_i, X)$) separately, and compare the JM procedure with the other multiple testing procedures designed for mediation analysis based on the calculated p-values. 

\begin{figure}
    \centering
    \includegraphics[width=0.8\textwidth]{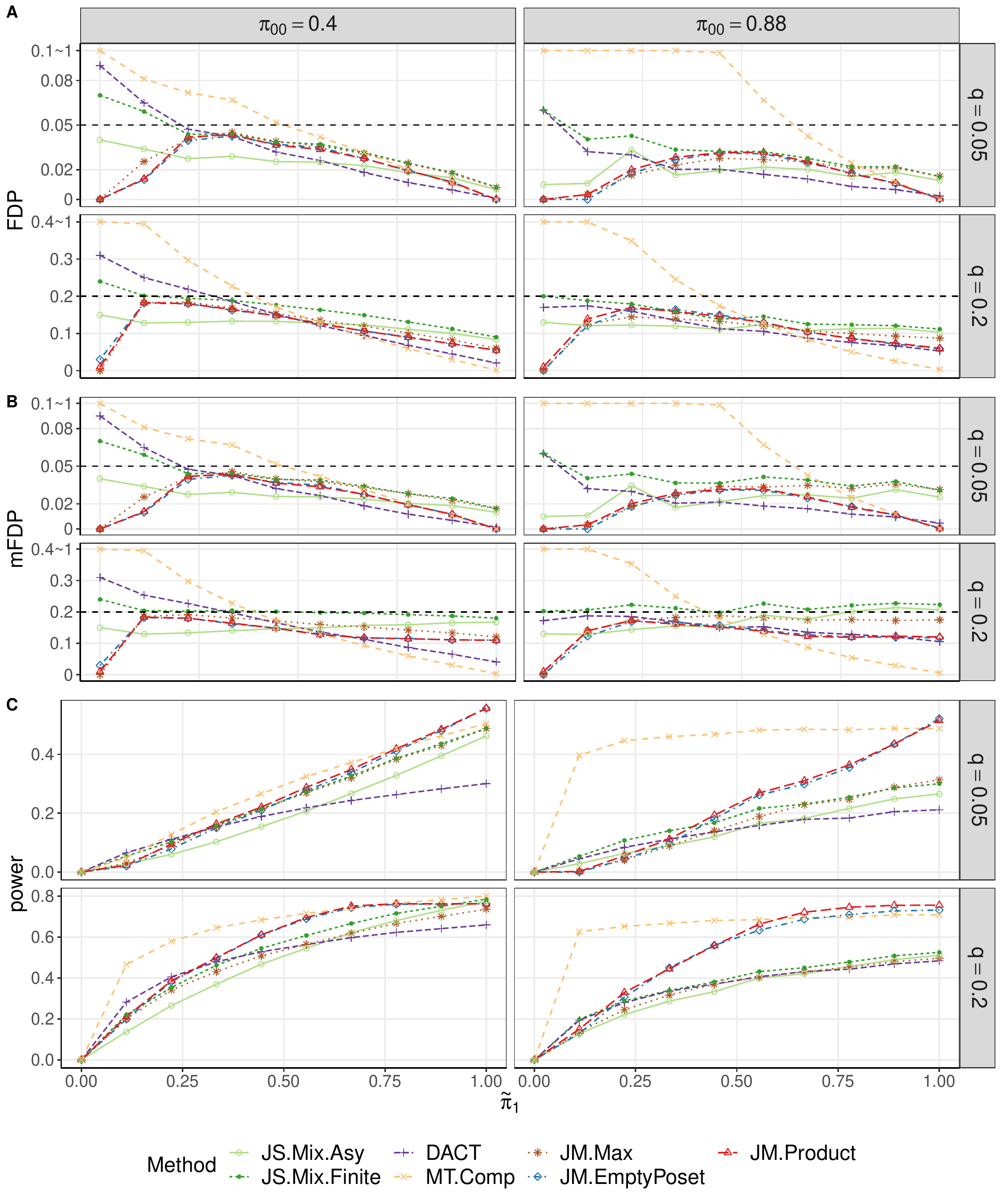}
        \caption{The FDPs, mFDPs, and powers of various methods in mediation analysis. The results are based on 100 independent replications. %{\color{red} Use ``power'' as the labels?}
        }
        \label{fig:StudyMed}
\end{figure}

Figure~\ref{fig:StudyMed} reports the FDPs, mFDPs, and powers when the target FDR levels $q$ are $0.05$ and $0.2$, respectively. Overall, JM.EmptyPoset and JM.Product deliver the highest power while controlling the empirical FDR, as well as the empirical mFDR, under the target FDR level. MT.Comp has severe FDR inflation when $\tilde{\pi}_1$ is small. The FDPs of DACT and JS.Mix.Finite are slightly inflated when $\tilde{\pi}_1$ is close to zero and $\pi_{00}=0.4$. 
The core idea of JS.Mix.Finite and JS.Mix.Asy is to approximate the mixture null distribution. JS.Mix.Finite realizes a better approximation by performing a finite-sample adjustment and hence delivers higher power. 
The power of JS.Mix.Finite is comparable to that of JM.Max when $q=0.05$ and $\tilde{\pi}_1$ is large. 
The power improvement of JM.EmptyPoset and JM.Product, compared to JM.Max owns to the better revealing ordering induced by the sequential unmasking rule: hypotheses likely to be under the null are revealed and excluded from the mask set earlier.
Additional simulation studies are provided in Section~\ref{sec:AddNumerical} of the supplement.

\subsection{Replicability analysis}
We evaluate the performance of various methods in detecting simultaneous signals in replicability analysis, where a hypothesis $i$ is considered under the alternative if and only if it is under the alternative in all $K$ experiments. Besides the three JM procedures (i.e., JM.Max, JM.EmptyPoset, and JM.Product), we also consider the BH procedure using the Bonferroni adjusted p-values (BonferroniBH, \citealp{Benjamini2008}), %replicability analysis with false discovery control (repfdr, \citealp{Heller2014,Heller2014repfdr}) {\color{red}fdr is highly inflated}, 
the adaptive filtering multiple testing procedure (AdaFilter, \citealp{Wang2022}), the conditional partial conjunction test based on the BH procedure (cPC, \citealp{Dickhaus2021}), and the simultaneous signal analysis approach (ssa, \citealp{Zhao2020}) in this numerical study. 
 
Our data-generating process is similar to that in \cite{Wang2022} with slight modification. We set $m=10,000$ and consider the number of experiments $K\in \{2,4,6,8\}$. 
Denote by $\pi^{\rm global}_{0}$ and $\pi_1$ the probabilities of being global null (all $K$ experiments are under the null) and alternative, respectively. 
Other null types i.e., $\{\btheta\in\{0,1\}^K:1\leq \sum_{k=1}^K\theta_{k}<K\}$ have equal probabilities, adding up to $1-\pi_0^{\rm global}-\pi_1$. 
We mimic the typical null and signal proportions in genetic studies by considering two values for $\pi_1$: $0.01$ and $0.03$, and two values for $\pi_0^{\rm global}$: $0.8$ and $0.95$. 

Given the latent state $\btheta_i$, we then generate the z-values from which we calculate the corresponding p-values. 
The z-values are independently sampled for the $K$ experiments. For the $i$th feature and the $k$th experiment, we generate $\mu^{0}_{ki}$ uniformly from $\{\pm3,\pm4,\pm5\}$ if $\theta_{ki}=1$ and let $\mu^0_{ki}=0$ otherwise. To allow the signal strength to vary across different experiments, we set the mean of the z-value $\mu_{ki}$ to be $\mu^0_k\{2-w_0-2k(1-w_0)/K\}$ for $1\leq k\leq K$, where $w_0\in \{0.5,1\}$. 
Finally, the correlation structure of the $m$ z-values within the same experiment is set as $I_{b\times b} \otimes \Sigma_{\rho}$, where $\otimes$ denotes the Kronecker product, $b$ is the number of blocks, and $\Sigma_{\rho}\in\Rb^{m/b\times m/b}$ is compound symmetry with the diagonal elements being 1 and the other elements being $\rho=0.5$. We consider two levels of dependency, i.e., $b=100$ for weak dependence and $b=10$ for strong dependence. 

\begin{figure}
     \centering
    \includegraphics[width=0.8\textwidth]{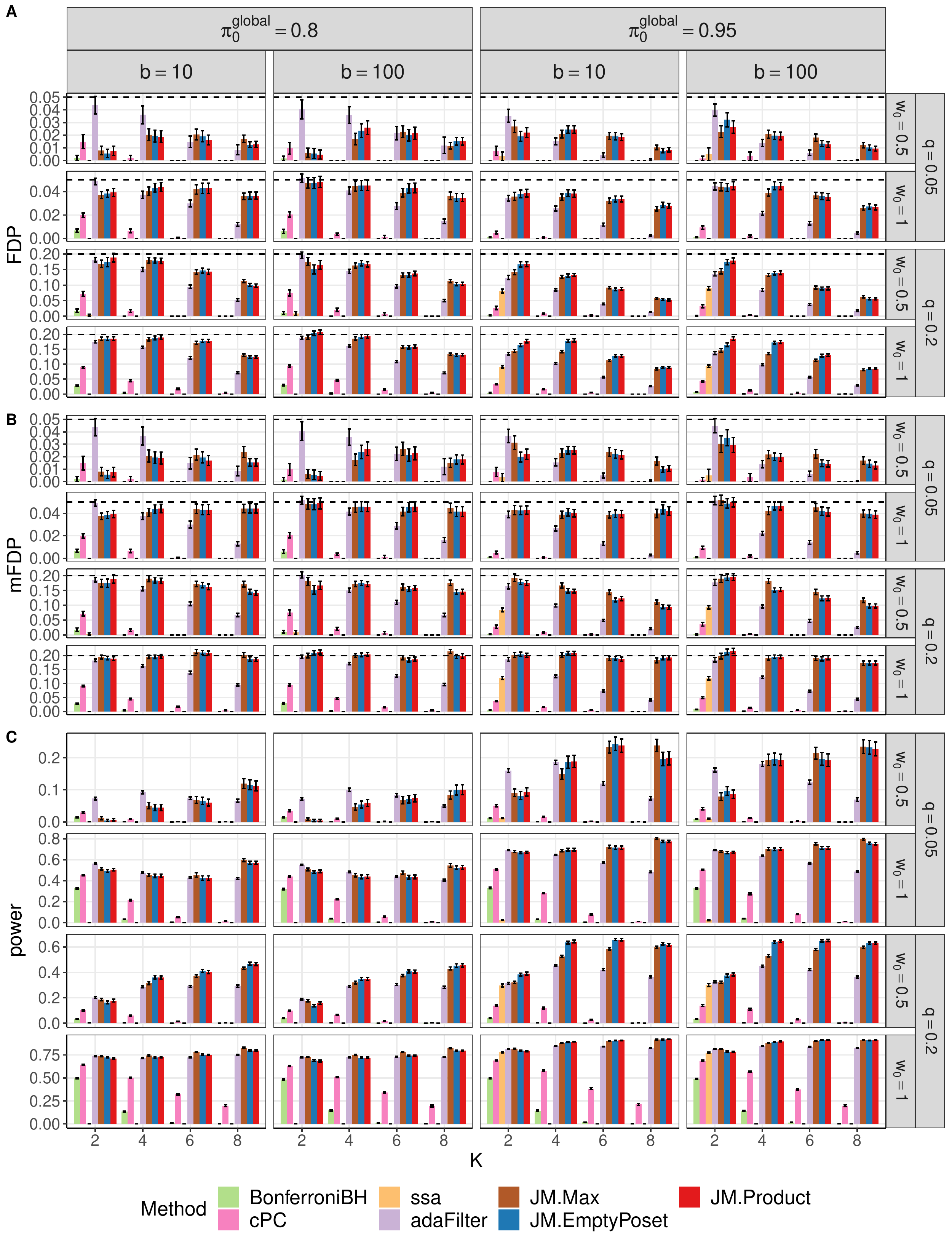}
        \caption{The FDPs, mFDPs, and powers of various methods in replicability analysis when $\pi_{1}=0.01$. The results are based on 100 independent replications.}
        \label{fig:StudyRep:1}
\end{figure}

\begin{figure}
     \centering
    \includegraphics[width=0.8\textwidth]{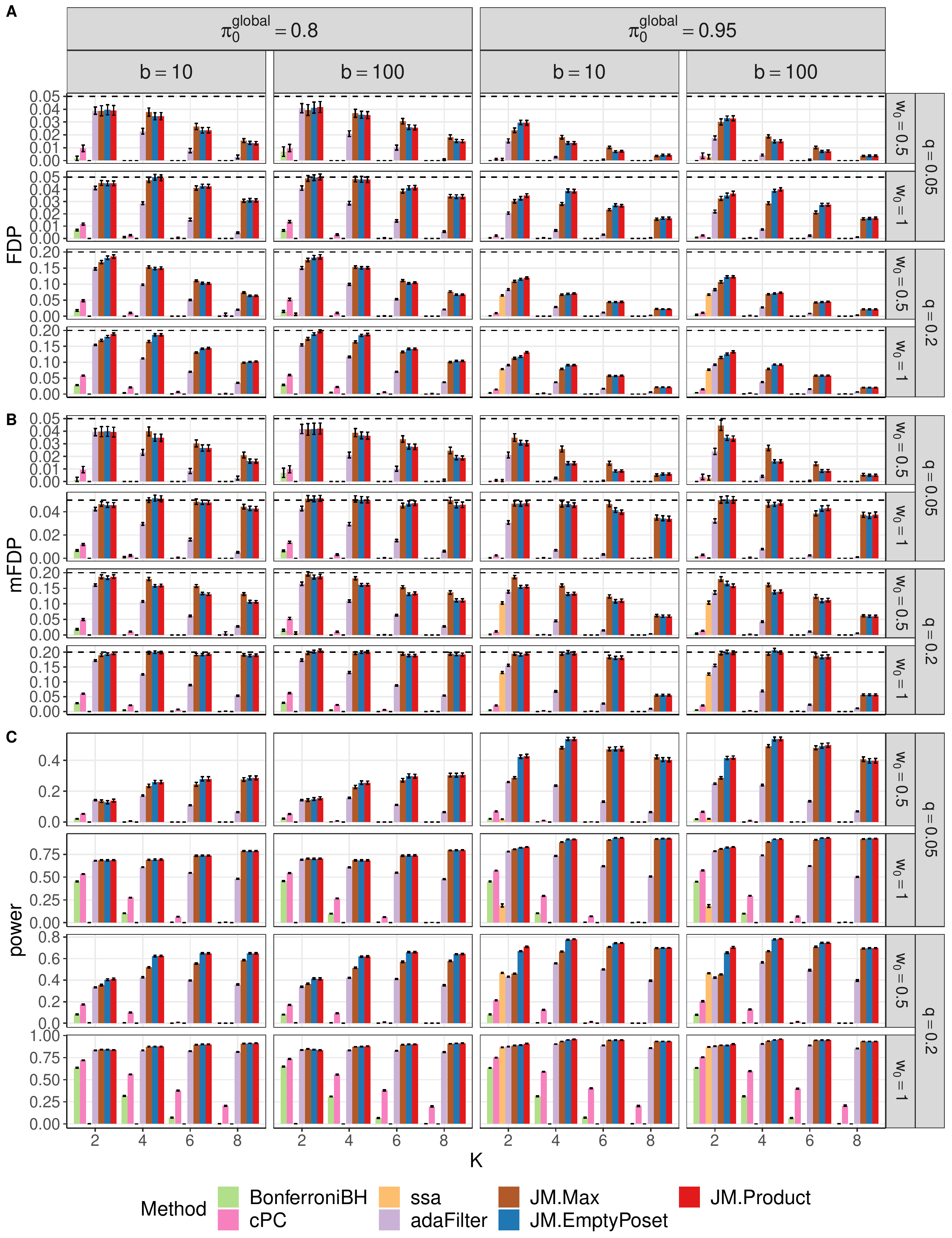}
    
        \caption{The FDPs, mFDPs, and powers of different methods in replicability analysis when $\pi_{1}=0.03$. The results are based on 100 independent replications.}
        \label{fig:StudyRep:2}
\end{figure}

Figures~\ref{fig:StudyRep:1} and \ref{fig:StudyRep:2} present the FDPs, mFDPs, and powers with the target FDR levels $q$ being $0.05$ and $0.2$. All methods successfully control the empirical FDR %{\color{red} This is not accurate. FDP is a random quantity. It is better to say empirical FDR, which is the average of the FDP over different simulation runs},
as well as the empirical mFDR, under the target FDR level. The empirical mFDR of the three JM procedures is close to the target level when $w_0=1$ and $K\leq 6$. As for the power, JM.EmptyPoset and JM.Product deliver the highest power when $\pi_1=0.03$. JM.Max discovers more signals than adaFilter when $K\geq 6$, which indicates that the FDP estimator in \eqref{equ:FDP.est} is less conservative when the number of experiments is large. JM.EmptyPoset and JM.Product are generally more powerful than JM.Max when $w_0=0.5$, which shows the usefulness of the sequential unmasking rule. Since the JM procedure is impossible to make fewer than $1/q$ rejections due to the form of our FDP estimator, adaFilter can outperform the three JM procedures when $q=0.05$ and $\pi_1=0.01$ (i.e., when alternative hypotheses are very sparse). 

\section{Real Data Applications}\label{sec:app}
\subsection{Cigarate Smoking, DNA methylation, and Lung Diseases}\label{sec:dna}
Recent research has identified several CpG sites whose DNA methylation mediates the negative effects of cigarette smoking on lung functions \citep{Zhang2016,Barfield2017,Sun2021}. We demonstrate the effectiveness of the JM procedure in detecting DNA methylation CpG sites being the mediators of smoking on lung diseases using the dataset studied in \cite{Liu2021}. The original dataset came from the Normative Aging Study, including the questionnaires, follow-up visits, and diagnostic data \citep{Bell1972}, as well as the DNA methylation on $m=484,613$ CpG sites measured using the Illumina Infinium HumanMethylation450 Beadchip on blood samples \citep{Bibikova2011}. The processed data include the estimated coefficients for the association between smoking and DNA methylation (denoted by $\gamma$), the association between the DNA methylation and lung function (denoted by $\beta$), and their standard errors and the corresponding p-values (denoted by $p_{\gamma}$ and $p_{\beta}$ respectively).

\begin{figure}[!h]
     \centering
    \includegraphics[width=0.95\textwidth]{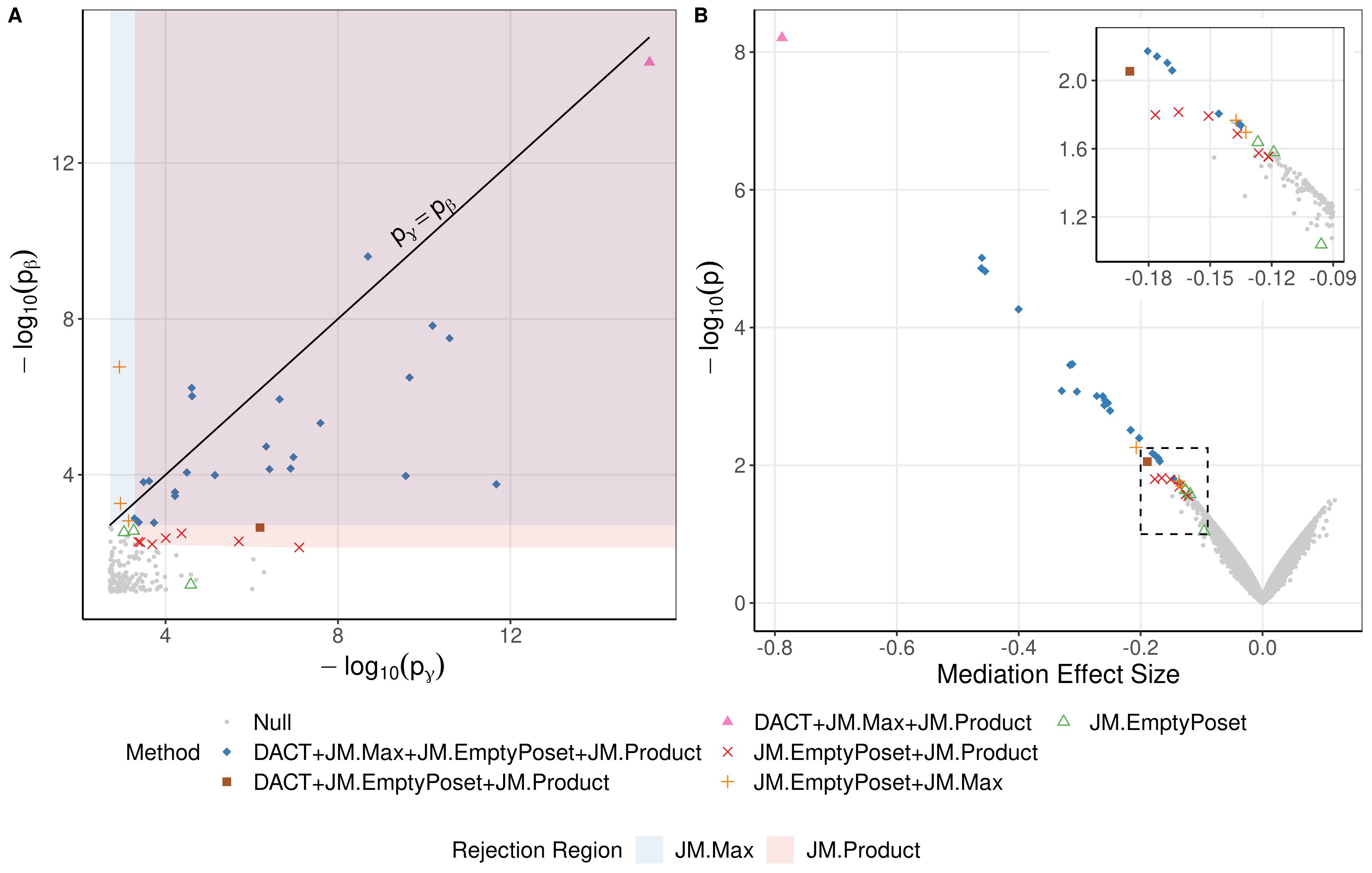}
    \caption{Panel A presents the pairs of p-values $(p_{\gamma},p_{\beta})$ with $p_\gamma\leq 0.002$ and $p_\beta\leq 0.1$ on the $-\log_{10}$ scale. The red and blue regions represent the rejection regions of JM.Product and JM.Max, respectively. Panel B depicts the volcano plot for the Sobel test, where the horizontal axis is the mediation effect size, and the vertical axis is the p-value calculated using the Sobel test. Points with different colors and shapes are the discoveries identified by different (combinations of) methods.}
        \label{fig:NAS}
\end{figure}

We applied the DACT method along with three variants of the JM procedure (JM.Max, JM.Product, and JM.EmptyPoset) to the dataset, setting the target FDR level at $0.2$. The discoveries made by each method are illustrated in Figure~\ref{fig:NAS}. In particular, DACT, JM.Max, JM.Product, and JM.EmptyPoset identified 21, 27, 32, and 37 CpG sites that mediate the effect of smoking on lung diseases, respectively. This result shows that the three JM procedures discovered more significant CpG sites compared to DACT. Furthermore, in Panel A, though finding the most sites, JM.EmptyPoset ignores the most significant CpG site (the pink solid triangle), which emphasizes the importance of including a partial order structure. Compared to JM.Max, JM.Product prefers the sites with the weaker association between DNA methylation and lung function but with a stronger association between smoking and DNA methylation. In fact, most points are below the 45-degree line, indicating a stronger association between smoking and DNA methylation. The rejection region of JM.Product also prefers a stronger smoking-methylation effect. Panel B shows that JM.Product and JM.EmptyPoset identify several CpG sites with large mediation effects that were not discovered by the other methods. 

\subsection{Genome-wide association study for Crohn’s Disease}\label{sec:crohn}

In this section, we study a data set\footnote{The data are available at \href{https://www.ibdgenetics.org}{ International Inflammatory Bowel Disease Genetics Consortium.}} previously analyzed in \cite{Franke2010} by applying the JM procedure to identify SNPs exhibiting replicable associations with the Crohn’s disease. In each experiment, the data comprises z-values for testing the association between SNPs and Crohn's disease. After retaining the SNPs that possess observations in all eight experiments, we have a $953,154\times 8$ data matrix corresponding to $953,154$ SNPs from 8 experiments. For each SNP, we test the null hypothesis that the SNP has no association with Crohn's disease in at least one experiment. 

\begin{figure}
     \centering
    \subfigure[UpSet plot\label{fig:CrohnGWAS:a}]{
    \includegraphics[width=0.95\textwidth]{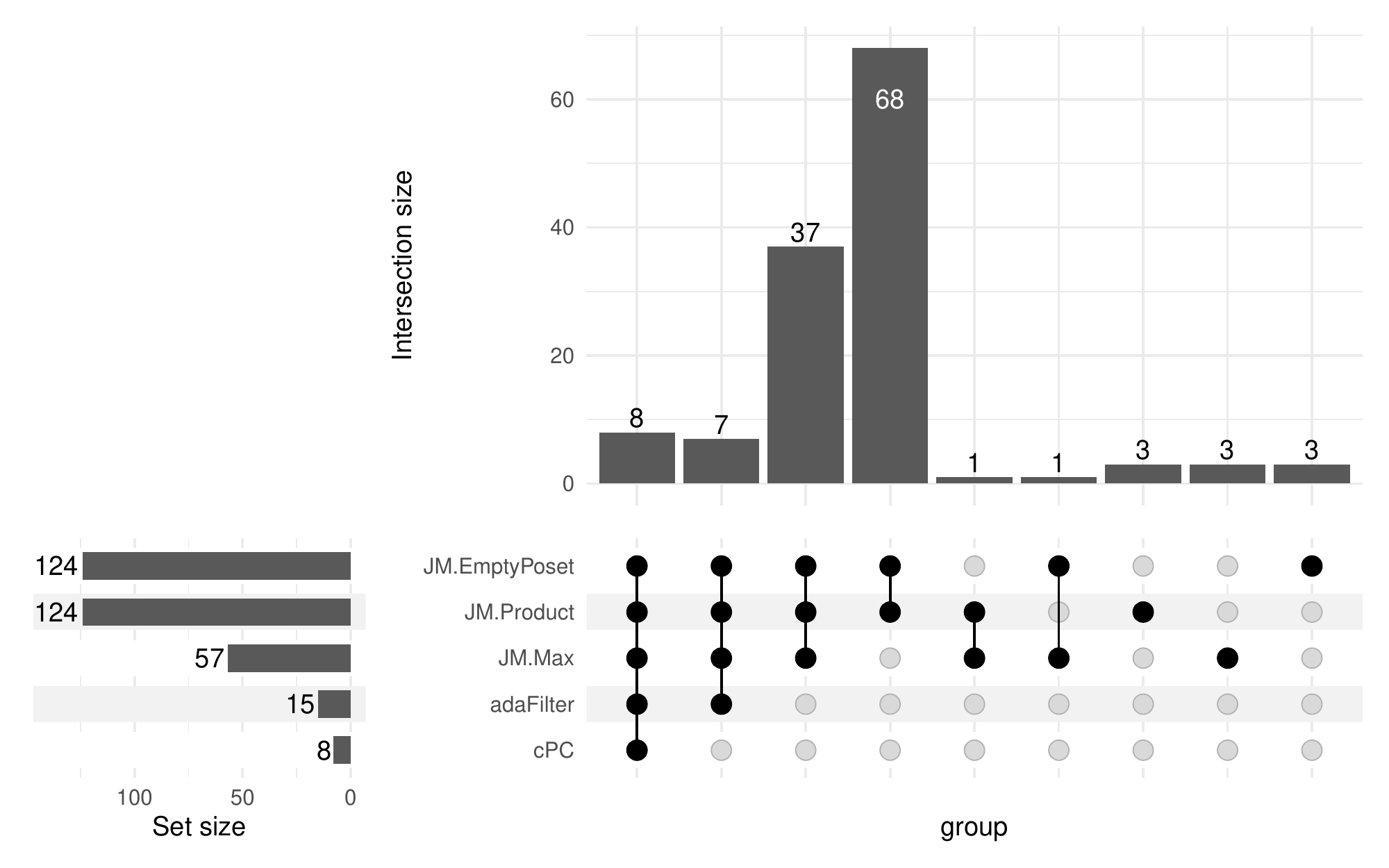}}
    \subfigure[Cumulative p-values.\label{fig:CrohnGWAS:b}]{
    \includegraphics[width=0.95\textwidth]{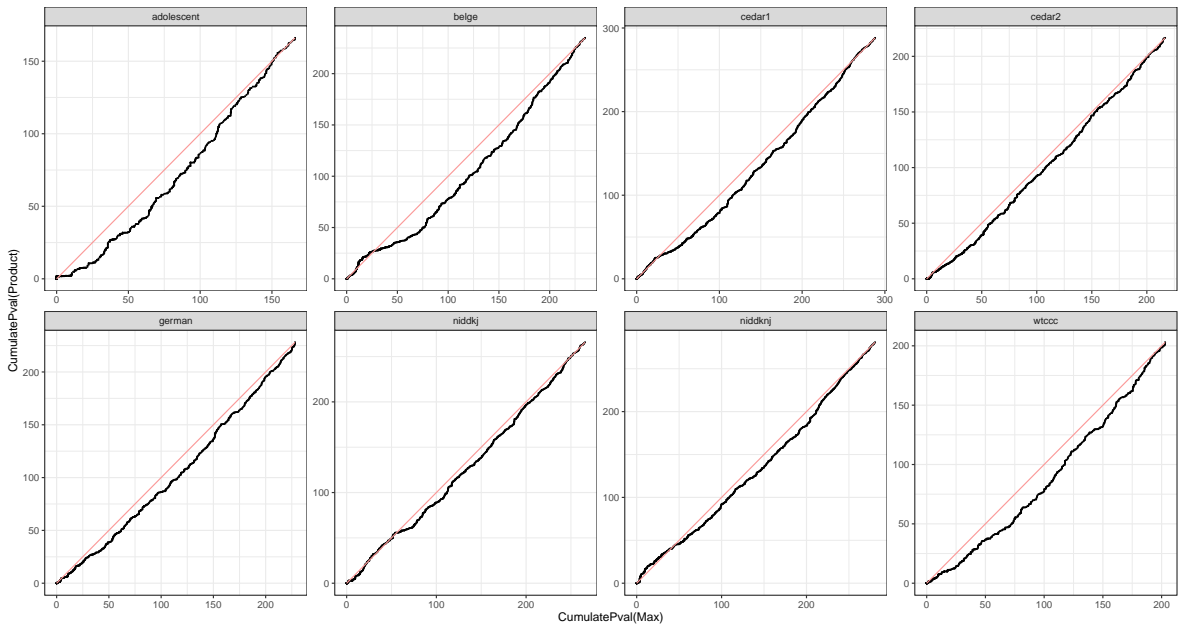}}
        \caption{Illustrations of Crohn's disease GWAS.}
        \label{fig:CrohnGWAS}
\end{figure}

Figure~\ref{fig:CrohnGWAS:a} presents the testing results for different methods when the target FDR level is $0.05$. Both JM.Product and JM.EmptyPoset discovered 124 Crohn's disease-related SNPs. In contrast, JM.Max found 57 SNPs, adaFilter identified 15 SNPs, while cPC only found eight SNPs. Note that JM.Product and JM.EmptyPoset additionally identified 68 significant SNPs that were not found by the other methods. 

To validate the findings by the JM procedure, we leave one of the eight studies out and implement JM.Product using the p-values from the remaining seven studies. We rank the SNPs based on their rejection orders in the JM procedure: a more significant SNP requires a lower target FDR level to be rejected and has a lower rank. 
As a comparison, we also rank the SNPs according to the maximum p-values of the remaining seven studies. 
Ranking using the maximum p-values is consistent with those using the
Boferrorni test \citep{Benjamini2008}, the adaFilter procedure \citep{Wang2022}, and the conditional partial conjunction test \citep{Dickhaus2021}. Motivated by the ordered testing problem \citep{GSell2016,Li2017,Ang2018}, we plot the cumulative sum of the p-values for the leave-one-out study based on these two different rankings in Figure~\ref{fig:CrohnGWAS:b}. We observe that the cumulative sum of the p-values based on the ranking produced by JM.Product is generally lower than that by the maximum p-values. This finding suggests that the SNPs ranked higher by JM.Product are more likely to be associated with Crohn's disease in a separate study, and hence corroborates the validity of the findings by the JM procedure to some extent.

\section{Discussions}\label{sec:discussion}
In this paper, we have proposed the JM procedure, a robust and powerful multiple testing method for identifying simultaneous signals. The JM procedure is based on a new FDP estimator built upon the mirror conservatism, and it involves the following key ingredients:
 \begin{itemize}%[topsep=0pt,itemsep=0pt,parsep=0pt]
     \item a rejection region and a set of mirror regions for estimating the number of false rejections;
     \item a sequential unmasking rule that uses partially masked p-values to determine the next p-value vector to unmask;
     \item a partial order that provides additional guidance on the unmasking orders;
     \item a sequential procedure that updates the FDP estimate, the rejection region, and the control region dynamically until the FDP estimate is below the target FDR level. 
 \end{itemize}   
We showed that the JM procedure provides finite sample FDR control under PCH. Additionally, we introduce a novel concept called modified FDR, which assigns weights to false discoveries based on the number of null components associated with each discovery. Remarkably, under appropriate assumptions, we established that the JM procedure also controls the modified FDR (mFDR) in finite samples.

We briefly mention a few future research directions. Firstly, it is worth exploring a more general masking scheme or control region that can enhance the performance of the JM procedure when the number of discoveries is small or when the null p-values are concentrated around a specific value \citep{Chao2021}. Secondly, while the JM procedure currently operates on p-values, it would be valuable to extend it to handle test statistics directly and control directional FDR, which is an error measure that incorporates sign information. We provide a brief summary of these first two ideas in Section~\ref{sec:GJM} of the supplement, along with more in-depth discussions. Thirdly, incorporating side information, such as spatial information in genetic studies, has the potential to enhance the power of multiple testing procedures and warrants further investigation. Finally, since the JM procedure offers stringent error control for testing PCH, it is intriguing to develop a less conservative procedure specifically tailored to such hypotheses, while still ensuring finite sample FDR control.

%Global null break the Assumption~\ref{ass:pi0}. 

%%%%%%%%%%%%%%%%%%%%%%%
%\bibliographystyle{jasa}
% I change the bib style to suppress url
%%%%%%%%%%%%%%%%%%%%%%%
\begin{spacing}{1}
\bibliographystyle{asa}
%\bibliographystyle{rss}
% \scriptsize
%\small
\setlength{\bibsep}{1.5pt}
\bibliography{reference}
\end{spacing}

\end{document}

\title{Supplementary Material for ``Joint Mirror Procedure: Controlling False Discovery Rate for Identifying Simultaneous Signals"}
   \author{Linsui Deng$^{1}$,
  Kejun He$^{1}$\thanks{Correspondence: \href{mailto:kejunhe@ruc.edu.cn}{kejunhe@ruc.edu.cn} and \href{mailto:zhangxiany@stat.tamu.edu}{zhangxiany@stat.tamu.edu}.}~,
  Xianyang Zhang$^{2}$\footnotemark[1] %\thanks{Zhang acknowledges support from NSF DMS-2113359, NIH 1R01GM144351-01 and NIH 1R21HG011662.}
  %\\
%^{1}$The Center for Applied Statistics, Institute of Statistics and Big Data, \\ Renmin University of China, Beijing, China \\
%$^{2}$Department of Statistics, Texas A\&M University, College Station, USA\\
}
  \footnotetext[1]{Institute of Statistics and Big Data, Renmin University of China, Beijing, China}
  \footnotetext[2]{Department of Statistics, Texas A\&M University, College Station, USA}
  \date{\today} 
  \maketitle
%\small 

The supplementary material is structured as follows:
Section~\ref{sec:discussionAlg2} discusses the detailed implementation of Algorithm~\ref{alg:JMposet} in the main paper. Section~\ref{sec:proof} presents the proofs of Theorems~\ref{thm:fdr}--\ref{thm:lmfdr} and Remark \ref{rmk:poarr} from the main paper. In Section~\ref{sec:GJM}, we explore two extensions of the JM procedure. The first extension introduces a JM procedure with a general masking scheme and provides proof of its finite sample FDR control. The second extension extends the JM procedure to handle z-values directly. Section~\ref{sec:AddNumerical} includes additional numerical results to complement those presented in the main paper.

\section{Discussions on Algorithm \ref{alg:JMposet}}\label{sec:discussionAlg2}
\begin{figure}
     \centering
    \includegraphics[width=\textwidth]{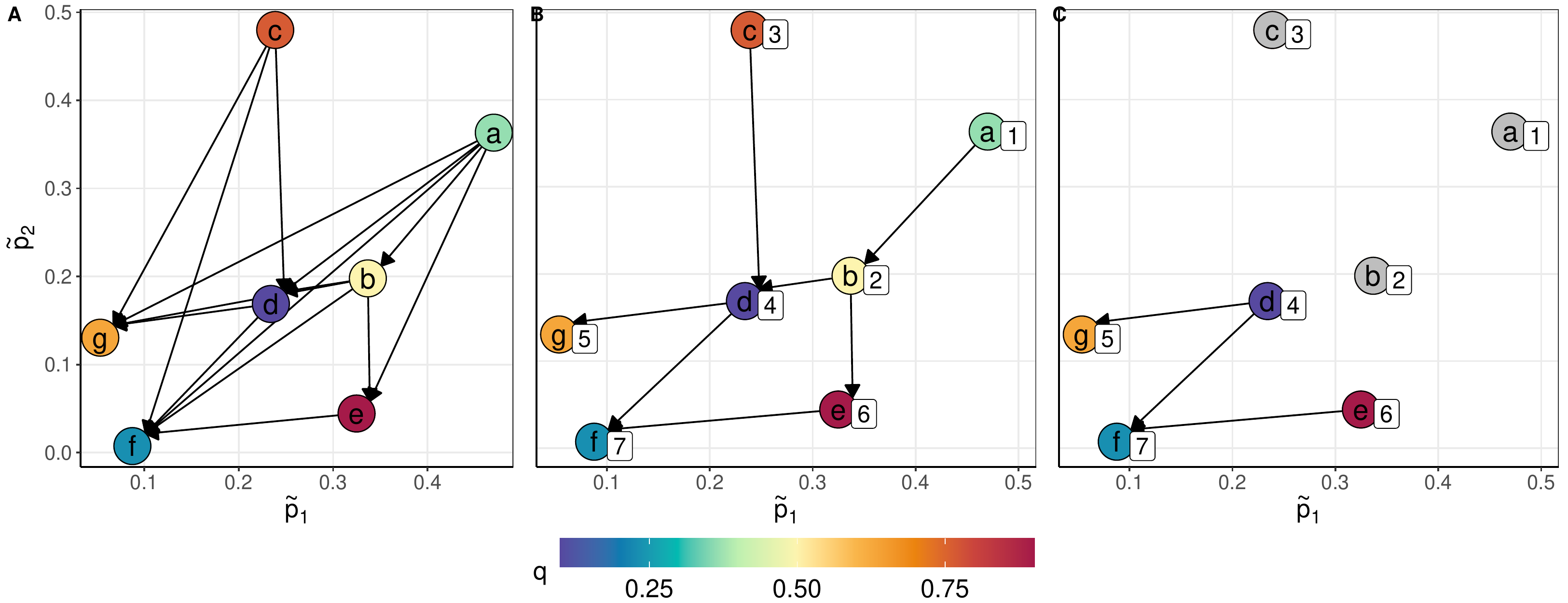}
    \caption{Panel A depicts the complete partial order set (poset). Panel B is the transitive reduction of Panel A. Colorful points and their edges in Panel C represent the transitive reduction of poset at Step 4. The color represents the conditional probability of a p-value locating at the rejection side defined in \eqref{equ:SeqUnmaskR} of the main paper. The labels in Panels B and C are the unmasking orders, e.g., vertex $a$ is the first one being revealed.}
        \label{fig:DAG}
\end{figure}

In this section, we describe how to update the maximal set $\Mcal_t^{maximal}$ in Algorithm~\ref{alg:JMposet}. We consider a poset consisting of seven masked bivariate p-values equipped with the product order $\prec$. Figure~\ref{fig:DAG} illustrates the poset $(\tPcal,\prec)$ with a directed acyclic graph $G=(V,E)$, where $V=\tPcal$ and $(b,a)\in E$ iff $a\prec b$. Panel A describes the partial order relationship completely, but the edges could include redundant information, e.g., with the edges $(b,a)$ and $(a,c)$, the edge $(b,c)$ can be inferred by the transitivity and hence can be removed. In other words, to store information efficiently, we adopt the idea of transitive reduction to remove redundant edges, which can recover the original directed acyclic graph (DAG) with the fewest edges. 
\begin{defn}[Transitive Reduction, \citealp{Aho1972}] A graph $G_t$ is a transitive reduction of a directed graph $G$ if it satisfies the following two conditions:  
\begin{enumerate}
    \item there is a directed path from $a$ to $b$ in $G$ iff there is a directed path from $a$ to $b$ in $G_t$; 
\item there is no graph with fewer edges than $G_t$ satisfying the first condition.
\end{enumerate}
\end{defn}
The computational cost of finding the transitive reduction is no more than $O(|V|^{\log_2 7})$; see \cite{Aho1972} for more details. Recall that our goal is to find the maximal set of the poset in each step, which is equivalent to finding the root set of the corresponding DAG. The process of updating the root set is described by Algorithm~\ref{alg:Kahn}, a well-known algorithm for searching a topological sorting \citep{Kahn1962}. Removing one root from the root set corresponds to revealing a hypothesis in the maximal set (line 5 of the Search Step in Algorithm~\ref{alg:JMposet} of the main paper). Updating the root set corresponds to finding the new hypotheses to be added to the maximal set (line 4 of the Search Step in Algorithm~\ref{alg:JMposet} of the main paper). The computational cost of updating the root set is approximately $O(|V|+|E|)$. Therefore, replacing the DAG with its transitive reduction can save much time. 

Finally, we exemplify the sequential unmasking rule using partial order with Figure~\ref{fig:DAG}. At Step 1, the maximal set is $\Mcal_1=\{a,c\}$. Since $q_a<q_c$, $a$ is more likely to locate at the control side, and therefore we remove it first. After removing $a$, we add $b$ into the maximal set because the in-degree of $b$ becomes zero. Then, the maximal set at Step 2 is $\Mcal_2=\{b,c\}$. We compare $q_b$ with $q_c$ to decide which one to be removed. Panel C describes the poset at Step 4, where the hypotheses $\{a,b,c\}$ have been unmasked at the previous steps. The maximal set of the remaining set is $\Mcal_{4}=\{d,e\}$. Since $q_d<q_e$, $d$ is revealed in Step 4. Similarly, Steps 5 to 7 reveal hypotheses $g$, $e$, and $f$ in order. %{\color{red} Should we use directed edges in the Figure?}

\begin{algorithm}[!h]
	\small
		\caption{Kahn's Algorithm \citep{Kahn1962}. \label{alg:Kahn}} %算法的名字
		\hspace*{0.02in} {\bf Input:} %算法的输入， \hspace*{0.02in}用来控制位置，同时利用 \\ 进行换行
	 An directed acyclic graph $G=(V,E)$;
		
		\hspace*{0.02in} {\bf Initialization:} 
		\begin{algorithmic}[1]
			\State Calculate the in-degrees $\{d_v\}_{v\in V}$ of all vertexes (i.e., the number of edges whose endpoint is $v$) ;
			\State Set the initial root set $V_r=\{v\in V: d_v=0\}$;
		\end{algorithmic}
		\hspace*{0.02in} {\bf Iterative Step:}
		\begin{algorithmic}[1]
			\While{$V_r\not=\emptyset$}
   \State Remove one root $v_r\in V_r$ from the root set;
   \State Update the in-degrees: For $v$ such that $(v_r,v)\in E$, $d_v=d_v -1$;
			\State Update the root set: For $v$ such that $(v_r,v)\in E$, add $v$ to $V_r$ if $d_v=0$.  
			\EndWhile
		\end{algorithmic}
	%Rejection region $\Rcal_{t}$.
\end{algorithm}

\begin{comment}
    
Compare to algorithm without constructing maximal set, the additional computational cost is finding the new hypothesis to be added to the maximal set. To do this, we firstly construct the transition reduction of $\tPcal_0$ using the method in \cite{Aho1972}, denoted as $(\Pcal_t^{red},\prec)$. The computation cost is dominated by the computation cost of finding the transitive closure, which is no more than $O(m^\alpha)$, for $\alpha\geq2$; see \cite{Hartmanis1971}. {\color{red}(In simulation, this step is indeed most time consuming)} Since the poset can be fully described by an acyclic directed graph through linking a directed edge $(b,a)$ if $a\prec b$, it transitive reduction is unique. The maximal set of poset is the root set of the acyclic directed graph. When removing an vertex from the root set, the vertexes possible to be added into the root set is its children. In fact, we conduct the topologic search on the  Hasse diagram.
%Intuitively, our algorithm generalizes the maximum left-right corner \citep{Barber2017,Ramdas2019,Katsevich2019}.
The order of including hypothesis induces a total order, which is actually topological sort because it is consistent with the partial order.

\end{comment}

\section{Proofs of the Theoretical Results}\label{sec:proof}
\subsection{Finite sample FDR control}
%Note that the JM procedure has two versions: JM procedure using rejection region and JM procedure unmasking rule. We prove Theorem~\ref{thm:fdr} for convenience the latter version as the two versions are equivalent. 
Throughout the proof, we fix $\kappa\in\{1,\cdots,K\}$ and consider testing the null hypothesis $\Hcal^{\kappa,partial}_0=\cup_{k=\kappa}^K\Hcal^{(\kappa)}$. Define 
\begin{equation}\label{eqn:biDef}
\begin{aligned}
    b_i&=\boldsymbol{1}\{\pf_i\in\Rcal_{-1}\},\\
    U_t&=\sum_{i\in\Mcal_{t}\cap\Hcal^{\kappa,partial}_{0}}\sum_{k=1}^K\boldsymbol{1}\{\pf_i\in\Acal^k_{-1}\}=\sum_{i\in\Mcal_{t}\cap\Hcal^{\kappa,partial}_{0}}(1-b_i),\\
    V_t&=\sum_{i\in\Mcal_{t}\cap\Hcal^{\kappa,partial}_{0}}\boldsymbol{1}\{\pf_i\in\Rcal_{-1}\}=\sum_{i\in\Mcal_{t}\cap\Hcal^{\kappa,partial}_{0}}b_i.
\end{aligned}
\end{equation}
Obviously, $U_t+V_t=|\Mcal_t\cap\Hcal^{\kappa,partial}_{0}|$. 
Recall that the filtration for determining the next hypothesis to be revealed is $\{\Fcal_t\}_{t=-1,0,\ldots}$, where $\Fcal_t=\sigma\left\{(\tpf_{t,i})_{i=1}^m,A_t,R_t,\Mcal_t,\Ucal_t\right\}
$. We further define a filtration $\{\Gcal_t\}_{t=-1,0,\ldots}$ with
$$
\Gcal_{t}=\left\{
\begin{aligned}
&\sigma\left(\{\tpf_{i}\}_{i=1}^m,\{\pf_i\}_{i\in\Hcal^{\kappa,partial}_1},\{\pf_i\}_{i\in\Ucal_{-1}\cap\Hcal^{\kappa,partial}_{0}},\Mcal_{-1},\Ucal_{-1},V_{-1},U_{-1}\right),  &\qquad t=-1,\\
&\sigma\left(\Gcal_{t-1},\{\pf_{i}\}_{i\in\Ucal_t\cap\Hcal^{\kappa,partial}_0},\Mcal_{t},\Ucal_t\right), &\qquad t\geq 0,
\end{aligned}
\right.
$$
where $\tpf_i=\Proj(\pf_i)$. 
We have that $\tau=\inf\{t:\widehat{\FDP}_t\leq q\}$ is a stopping time with respect to $\{\Gcal_t\}$ because $\widehat{\FDP}_t$ is determined by $A_t$ and $R_t$ whose $\sigma$-field is included in $\Fcal_t$
and $\Fcal_{t}\subseteq\Gcal_{t}$. We prove Theorem~\ref{thm:fdr} of the main paper using the following two lemmas. 

\begin{lemma}\label{lemma:fdr}
    Fix $\kappa\in\{1,\cdots,K\}$. Suppose that the null p-values $\{\pf_{i}\}_{i\in\Hcal^{\kappa,partial}_0}$ are mutually independent, independent of $\{\pf_{i}\}_{i\in\Hcal^{\kappa,partial}_1}$, and conditionally mirror conservative. 
    %\kejun{For ``independent of each other'', you mean mutually independent or pairwise independent?}\linsuin{mutually independent}
    %Then, we have $\Pb\left(b_i=1\mid\Gcal_{-1}\right)\leq 1/2$ for all $i\in\Hcal_0$.
    Then $\{b_i\}_{i\in\Hcal^{\kappa,partial}_0}$ defined in \eqref{eqn:biDef} are independent Bernoulli random variables conditional on $\Gcal_{-1}$ with $$\Pb\left(b_i=1\mid\Gcal_{-1}\right)\leq 1/(\kappa+1)$$
    for all $i\in\Hcal_0^{\kappa,partial}$.
\end{lemma}

\begin{lemma}\label{lemma:martingale}
    Suppose $b_1,\cdots,b_m$ are independent Bernoulli random variable conditional on $\Gcal_{-1}$ with $\Pb(b_i=1\mid\Gcal_{-1})=\rho_i\leq\rho<1$ almost surely. Consider a sequence of masked sets $\Mcal_{-1}\supseteq\Mcal_0\supseteq\cdots$. If $\Mcal_t$ is measurable with respect to $\Gcal_{t}$ and $\tau$ is an almost surely finite stopping time with respect to the filtration $\{\Gcal_t\}_{t=0,1,\ldots}$, then we have \begin{equation}\label{equ:res_lemma2}
    \Eb\left(\frac{1+|\Mcal_\tau\cap\Hcal^{\kappa,partial}_0|}{U_\tau+1}\Bigg|\Gcal_{-1}\right)\leq 1/(1-\rho).
    \end{equation}
\end{lemma}

\allowdisplaybreaks
We now present the proof of Theorem \ref{thm:fdr}.
\begin{proof}[Proof of Theorem \ref{thm:fdr}]
Similar to the derivations in Section~\ref{sec:JM} of the main paper, we have 
\begin{align*}
\FDP_\tau
&=\frac{V_\tau}{R_\tau\vee 1}
=\frac{A_\tau+1}{R_\tau\vee 1}\frac{V_\tau}{A_\tau+1} \\
&\leq 
\widehat{\FDP}_\tau
\frac{V_\tau}{U_\tau+1}\\ 
&\leq
q \frac{V_\tau}{U_\tau+1} =
q
\left(\frac{1+|\Mcal_\tau\cap\Hcal^{\kappa,partial}_0|}{U_\tau+1}-1\right),  
\end{align*}
where the two inequalities in the second and third lines hold because of $A_\tau\geq U_\tau$ and the  definition of $\tau$. Using the above inequality, Theorem~\ref{thm:fdr} immediately stands by applying Lemmas~\ref{lemma:fdr} and \ref{lemma:martingale}. More precisely,
\begin{align*}
    \FDR
& = \Eb\left(\FDP_\tau\right)
\\  & \leq
q\Eb\left(\frac{V_\tau}{U_\tau+1}\right)\\
 & =
q\Eb
\left\{\Eb
\left(\frac{1+|\Mcal_\tau\cap\Hcal^{\kappa,partial}_0|}{U_\tau+1}-1\Bigg|\Gcal_{-1}\right)\right\} \\ 
& \leq q\left(\frac{\kappa+1}{\kappa}-1\right) = q/\kappa,
\end{align*}
which completes the proof.
\end{proof}

Next, we prove Lemmas~\ref{lemma:fdr} and \ref{lemma:martingale}. 
\begin{proof}[Proof of Lemma~\ref{lemma:fdr}]
Since $\Rcal_{-1}=[0,1/2)^K$ is non-random, $\{b_i\}_{i\in\Hcal_0}$ are independent Bernoulli random variables due to the definition in \eqref{eqn:biDef}. 
    To prove the bound for the conditional probability, we first note that $b_i=0$ for $i\in\Hcal^{\kappa,partial}_0\cap\Ucal_{-1}$ by the definition of $\Ucal_{-1}$. Therefore, we only need to prove $\Pb\left(b_i=1\mid\Gcal_{-1}\right)\leq 1/(\kappa+1)$ for $i\in\Hcal^{\kappa,partial}_0\cap\Mcal_{-1}$. Since $i\in\Mcal_{-1}$ is equivalent to $\pf_i\in\cup_{k=1}^K\Acal_{-1}^k\cup\Rcal_{-1}$, it is sufficient to show  
\begin{equation}\label{equ:p_rej}
\Pb_i\left(\pf\in\Rcal_{-1}\mid \tpf=\tpf_i,\pf\in\cup_{k=1}^K\Acal_{-1}^k\cup\Rcal_{-1}\right)\leq 1/(\kappa+1)
\end{equation}
for $i\in\Hcal^{\kappa,partial}_0$, where $\Pb_{i}$ is the probability distribution of $\pf_i$. 
Fixing $i\in\Hcal^{\kappa,partial}_0$ and for any $\ttf\in\Rcal_{-1}$, it is can be shown that 
\begin{equation}\label{equ:sumtoone}
\begin{split}
    &\Pb_i\left(\pf\in\Rcal_{-1}\mid \tpf=\ttf,\pf\in\cup_{k=1}^K\Acal_{-1}^k\cup\Rcal_{-1}\right)
    \\ & \qquad \qquad +
\sum_{k=1}^K\Pb_i\left(\pf\in\Acal^k_{-1}\mid \tpf=\ttf,\pf\in\cup_{k=1}^K\Acal_{-1}^k\cup\Rcal_{-1}\right)=1,
\end{split}
\end{equation}
because $\Rcal_{-1}$ and $\{\Acal^k_{-1}\}_{k=1}^K$ are mutually exclusive and their union covers the whole space. Since the null p-values are conditionally mirror conservative, we have
\begin{align}\label{eq-mirror}
\kappa\Pb_{i}\left(\pf\in\Rcal \right)\leq 
\sum_{k\in\Scal_{0i}}\Pb_{ i}\left(\pf\in\Acal^k\right)
=\Pb_{ i}\left(\pf\in\cup_{k\in\Scal_{0i}}\Acal^k\right)
\leq 
\Pb_{ i}\left(\pf\in\cup_{k=1}^K\Acal^k\right),
\end{align}
where $\Scal_{0i}=\{1\leq k\leq K:\theta_{ki}=0\}$. 
Denote by $\Rcal_{\Delta 
}=\{\tf:\|\tf-\ttf\|_2\leq \Delta\}\cap\Rcal_{-1}$ with $\|\cdot\|_2$ being Euclidean norm and $\Acal_{\Delta}^k:=\Acal^k(\Rcal_{\Delta})$. %{\color{red} The notation is a bit confusing here. Note that $\Rcal_{\Delta}$ depends on $\ttf$ and $\Delta$. $t$ is merely a dummy variable. The subscript $\Delta t$ is confusing here}. 
Taking $\Rcal=\Rcal_{\Delta}$ and $\Acal^k=\Acal^k_{\Delta}$, then (\ref{eq-mirror}) becomes
\begin{align*}
    &\kappa\Pb_i\left(\pf\in\Rcal_{-1} ,\tpf\in\Rcal_{\Delta
},\pf\in\cup_{k=1}^K\Acal_{-1}^k\cup\Rcal_{-1}\right)\\
    & \qquad \qquad \leq 
    \sum_{k=1}^K\Pb_i\left(\pf\in\Acal^k_{-1},\tpf\in\Rcal_{\Delta
    },\pf\in\cup_{k=1}^K\Acal_{-1}^k\cup\Rcal_{-1}\right),
\end{align*}
because $\{\tpf\in\Rcal_{\Delta
},\pf\in\Rcal_{-1}\}$ and $\{\tpf\in\Rcal_{\Delta
},\pf\in\Acal^{k}_{-1}\}$ are equivalent to $\{\pf\in\Rcal_{\Delta
}\}$ and $\{\pf\in\Acal_{\Delta}^k\}$, respectively.
If $\Pb_i\left(\tpf\in\Rcal_{\Delta
},\pf\in\cup_{k=1}^K\Acal_{-1}^k\cup\Rcal_{-1}\right)>0$, then 
\begin{align*}
    &\kappa\Pb_i\left(\pf\in\Rcal_{-1} \mid \tpf\in\Rcal_{\Delta
    },\pf\in\cup_{k=1}^K\Acal_{-1}^k\cup\Rcal_{-1}\right)\\
    & \qquad \qquad \leq 
    \sum_{k=1}^K\Pb_i\left(\pf\in\Acal^k_{-1}\mid\tpf\in\Rcal_{\Delta
    },\pf\in\cup_{k=1}^K\Acal_{-1}^k\cup\Rcal_{-1}\right).
\end{align*}
Letting $\|\Delta\|_2\rightarrow0$ and taking $\ttf=\tpf_i$, we have 
\begin{equation}\label{equ:RAcomp}
    \begin{aligned}
        &\kappa\Pb_i\left(\pf\in\Rcal_{-1}\mid \tpf=\tpf_i,\pf\in\cup_{k=1}^K\Acal_{-1}^k\cup\Rcal_{-1}\right)
    \\ & \qquad \qquad \leq
    \sum_{k=1}^K \Pb_i\left(\pf\in\Acal^{k}_{-1}\mid \tpf=\tpf_i,\pf\in\cup_{k=1}^K\Acal_{-1}^k\cup\Rcal_{-1}\right).
    \end{aligned}
\end{equation}
Finally, we can obtain \eqref{equ:p_rej} by combining the above formula with \eqref{equ:sumtoone}.
\end{proof}

\begin{proof}[Proof of Lemma~\ref{lemma:martingale}]
    This lemma holds by slightly modifying the proof of Lemma~2 in \cite{Lei2016}. We highlight the key difference here and omit the detailed proof. %Without loss of generality, we can assume that $\Pb(b_i=1\mid\Gcal_{-1})=\rho$.
    Conditional on $\Gcal_{-1}$, we can construct conditionally independent random variables $\{\tb_i\}$ and $\{\bar{b}_i\}$ such that $\bar{b}_i\sim\text{Bernoulli}(\rho)$, $\tb_i\sim\text{Bernoulli}(\rho_i)$ and $\tb_i\leq \bar{b}_i$ almost surely. By construction, the conditional distribution of $(\tb_1,\cdots,\tb_m)$ is identical to that of $(b_1,\cdots,b_m)$. Then, LHS of \eqref{equ:res_lemma2} can be bounded by
    $$\Eb\left(\frac{1+|\Mcal_\tau\cap\Hcal^{\kappa,partial}_0|}{\sum_{i\in\Mcal_{t}\cap\Hcal^{\kappa,partial}_{0}}(1-\tb_i)+1}\Bigg|\Gcal_{-1}\right)
    \leq \Eb\left(\frac{1+|\Mcal_\tau\cap\Hcal^{\kappa,partial}_0|}{\sum_{i\in\Mcal_{t}\cap\Hcal^{\kappa,partial}_{0}}(1-\bar{b}_i)+1}\Bigg|\Gcal_{-1}\right).
    $$
    In this way, we can work with $\bar{b}_i$ whose conditional distributions are identical.
    Interested readers are referred to the proof of Lemma 2 in \cite{Lei2016} for more details about the construction.
    
    %{\color{red} transform the proof into constructed random variables that are conditionally independent and identically follows $\Ber(\rho)$. This is not clear if the readers did not read \cite{Lei2016}.} 
    
    A key difference compared to \cite{Lei2016} is that %(\rNum{1}) $b_i=1$ denotes the event that the p-value is in the rejection region rather than in the control regions; (\rNum{2}) 
    $\Gcal_t$ includes the information of $\{\pf_i\}_{i\in\Ucal_{t}\cap\Hcal^{\kappa,partial}_0}$ in our case. 
    %$\{b_i\}_{i\in\Ucal_{t}\cap\Hcal^{\kappa,partial}_0}$ with the information about . 
    %For (\rNum{1}), we can simply change $b_i$ to $1-b_i$ to be consistent with the notation in \cite{Lei2016}. For (\rNum{2}), 
    Knowing $b_i$ and $\tpf_{i}$ is sufficient to recover $\pf_i$ in \cite{Lei2016}. However, we also need to know the mirror region in which the p-value $\pf_i$ locates to recover $\pf_i$. The proof in \cite{Lei2016} is still valid with this modification because $\{\Gcal_{k}\}_{k=-1,0,\cdots}$ is a filtration and $\Pb(b_i=1\mid\Gcal_{t})=V_t/(U_t+V_t)$ for $i\in\Ucal_t\cap\Hcal_0^{\kappa,partial}$.  
    
    %{\color{red} $b_i$'s are not identically distributed. Would that be an issue for showing $\Pb(b_i=1\mid\Gcal_{t})=V_t/(U_t+V_t)$?} 
\end{proof}
\subsection{Finite sample mFDR control}\label{sec:mFDRproof}
In this section, we prove Theorem~\ref{thm:mfdr}. 
\begin{proof}[Proof of Theorem~\ref{thm:mfdr}]
    The proof is inspired by the leave-one-out technique introduced in \cite{Barber2020}. In our context, the leave-one-out p-value vectors are defined as $\{\pf_{(-i)},\tpf_i\}:=\{\pf_1,\cdots,\pf_{i-1},\tpf_i,\pf_{i+1},\cdots,\pf_m\}$ where the $i$th p-value $\pf_i$ vector is replaced by $\tpf_i=\Proj(\pf_i)$. 
 %and provide an equivalent expression for JM procedure when $\Rcal_t$ is chosen subject to (i) the shrinkage principle and (ii) $\Rcal_{t}\in\Fcal_{-1}$
 
    To begin with, we introduce some notation. Given $(\tpf_{i})_{i=1}^m$, the candidate rejection regions are denoted by $\{\Rcal_t\}_{t=-1,0,1,\cdots}$. We express the FDP estimator and rejection region associating with the p-values explicitly as follows:
    \begin{itemize}
        \item the FDP estimator: $\widehat{\FDP}^{\{\pf_i\}_{i=1}^m}(\Rcal)$;
        \item the final rejection region: $\Rcal^{\{\pf_i\}_{i=1}^m}=\cup_{t}\{\Rcal_t: \widehat{\FDP}^{\{\pf_i\}_{i=1}^m}(\Rcal_t)\leq q\}$, which is equivalent to $\Rcal_{\tau^{\{\pf_i\}_{i=1}^m}}$, where $\tau^{\{\pf_i\}_{i=1}^m}=\inf\{t>0:\widehat{\FDP}^{\{\pf_i\}_{i=1}^m}(\Rcal_t)\leq q\}$.
    \end{itemize}
    Similar notation can be defined for the leave-one-out p-values $\{\pf_{(-i)},\tpf_i\}$. As $\Rcal_t\in\sigma\{(\tpf_i)_{i=1}^m\}$, the candidate rejection sets $\{\Rcal_t\}_{t=-1,0,1,\cdots}$ for $\{\pf_{(-i)},\tpf_i\}$ remain unchanged. For simplicity, we denote:  
    \begin{itemize}
        \item the FDP estimator based on $\{\pf_{(-i)},\tpf_i\}$ as $\widehat{\FDP}^{(-i)}(\Rcal):=\widehat{\FDP}^{\{\pf_{(-i)},\tpf_i\}}(\Rcal)$;
        \item the final rejection region as $\Rcal^{(-i)}:=\Rcal^{\{\pf_{(-i)},\tpf_i\}}=\cup_{t}\{\Rcal_t: \widehat{\FDP}^{(-i)}(\Rcal_t)\leq q\}$.
    \end{itemize}
    Let $\kappa_i=|\Scal_{0i}|$. The mFDP, defined in Definition \ref{def:mfdr}, can be equivalently expressed as
    $$\mFDP = 
\frac{
\sum_{\kappa=1}^K \kappa|\whScal\cap\Hcal^{(\kappa)}|
}{|\whScal|\vee 1} 
=\frac{\sum_{i\in\Hcal_0}\kappa_i\boldsymbol{1}\{\pf_i\in\Rcal_{-1},\tpf_i\in\Rcal^{\{\pf_i\}_{i=1}^m})\}}{\sum_{i=1}^m\boldsymbol{1}\{\pf_i\in\Rcal_{-1},\tpf_i\in\Rcal^{\{\pf_i\}_{i=1}^m})\}\vee 1},
$$
where $i$th hypothesis is selected whenever $\pf_i\in\Rcal_{-1}$ and $\tpf_i\in\Rcal^{\{\pf_i\}_{i=1}^m}$. By the definition of $\Rcal^{\{\pf_i\}_{i=1}^m}$, the mFDP can be bounded by
    $$
\begin{aligned}
\mFDP=&
\frac{1+\sum_{i=1}^m\boldsymbol{1}\{\pf_i\in\Acal_{-1},\tpf_i\in\Rcal^{\{\pf_i\}_{i=1}^m}\}}{\sum_{i=1}^m\boldsymbol{1}\{\pf_i\in\Rcal_{-1},\tpf_i\in\Rcal^{\{\pf_i\}_{i=1}^m}\}\vee 1}\frac{\sum_{i\in\Hcal_0}\kappa_i\boldsymbol{1}\{\pf_i\in\Rcal_{-1},\tpf_i\in\Rcal^{\{\pf_i\}_{i=1}^m}\}}{1+\sum_{i=1}^m\boldsymbol{1}\{\pf_i\in\Acal_{-1},\tpf_i\in\Rcal^{\{\pf_i\}_{i=1}^m}\}}\\
\leq & q \frac{\sum_{i\in\Hcal_0}\kappa_i\boldsymbol{1}\{\pf_i\in\Rcal_{-1},\tpf_i\in\Rcal^{\{\pf_i\}_{i=1}^m}\}}{1+\sum_{i\in\Hcal_0}\boldsymbol{1}\{\pf_i\in\Acal_{-1},\tpf_i\in\Rcal^{\{\pf_i\}_{i=1}^m}\}}
:= q \Ecal(\Rcal^{\{\pf_i\}_{i=1}^m}),
\end{aligned}
$$
where $\Acal_{-1}=\cup_{k=1}^K\Acal_{-1}^k$ is the control side. 
The next step is to prove $\Eb\left\{\Ecal(\Rcal^{\{\pf_i\}_{i=1}^m})\right\}\leq 1$. We bound $\Eb\left\{\Ecal(\Rcal^{\{\pf_i\}_{i=1}^m})\right\}$ by
$$
\begin{aligned}
\Eb\left\{\Ecal(\Rcal^{\{\pf_i\}_{i=1}^m})\right\}
&\stackrel{(1)}{=}\sum_{i\in\Hcal_0}\Eb\left(\frac{\kappa_i\boldsymbol{1}\{\pf_i\in\Rcal_{-1},\tpf_i\in\Rcal^{\{\pf_i\}_{i=1}^m}\}}{1+\sum_{j\in\Hcal_0,j\not=i}\boldsymbol{1}\{\pf_j\in\Acal_{-1},\tpf_j\in\Rcal^{\{\pf_i\}_{i=1}^m}\}}
\right)\\
&\stackrel{(2)}{=}\sum_{i\in\Hcal_0}\Eb\left(\frac{\kappa_i\boldsymbol{1}\{\pf_i\in\Rcal_{-1}\}\boldsymbol{1}\{\tpf_i\in\Rcal^{(-i)}\}}{1+\sum_{j\in\Hcal_0,j\not=i}\boldsymbol{1}\{\pf_j\in\Acal_{-1},\tpf_j\in\Rcal^{(-i)}\}}
\right)\\
&\stackrel{(3)}{=}\sum_{i\in\Hcal_0}\Eb\left(\frac{\kappa_i\Pb\{\pf_i\in\Rcal_{-1}\mid\tpf_i,\pf_{(-i)}\}\boldsymbol{1}\{\tpf_i\in\Rcal^{(-i)}\}}{1+\sum_{j\in\Hcal_0,j\not=i}\boldsymbol{1}\{\pf_j\in\Acal_{-1},\tpf_j\in\Rcal^{(-i)}\}}
\right)\\
&\stackrel{(4)}{\leq}\sum_{i\in\Hcal_0}\Eb\left(\frac{\Pb\{\pf_i\in\Acal_{-1}\mid\tpf_i,\pf_{(-i)}\}\boldsymbol{1}\{\tpf_i\in\Rcal^{(-i)}\}}{1+\sum_{j\in\Hcal_0,j\not=i}\boldsymbol{1}\{\pf_j\in\Acal_{-1},\tpf_j\in\Rcal^{(-i)}\}}
\right)\\
&\stackrel{(5)}{=}\Eb\left(\sum_{i\in\Hcal_0}\frac{\boldsymbol{1}\{\pf_i\in\Acal_{-1},\tpf_i\in\Rcal^{(-i)}\}}{1+\sum_{j\in\Hcal_0,j\not=i}\boldsymbol{1}\{\pf_j\in\Acal_{-1},\tpf_j\in\Rcal^{(-i)}\}}
\right),
\end{aligned}
$$
where the above derivations stand because (1) if $\pf_i\in\Rcal_{-1}$ then $\pf_i\not\in\Acal_{-1}$; (2) if $\pf_i\in\Rcal_{-1}$ then $\pf_i=\tpf_i$ and hence $\Rcal^{(-i)}=\Rcal^{\{\pf_{(-i)},\tpf_i\}}=\Rcal^{\{\pf_{(-i)},\pf_i\}}=\Rcal^{\{\pf_i\}_{i=1}^m}$; (3) $\Rcal^{(-i)}$, $\tpf_i$ and $\{\pf_j\}_{j\in\Hcal_0,j\not=i}$ are measurable with respect to $\sigma(\tpf_i,\pf_{(-i)})$; (4) $\pf_i$ is independent of $\pf_{(-i)}$ and using \eqref{equ:RAcomp} in the proof of Lemma~\ref{lemma:fdr}; (5) it follows by applying the tower law. 
Subsequently, we can prove that
$$
\begin{aligned}
\sum_{i\in\Hcal_0}\frac{\boldsymbol{1}\{\pf_i\in\Acal_{-1},\tpf_i\in\Rcal^{(-i)}\}}{1+\sum_{j\in\Hcal_0,j\not=i}\boldsymbol{1}\{\pf_j\in\Acal_{-1},\tpf_j\in\Rcal^{(-i)}\}}
\stackrel{(\star)}{=}&
\sum_{i\in\Hcal_0}\frac{\boldsymbol{1}\{\pf_i\in\Acal_{-1},\tpf_i\in\Rcal^{(-i)}\}}{1+\sum_{j\in\Hcal_0,j\not=i}\boldsymbol{1}\{\pf_j\in\Acal_{-1},\tpf_j\in\Rcal^{(-j)}\}}\\
=&
\sum_{i\in\Hcal_0}\frac{\boldsymbol{1}\{\pf_i\in\Acal_{-1},\tpf_i\in\Rcal^{(-i)}\}}{\sum_{j\in\Hcal_0}\boldsymbol{1}\{\pf_j\in\Acal_{-1},\tpf_j\in\Rcal^{(-j)}\}\vee 1} 
\leq 1,
\end{aligned}
$$
where ($\star$) replaces the $\Rcal^{(-i)}$ in the denominator with $\Rcal^{(-j)}$ and will be proved later; the second equality holds by checking the values of each term with $\boldsymbol{1}\{\pf_i\in\Acal_{-1},\tpf_i\in\Rcal^{(-i)}\}$ being zero or one. 
Combining the above derivation, we conclude that $\mFDR\leq q\Eb(\Ecal(\Rcal^{\{\pf_i\}_{i=1}^m}))\leq q$.

We now justify ($\star$) to complete the proof. Specifically, under the condition $\pf_i\in\Acal_{-1},\tpf_i\in\Rcal^{(-i)}$, we aim to show that $\pf_j\in\Acal_{-1},\tpf_j\in\Rcal^{(-i)}\Longleftrightarrow\pf_j\in\Acal_{-1},\tpf_j\in\Rcal^{(-j)}$. 

First, recall that the final rejection region using $\{\pf_{(-i)},\tpf_i\}$ is defined as $\Rcal^{(-i)}:=\Rcal^{\{\pf_{(-i)},\tpf_i\}}=\cup_{t}\{\Rcal_t: \widehat{\FDP}^{(-i)}(\Rcal_t)\leq q\}$. Since the candidate rejection regions $\{\Rcal_t\}_{t=-1,0,1,\cdots}$ are based on $\sigma\left\{(\tpf_{i})_{i=1}^m\right\}$ and $\Rcal^{(-i)}$ should be one of them. The same discussion holds for $\Rcal^{(-j)}$. Thus, $\Rcal^{(-i)}$ and $\Rcal^{(-j)}$ are nested, i.e.,
\begin{equation}\label{equ:subset}
    \text{either}\quad \Rcal^{(-i)}\subseteq\Rcal^{(-j)}  \quad \text{or}\quad \Rcal^{(-j)}\subseteq\Rcal^{(-i)}\quad
    \text{for any }i,j=1,\cdots,m.
\end{equation}

Next, we prove the direction ``$\Longrightarrow$". Using the final rejection region $\Rcal^{(-i)}$, the FDP estimates using $\{\pf_{(-j)},\tpf_j\}$ and $\{\pf_{(-i)},\tpf_i\}$ are the same: 
    $$\begin{aligned}
        &\widehat{\FDP}^{(-j)}(\Rcal^{(-i)})\\
        \stackrel{(1)}{=}&
        \frac{1+\sum_{l\in[m],l\not=j}\boldsymbol{1}\{\pf_l\in\Acal_{-1},\tpf_l\in\Rcal^{(-i)}\}+\boldsymbol{1}\{\tpf_j\in\Acal_{-1},\tpf_j\in\Rcal^{(-i)}\} }{\sum_{l\in[m],l\not=j}\boldsymbol{1}\{\pf_l\in\Rcal_{-1},\tpf_l\in\Rcal^{(-i)}\}\ + \boldsymbol{1}\{\tpf_j\in\Rcal_{-1},\tpf_j\in\Rcal^{(-i)}\}}\\
         \stackrel{(2)}{=}&\frac{1+\sum_{l\in[m],l\not=i,j}\boldsymbol{1}\{\pf_l\in\Acal_{-1},\tpf_l\in\Rcal^{(-i)}\}+\boldsymbol{1}\{\pf_i\in\Acal_{-1},\tpf_i\in\Rcal^{(-i)}\}+ \boldsymbol{1}\{\tpf_j\in\Acal_{-1},\tpf_j\in\Rcal^{(-i)}\}}{\sum_{l\in[m],l\not=i,j}\boldsymbol{1}\{\pf_l\in\Rcal_{-1},\tpf_l\in\Rcal^{(-i)}\}+\boldsymbol{1}\{\pf_i\in\Rcal_{-1},\tpf_i\in\Rcal^{(-i)}\} +  \boldsymbol{1}\{\tpf_j\in\Rcal_{-1},\tpf_j\in\Rcal^{(-i)}\}}\\
         \stackrel{(3)}{=}&\frac{1+\sum_{l\in[m],l\not=i,j}\boldsymbol{1}\{\pf_l\in\Acal_{-1},\tpf_l\in\Rcal^{(-i)}\}+1+0}{\sum_{l\in[m],l\not=i,j}\boldsymbol{1}\{\pf_l\in\Rcal_{-1},\tpf_l\in\Rcal^{(-i)}\}+0+1}\\
         \stackrel{(4)}{=}&\frac{1+\sum_{l\in[m],l\not=i,j}\boldsymbol{1}\{\pf_l\in\Acal_{-1},\tpf_l\in\Rcal^{(-i)}\}+\boldsymbol{1}\{\pf_j\in\Acal_{-1},\tpf_j\in\Rcal^{(-i)}\}+ \boldsymbol{1}\{\tpf_i\in\Acal_{-1},\tpf_i\in\Rcal^{(-i)}\}}{\sum_{l\in[m],l\not=i,j}\boldsymbol{1}\{\pf_l\in\Rcal_{-1},\tpf_l\in\Rcal^{(-i)}\}+\boldsymbol{1}\{\pf_j\in\Rcal_{-1},\tpf_j\in\Rcal^{(-i)}\} +  \boldsymbol{1}\{\tpf_i\in\Rcal_{-1},\tpf_i\in\Rcal^{(-i)}\}}\\
         \stackrel{(5)}{=}&\frac{1+\sum_{l\in[m],l\not=i}\boldsymbol{1}\{\pf_l\in\Acal_{-1},\tpf_l\in\Rcal^{(-i)}\}+ \boldsymbol{1}\{\tpf_i\in\Acal_{-1},\tpf_i\in\Rcal^{(-i)}\}}{\sum_{l\in[m],l\not=i}\boldsymbol{1}\{\pf_l\in\Rcal_{-1},\tpf_l\in\Rcal^{(-i)}\}+  \boldsymbol{1}\{\tpf_i\in\Rcal_{-1},\tpf_i\in\Rcal^{(-i)}\}}\\
         \stackrel{(6)}{=} &\widehat{\FDP}^{(-i)}(\Rcal^{(-i)})
        \leq q,
    \end{aligned}$$
    where (1) and (6) stand according to the definitions of $\widehat{\FDP}^{(-j)}(\Rcal)$ and  $\widehat{\FDP}^{(-i)}(\Rcal)$; (2) and (5) hold by extracting the $i$th term and merging the $j$th term into the summation respectively; (3) and (4) use the condition $\pf_i\in\Acal_{-1},\tpf_i\in\Rcal^{(-i)}\subseteq \Rcal_{-1}$, $\pf_j\in\Acal_{-1}$ and $\tpf_j\in \Rcal^{(-i)}\subseteq \Rcal_{-1}$ and notice that $\Acal_{-1}$ and $\Rcal_{-1}$ are disjoint; the last inequality stands by the definition of $\Rcal^{(-i)}$. Therefore, $\Rcal^{(-i)}$ is a candidate rejection set satisfying $\widehat{\FDP}^{(-j)}(\Rcal^{(-i)})\leq q$. By the definition of $\Rcal^{(-j)}$, $\Rcal^{(-i)}\subseteq  \Rcal^{(-j)}$ and thus $\tpf_j\in\Rcal^{(-j)}$.

    Finally, we prove the direction ``$\Longleftarrow$". If $\tpf_j\in\Rcal^{(-j)}$ but $\tpf_j\not\in\Rcal^{(-i)}$, we have $\Rcal^{(-i)}\subsetneq\Rcal^{(-j)}$ according to \eqref{equ:subset}. Hence, we have $\pf_j\in\Acal_{-1},\tpf_j\in\Rcal^{(-j)}$, $\pf_i\in\Acal_{-1}$ and $\tpf_i\in\Rcal^{(-i)}\subsetneq\Rcal^{(-j)}$. Using the same argument, we have
    $$\widehat{\FDP}^{(-i)}\left(\Rcal^{(-j)}\right)
    =\widehat{\FDP}^{(-j)}\left(\Rcal^{(-j)}\right)
        \leq q,$$
        which implies that $\Rcal^{(-j)}\subseteq\Rcal^{(-i)}$ and contradicts with $\tpf_j\not\in\Rcal^{(-i)}$. Therefore, we complete the proof of ($\star$).
\end{proof}

\subsection{Optimality}
In this section, we prove Theorems~\ref{thm:pow} and \ref{thm:lmfdr}.
\begin{proof}[Proof of Theorem \ref{thm:pow}]
As those p-values in $\Ucal_{0}$ do not play a role in our procedure, we proceed with the proof under the condition $i\in\Mcal_{0}$.  
Recall that our stopping condition is $\widehat{\FDP}_{t}\leq q$. The sequential unmasking rule defined in \eqref{equ:SeqUnmaskR} of the main paper reveals the p-value vector of one hypothesis at one time. Thus, at step $t$, we have $A_t+R_t = |\Mcal_0|-t$. Simple derivation shows that the stopping condition is equivalent to $A_t\leq (q|\Mcal_0|-qt-1)/(1+q)$.

Define $B_i=\boldsymbol{1}\{\pf_i\in\cup_{k=1}^K\Acal_{-1}^K\}$. It is straightforward to see that $B_i$ independently follows $\text{Bernoulli}(1-q_i)$ for $i\in\Mcal_0$ since 
$$\mathbb{P}(\pf_i\in\cup_{k=1}^K\Acal_{-1}^k|\mathcal{G}_0)=\mathbb{P}(\pf\in\cup_{k=1}^K\Acal_{-1}^k|\tilde{\pf}=\tilde{\pf}_i)=1-\mathbb{P}(\pf\in\mathcal{R}_{-1}|\tilde{\pf}=\tilde{\pf}_i)=1-q_i \, .$$
By Theorem A.2 in \cite{Chao2021} with $S_t=A_t=\sum_{i\in\Mcal_t} B_i$ and $s_t=(q|\Mcal_0|-qt-1)/(1+q)$, the most powerful procedure, minimizing $\tau$ so that maximizing $$R_\tau=|\Mcal_0|-\tau-A_\tau=|\Mcal_0|-\tau-\frac{q|\Mcal_0|-q\tau-1}{1+q}=|\Mcal_0|-\frac{\tau}{1+q}-\frac{q|\Mcal_0|-1}{1+q},$$  
is to reveal the masked hypotheses in the decreasing order of $1-q_i$, i.e., the increasing order of $q_i$.
\end{proof}

\begin{proof}[Proof of Theorem \ref{thm:lmfdr}]
Under Assumptions (a) and (b) on the distribution of $\pf$, %\kejun{which three assumption(s) in particular?}\linsuin{There are two assumptions, both of which are useful for deriving the following equation.} 
we claim that the density of $\pf$ given $\btheta$ is
\begin{equation}\label{equ:pDen}
f_{\btheta}(\tf)=f_{\btheta}(\Proj(\tf))\times \left( 1-\max_{k\in\{1,\ldots,K\}}\boldsymbol{1}\{\theta_k=1,t_k>1/2\}\right).
\end{equation}
Given the signal indicator $\btheta$, the density of $\pf$ conditional on the event $\pf\in\cup_{k=1}^K\Acal_{-1}^k\cup\Rcal_{-1}$ is given by 
\begin{equation}\label{equ:f_dcomp}
\begin{aligned}
& f_{\btheta}(\tf\mid\pf\in\cup_{k=1}^K\Acal_{-1}^k\cup\Rcal_{-1}) \\
&  \qquad \propto f_{\btheta}(\tf)\boldsymbol{1}\{\tf\in\cup_{k=1}^K\Acal_{-1}^k\cup\Rcal_{-1}\}\\
&  \qquad = f_{\btheta}(\tf)\boldsymbol{1}\{\tf\in\Rcal_{-1}\}+\sum_{k=1}^Kf_{\btheta}(\tf)\boldsymbol{1}\{\tf\in\Acal_{-1}^k\}\\
&  \qquad =
f_{\btheta}(\tf)\boldsymbol{1}\{\tf\in \Rcal_{-1}\}+f_{\btheta}(\Proj(\tf))\times \left(\sum_{k=1}^K \boldsymbol{1}\{\theta_k=0,\tf\in\Acal^k_{-1}\}\right).
\end{aligned}
\end{equation}
To see why the last equality in \eqref{equ:f_dcomp} holds, we notice that $\tf\in\Acal_{-1}^k$ implies $t_{l}<1/2$ for all $l\not=k$ and hence $\boldsymbol{1}\{\theta_l=1,t_l>1/2\}=0$ for $l\not=k$. Moreover, we have $t_k>1/2$. Then, for $\tf\in\Acal_{-1}^k$, we obtain $$\left( 1-\max_{l\in\{1,\ldots,K\}}\boldsymbol{1}\{\theta_l=1,t_l>1/2\}\right)=1-\boldsymbol{1}\{\theta_k=1,t_k>1/2\}=\boldsymbol{1}\{\theta_k=0\}.$$ 
The last equality holds according to \eqref{equ:pDen}.

We now examine the denominator of \eqref{equ:SeqUnmaskR}. Recalling $g(\ttf)$ defined in \eqref{eq-unmask} and rearranging the RHS of \eqref{eq-unmask}, we get
\begin{align}
g(\ttf) &\propto \sum_{\tf:\Proj(\tf)=\ttf} f(\tf)\boldsymbol{1}\{\tf\in \cup_{k=1}^K\Acal^k_{-1}\cup\Rcal_{-1}\} \nonumber \\
& = 
\sum_{\tf:\Proj(\tf)=\ttf} \left\{\sum_{\btheta\in\{0,1\}^K}\pi_{\btheta}f_{\btheta}(\tf)\right\}\boldsymbol{1}\{\tf\in \cup_{k=1}^K\Acal^k_{-1}\cup\Rcal_{-1}\} \nonumber \\
&= \sum_{\btheta\in\{0,1\}^K}
\sum_{\tf:\Proj(\tf)=\ttf} \pi_{\btheta}f_{\btheta}(\Proj(\tf))\times \left(\boldsymbol{1}\{\tf\in\Rcal_{-1}\}+\sum_{k=1}^K \boldsymbol{1}\{\theta_k=0,\tf\in\Acal^k_{-1}\}\right),\label{eq-last}
\end{align}
where the first equality is due to \eqref{equ:twoNmodel} and the second equality follows by changing the order of the summations. Noting that $\ttf=\Proj(\tf)$, (\ref{eq-last}) becomes
\begin{align*}
&f(\ttf)+\sum_{\btheta\in\{0,1\}^K}\pi_{\btheta}
\sum_{\tf:\Proj(\tf)=\ttf,\tf\not=\ttf}f_{\btheta}(\ttf)\sum_{k=1}^K \boldsymbol{1}\{\theta_k=0,\tf\in\Acal^k_{-1}\}\\
& \qquad \qquad = f(\ttf)+\sum_{\btheta\in\{0,1\}^K}\pi_{\btheta}f_{\btheta}(\ttf) 
\sum_{k=1}^K\boldsymbol{1}\{\theta_k=0\}\\
& \qquad \qquad = f(\ttf)+\sum_{\btheta\not=(1,\cdots,1)}\kappa_{\btheta}\pi_{\btheta}f_{\btheta}(\ttf) \, ,
\end{align*}
where the first line uses the technique for deriving \eqref{equ:f_dcomp}, the second line stands because each $\tf$ can only locate at exactly one mirror side, and the last equality is because of the definition $\kappa_{\btheta}=\sum_{k=1}^K(1-\theta_k)$. Therefore, $f(\ttf)/g(\ttf)$ is proportional to 
$$\frac{f(\ttf)}{f(\ttf)+\sum_{\btheta\not=(1,\cdots,1)}\kappa_{\btheta}\pi_{\btheta}f_{\btheta}(\ttf)},$$
the reciprocal of which is equal to
$$\frac{f(\ttf)+\sum_{\btheta\not=(1,\cdots,1)}\kappa_{\btheta}\pi_{\btheta}f_{\btheta}(\ttf)}{f(\ttf)}
=1+\frac{\sum_{\btheta\not=(1,\cdots,1)}\kappa_{\btheta}\pi_{\btheta}f_{\btheta}(\ttf)}{f(\ttf)} \, .$$

We now prove the claim in \eqref{equ:pDen}. Our argument consists of two steps. First, we examine the coordinates of the p-value vector under the alternative. If $\theta_k=1$, then $p_k$ has to be below $1/2$ by (a) and hence $f_{\btheta}(\tf)=0$ for $t_k>1/2$. In other words, $f_{\btheta}(\tf)=0$ if and only if $\theta_k=1$ and $t_k>1/2$ is true for at least one $k\in\{1,\ldots,K\}$. Therefore, we can multiply the density by $1-\max_{k\in\{1,\ldots,K\}} \boldsymbol{1}\{\theta_k=1,t_k>1/2\}$. Next, we examine the coordinates under the null. If $\theta_k=0$, then $p_k$ is symmetric about 1/2 conditional on the rest of the p-values by (b), and hence 
\begin{align*}
    f_{\btheta}(\tf)
& = f_{\btheta,k\mid -k}(t_k\mid\pf_{-k}
=\tf_{-k})f_{\btheta,-k}(\tf_{-k})\\
&= f_{\btheta,k\mid -k}(1-t_k\mid\pf_{-k}=\tf_{-k})f_{\btheta,-k}(\tf_{-k})=f_{\btheta}(t_1,\cdots,t_{k-1},1-t_{k},t_{k+1},\cdots,t_K),
\end{align*}
where $\tf_{-k}=(t_1,\cdots,t_{k-1},t_{k+1},\cdots,t_K)$, $f_{\btheta,-k}(\cdot)$ is the density of $\pf_{-k}$ conditional on $\btheta$, and $f_{\btheta,k\mid -k}$ is the density of $p_k$ conditional on $\pf_{-k}$ and $\btheta$. Hence the density remains unchanged if we replace $t_k$ by $1-t_k$. Combining the arguments, we conclude that claim \eqref{equ:pDen} is true.
\end{proof}

\subsection{Partial-order-assisted rejection region}
In this section, we present a brief proof for Remark \ref{rmk:poarr} of the main paper.
%\begin{proof}[Proof of Remark \ref{rmk:poarr}.] {\color{red} 

First, note that Algorithm~\ref{alg:JM} implies two two basic principles for the rejection regions: (\rNum{1}) %As $t$ increases, 
$\Rcal_t$ shrinks as $t$ increases because $\tPcal_t$ in Algorithm~\ref{alg:JM} shrinks as well; (\rNum{2}) $\hi_t^*$ is determined by $\Fcal_t$ so that $\Rcal_t\in\Fcal_t$. 

Second, employing these $\Rcal_t$ in Algorithm~\ref{alg:JM}, in turn, results in the same unmask set $\Mcal_t$ as in Algorithm~\ref{alg:JMposet}, i.e., $i\in\Mcal_t$ iff $\tpf_i\in\Rcal_t$ (which is also equivalent to $\pf_i\in \cup_{k=1}^K\Acal_t^k\cup\Rcal_t$ with $\Rcal_t$ defined in Remark~\ref{rmk:poarr}). Thus, we have the following two results. 
\begin{itemize}
    \item For any $i\in\Mcal_t$, we have $\tpf_i\in\tPcal_t\subseteq\Rcal_t$ according to the definition of $\Rcal_t$. 
    \item For any $i\not\in\Mcal_t$, hypotheses $i$ is revealed at step $s$ for some $s< t$. If $\tpf_i\in\Rcal_t$, then $\tpf_i\prec\tpf_j$ for some $j\in\Mcal_t$. However, $j\in\Mcal_t$ implies $\tpf_j\in\tPcal_t\subset\tPcal_s$ for all $s< t$. Thus, $\tpf_i$ would not be a maximal element of $\tPcal_s$ for any $s<t$, which contradicts with the assumption that hypothesis $i$ is revealed at step $s$. Therefore, $\tpf_i\not\in\Rcal_t$.
\end{itemize}
%The proof is completed by combining the above arguments. 
It is worth mentioning that the presented argument is also applicable to the JM procedure with an unmasking rule using partial order, i.e., one can use a preferred method to choose $\hi_t^*$ based solely on the information in $\Fcal_t$. 
%}\end{proof}

\section{Generalized JM Procedure: Two extensions}\label{sec:GJM}
\subsection{JM procedure with a general masking scheme}\label{sec:JMG}
Similar to \cite{Lei2016}, the JM procedure may lose power when (\rNum{1}) the discovery set is small or (\rNum{2}) the null p-values concentrate around one. Motivated by \cite{Chao2021}, we generalize the masking scheme to resolve these two issues. To be precise, we define $\Proj^h(\tf)=(h(t_1),\cdots,h(t_K))$, where 
$$
    h(t)=\left\{
    \begin{aligned}
        &(\nu-t)/\zeta,\quad &t\in(\lambda,\nu]\\
        &t,\quad &\text{otherwise}
    \end{aligned}
    \right.
$$
for $t\in[0,1]$, and $\zeta = (\nu-\lambda)/\alpha_m$, $0<\alpha_m<\lambda<\nu$, the ratio of probabilities located at the rejection and control sides. The corresponding rejection and control sides are $\Rcal_{-1}=[0,\alpha_m)^K$ and 
$$
    \Acal^k_{-1} = \{\tf:t_k\in(\lambda,\nu]\text{ and }t_l\in[0,\alpha_m)\text{ for }l\not=k\}
$$
for $k=1,\ldots,K$. Then we can generalize the JM procedure using this masking scheme by replacing $\Proj$ with $\Proj^h$ and the FDP bound with
$$\widehat{\FDP}^M_t=\frac{1+A_t}{\zeta (R_t\vee 1)},$$
where $R_t$ and $A_t$ are the numbers of discoveries and controls, respectively.  When we take $\nu=1$ and $\alpha_m=\lambda=1/2$, the procedure reduces to the original JM procedure. To see why the generalized procedure may avoid power loss, we take $\kappa=1$ as an example. For (\rNum{1}), the JM procedure requires at least $1/ q$ rejections, while the generalized JM procedure can make as small as $1/(q\zeta)$ rejections. For  (\rNum{2}), the generalized JM procedure can choose $\nu<1$ to exclude the hypotheses whose p-values are close to one from constructing the controls.

We now show that the generalized JM procedure also controls FDR in finite samples. We require a condition about the distribution of the null p-values, which is more stringent than the conditional mirror conservatism. 

\begin{defn}[Conditional non-decreasing density]
Let $\mathcal{S}_0=\{1\leq k\leq K: \theta_k=0\}$. We say that $\pf$ has a conditional non-decreasing density if the conditional density of $p_k$ given $\pf_{-k}=\tf_{-k}=(t_1,\cdots,t_{k-1},t_{k+1},\cdots,t_K)$, denoted by 
$f_{k\mid -k}(t_k\mid \pf_{-k}=\tf_{-k})$, is non-decreasing w.r.t. $t_k$ for any $\tf_{-k}$ and $k\in\mathcal{S}_0$. %\linsuin{Although we just need one of component (not all component) is conditionally mirror conservative, we need to include more notations to clarify. To make our discussion simple and direct, I keep the statement all components.}
\end{defn}
With this definition, we can derive a result similar to Lemma~\ref{lemma:fdr}. 
\begin{lemma}\label{lemma:fdrG}
Fix $\kappa\in\{1,\cdots,K\}$. Suppose that the null p-values $\{\pf_{i}\}_{i\in\Hcal^{\kappa,partial}_0}$ are independent of each other and have conditional non-decreasing densities. %Then, we have $\Pb\left(b_i=1\mid\Gcal_{-1}\right)\leq 1/2$ for all $i\in\Hcal_0$.
Furthermore, $\{b_i\}_{i\in\Hcal^{\kappa,partial}_0}$ are independent Bernoulli random variables conditioned on $\Gcal_{-1}$ with $$\Pb\left(b_i=1\mid\Gcal_{-1}\right)\leq 1/(\kappa\zeta+1)$$
for all $i\in\Hcal_0^{\kappa,partial}$.
\end{lemma}

\begin{proof}[Proof of Lemma~\ref{lemma:fdrG}]
Following the proof of Lemma~\ref{lemma:fdr}, we only need to show a variant of \eqref{equ:p_rej},  
\begin{equation}
    \label{equ:p_rej_g}
\Pb_i\left(\pf\in\Rcal_{-1}\mid \tpf=\tpf_i,\pf\in\cup_{k=1}^K\Acal_{-1}^k\cup\Rcal_{-1}\right)\leq 1/(\kappa\zeta+1)
\end{equation}
for $i\in\Hcal_0^{\kappa,partial}\cap\Mcal_{-1}$. Define $\Pb_i$ and $f_{k\mid-k,i}$ as the probability distribution of $\pf_i$ and conditional density of its $k$th coordinate given the remaining coordinates for $i\in\Hcal_0^{\kappa,partial}$. Setting $h_k=(\nu-p_k)/\zeta$, the conditional density of $h_k\mid \pf_{-k}=\tf_{-k}$ is 
$$
f^h_{k\mid -k,i}(h\mid \pf_{-k}=\tf_{-k})
=
\zeta f_{k\mid -k,i}(t\mid \pf_{-k}=\tf_{-k}),
$$
where $t=\nu-\zeta h$. 
For $\ttf\in\Rcal_{-1}$ and $k\in\Scal_{0i}$, we have
$$
\begin{aligned}
\frac{\Pb_i\left(\pf\in\Rcal_{-1} ,\tpf=\ttf\mid\pf\in\cup_{k=1}^K\Acal_{-1}^k\cup\Rcal_{-1}\right)}{\Pb_i\left(\pf\in\Acal^k_{-1}, \tpf=\ttf\mid\pf\in\cup_{k=1}^K\Acal_{-1}^k\cup\Rcal_{-1}\right)}
=&
\frac{f_{k\mid -k,i}(\tt_k\mid \pf_{-k}=\ttf_{-k})}{f^h_{k\mid -k,i}(\tt_k\mid \pf_{-k}=\ttf_{-k})}\\
= &
\frac{f_{k\mid -k,i}(\tt_k\mid \pf_{-k}=\ttf_{-k})}{\zeta f_{k\mid -k,i}(\nu-\zeta \tt_k\mid \pf_{-k}=\ttf_{-k})}
\leq 1/\zeta,
\end{aligned}
$$
where the last inequality is because $\pf_i$ has conditional non-decreasing density and $\tt_k\leq \nu-\zeta\tt_k$ for $0<\tt_k<\alpha_m$. Similar to Lemma~\ref{lemma:fdr}, we obtain
$$\frac{\kappa\zeta\Pb_i\left(\pf\in\Rcal_{-1} ,\tpf=\ttf\mid\pf\in\cup_{k=1}^K\Acal_{-1}^k\cup\Rcal_{-1}\right)}{\sum_{k=1}^K\Pb_i\left(\pf\in\Acal^k_{-1}, \tpf=\ttf\mid\pf\in\cup_{k=1}^K\Acal_{-1}^k\cup\Rcal_{-1}\right)}\leq 1
$$ 
and \eqref{equ:p_rej_g} stands.
\end{proof}

\begin{thm}[Finite sample FDR control]\label{thm:fdrG}
Consider the problem of testing $\Hcal^{\kappa,partial}_0$ for any $\kappa=1,\ldots,K$, where the corresponding FDR
is defined in (\ref{equ:FDRH0}) with $\mathcal{H}_0$ being replaced by $\Hcal^{\kappa,partial}_0$.  
Suppose that the null p-values $\{\pf_{i}\}_{i\in\Hcal^{\kappa,partial}_0}$ are independent of each other and of the non-null
p-values $\{\pf_{i}\}_{i\in\Hcal^{\kappa,partial}_1}$. Suppose $\pf_{i}$ has a conditional non-decreasing density for all $i\in\Hcal^{\kappa,partial}_0$. Then, the JM procedure with a general masking scheme controls the
$\FDR$ at level $q/\kappa$, or equivalently controls the $\kappa\FDR$ at level $q$.  %As a consequence, when $\kappa=1$ (i.e., under the composite null $\mathcal{H}_0$), the JM procedure with a general masking scheme controls the FDR at level $q$.
\end{thm}

\begin{proof}[Proof of Theorem~\ref{thm:fdrG}]
    Similar to the proof of Theorem~\ref{thm:fdr}, we have 
    $$
\FDP_\tau
=\frac{V_\tau}{R_\tau\vee 1}
=\zeta\frac{A_\tau+1}{\zeta (R_\tau\vee 1)}\frac{V_\tau}{A_\tau+1}
\leq 
\zeta 
q
\frac{V_\tau}{U_\tau+1}
=
\zeta q
\left(\frac{1+|\Mcal_\tau\cap\Hcal^{\kappa,partial}_0|}{U_\tau+1}-1\right).
$$ Applying Lemmas~\ref{lemma:martingale} and \ref{lemma:fdrG}, we get
$$
    \FDR
=\Eb\left(\FDP_\tau\right)
\leq 
\zeta q\Eb\left(\frac{V_\tau}{U_\tau+1}\right)
\leq 
%\zeta q\Eb\left\{\Eb\left(\frac{1+|\Mcal_\tau\cap\Hcal^{\kappa,partial}_0|}{U_\tau+1}-1\mid\Gcal_{-1}\right)\right\}\leq 
\zeta q\left(\frac{\zeta \kappa+1}{\zeta \kappa}-1\right) = q/\kappa,
$$
which completes the proof.
\end{proof}

%We next prove Lemma~\ref{lemma:fdrG} to complete the proof of Theorem~\ref{thm:fdrG}. 

\subsection{JM procedure with z-values}\label{sec:JMZ}
In some replicability studies, researchers require both the signals and their signs to be repetitive. When making inferences with two-sided p-values, the sign information is lost. A remedy is to use the one-sided p-values for testing the two one-sided hypotheses %{\color{red} What is a one-sided p-value for a positive sign?} 
separately and combine the results \citep{Wang2022}.
%{\color{red} Can we rewrite this paragraph to provide more background?The signs of signals identified by the JM procedure might be different when considering two-sided p-values. If the interested sign is specified, the straightforward approach to require consistent signs is taking one-sided p-values.} 
Here we introduce a z-value-based approach to control the directional false discovery rate (dFDR), which has been considered in the literature with a single experiment \citep{Leung2021,Leung2022}. 

Consider a sequence of z-values arising from $K$ experiments $\Zf_i=(Z_{1i},\cdots,Z_{Ki})$ for $i\in[m]$, where $Z_{ki}$ are independently generated from the normal distribution $\Ncal(\mu_{ki},\sigma^2)$. We define the alternative set with positive and negative signs as $\Hcal_{1}^{+}=\{i:\mu_{ki}>0, \text{ for all } k\}$ and $\Hcal_{1}^{-}=\{i:\mu_{ki}<0, \text{ for all } k\}$, respectively. The rest belongs to the null set, $\Hcal_0=[m]\backslash(\Hcal_1^+\cup\Hcal_1^-)$. Let $s_i=1$ for $i\in\Hcal_{1}^+$, $s_i=-1$ for $i\in\Hcal_{1}^-$, and $s_i=0$ for $i\in\Hcal_{0}$. Our goal is to decide the sign of each hypothesis, say by $\whScal^d=\{\hs_i\}_{i=1}^m$ with $\hs_i\in \{-1,0,1\}$, while controlling the directed FDR given by 
$$
\dFDR(\whScal^d)=\mathbb{E}\left[\frac{\sum_{i=1}^m\boldsymbol{1}\{\hs_i\not=0, \hs_i\not=s_i\}}{(\sum_{i=1}^m\boldsymbol{1}\{
\hs_i\not=0\})\vee 1}\right]
$$
at a target dFDR level $q\in(0,1)$. The main idea of the z-value-based approach is to construct two rejection regions for the repeated positive (negative) signs $\Rcal^+$ ($\Rcal^{-}$) and define the sign estimate as
$$
\hs_i=\left\{
\begin{aligned}
1,& \qquad\Zf_i\in\Rcal^+,\\
-1,& \qquad\Zf_i\in\Rcal^-,\\
0,& \qquad\text{otherwise}.
\end{aligned}
\right. 
$$
 For simplicity, we let  $\Rcal^+\subset(0,\infty)^K$ and $\Rcal^-=\{(-t_1,\cdots,-t_K):\tf\in\Rcal^+\}$. Rearranging the number of false sign assignments, we obtain 
$$
\begin{aligned}
\sum_{i=1}^m\boldsymbol{1}\{\hs_i\not=0,\hs_i\not=s_i\}
=&\sum_{i\in\Hcal_0}\boldsymbol{1}\{\hs_i\not=0\}
+
\sum_{i\in\Hcal^+_1}\boldsymbol{1}\{\hs_i=-1\}
+
\sum_{i\in\Hcal^-_1}\boldsymbol{1}\{\hs_i=1\}\\
=&
\sum_{i\in\Hcal_0\cup\Hcal_1^-}\boldsymbol{1}\{\hs_i=1\}
+
\sum_{i\in\Hcal_0\cup \Hcal_1^+}\boldsymbol{1}\{\hs_i=-1\}\\
=&
\sum_{i\in\Hcal_0\cup \Hcal_1^-}\boldsymbol{1}\{\Zf_i\in\Rcal^+\}
+
\sum_{i\in\Hcal_0\cup \Hcal_1^+}\boldsymbol{1}\{\Zf_i\in\Rcal^-\}.\\
\end{aligned}
$$
In this case, we define 2$K$ mirror regions as
$$
\begin{aligned}
\Acal^{k,+}&=\{(t_1,\cdots,t_{k-1},-t_{k},t_{k+1},t_K):\tf\in\Rcal^+\}\\
\Acal^{k,-}&=\{(-t_1,\cdots,-t_{k-1},t_{k},-t_{k+1},-t_K):\tf\in\Rcal^+\}\\
\end{aligned}
$$
for $k=1,\ldots,K$. For any $i\in\Hcal_0\cup \Hcal_1^-$, we have at least one element in $\{\mu_{ki}\}_{k=1}^K$, say $\mu_{k_0 i}$, less or equal than zero, which indicates that the corresponding z-value is more likely to locate at $\Acal^{k_0,+}$ than at $\Rcal^+$. Therefore, the number of false sign assignments can be approximately upper bounded by 
$$ 
\sum_{i\in\Hcal_0\cup \Hcal_1^-}\sum_{k=1}^K\boldsymbol{1}\{\Zf_i\in\Acal^{k,+}\}
+
\sum_{i\in\Hcal_0\cup \Hcal_1^+}\sum_{k=1}^K\boldsymbol{1}\{\Zf_i\in\Acal^{k,-}\}. 
$$
Subsequently, we propose a conservative estimate of the directed FDR given by
$$
\widehat{\dFDP}(\Rcal^+,\Rcal^-)=\frac{1+\sum_{i=1}^m\sum_{k=1}^K\boldsymbol{1}\{\Zf_i\in\Acal^{k,+}\cup\Acal^{k,-}\}}{\sum_{i=1}^m\boldsymbol{1}\{\Zf_i\in\Rcal^{+}\cup\Rcal^{-}\}\vee 1}.$$
Finally, the dFDR controlling procedure for repetitive signals can be established in the same way as in the main paper. If one has the prior knowledge that more signals are of the positive sign, then it makes sense to enlarge $\Rcal^+$. More generally, when there is potentially useful side information, it is interesting to design adaptive rejection regions to incorporate the information.
%{\color{blue} From the z-value perspective, there are far more interesting directions. For instance, when side information is available, a covariate assistant rejection region is preferred, in which case $\Rcal^+$ and $\Rcal^-$ might be asymmetric. }

\section{Additional Numerical Results}\label{sec:AddNumerical}
In this section, we further investigate the performance of the JM procedure in mediation analysis and compare it with some existing alternatives as discussed in Section~\ref{sec:simuMed} of the main paper. We follow the settings in Section~\ref{sec:simuMed} with some modifications: Under $\Hcal_{00}, \alpha_i=0$ and $\beta_i=0$; under $\Hcal_{10}, \alpha_i=0.5$ and $\beta_i=0$; under $\Hcal_{01},$ $\alpha_i=0$ and $\beta_i=0.75$; under $\Hcal_{11}$, $\alpha_i=0.5$ and $\beta_i=0.75$.  Motivated by \cite{Dai2020}, we consider five hypothesis configurations: Global Null (GNull), Sparse Null (SNull), Dense Null (DNull), Sparse Alternative (SAlter), Dense Alternative (DAlter). Table~\ref{tab:config} presents the proportions of the four different types of null hypotheses under different configurations. 

Tables~\ref{tab:FDPApp}--\ref{tab:POWApp} respectively summarize the FDPs, mFDPs, and powers when the target FDR levels $q$ are $0.05$ and $0.2$. The FDPs and mFDPs of the three JM procedures are either below or around the target FDR levels. They are extremely conservative under the three null configurations (i.e., GNull, SNull, and DNull). We notice that their mFDPs are around the target FDR levels under the two types of alternatives (i.e., SAlter and DAlter), while their FDPs are below the target FDR levels under SAlter. %These observations suggest the JM procedure controls the mFDR. 
The other methods all suffer from FDR inflation to some extent. 
For example, we observe FDR inflation for JS.Mix.Asy under GNull, 
DACT under all configurations except for GNull, and JS.Mix.Finite and MT.Comp under all the configurations. As the mFDP counts the number of false discoveries for the null hypotheses in $\Hcal_{00}$ twice, it is not surprising that the mFDPs of JS.Mix.Asy and JS.Mix.Finite are around $2q=0.4$ under GNull when $q=0.2$. We exclude MT.Comp in the power comparison due to its severe FDR inflation. DACT has the most discoveries except for the case of DAlter and $q=0.05$. JM.Product has competitive performance and achieves the highest power under DAlter. 

\begin{table}[h!]
\centering
\begin{tabular}{ccccc}
 \toprule
\multicolumn{1}{c}{Hypothesis Configuration} & $\pi_{00}$&$\pi_{01}$&$\pi_{10}$&$\pi_{11}$\\ 
  \midrule
  GNull & 1&0&0&0\\
  SNull &  0.90&0.05&0.05&0\\
  DNull & 0.60&0.20&0.20&0\\
  SAlter & 0.88&0.05&0.05&0.02\\
  DAlter & 0.40&0.20&0.20&0.20\\
 \bottomrule
\end{tabular}
\caption{The null proportions under different configurations.}\label{tab:config} 
\end{table}

\begin{table}[h!]
\centering
\resizebox{\linewidth}{!}{
\begin{tabular}{cllllll}
 \toprule
  Target &   \multirow{3}*{Method} &    \multicolumn{5}{c}{Hypothesis Configuration}\\ 
   \cmidrule(lr){3-7}
 FDR&  & \multirow{2}*{GNull} &  \multirow{2}*{SNull} &  \multirow{2}*{DNull} &  \multirow{2}*{SAlter} &  \multirow{2}*{DAlter} \\ 
Level &&&&&&\\
  \hline
  \multirow{7}*{$q=0.05$} & JS.Mix.Asy & 0.070(0.162) & 0.030(0.108) & 0.030(0.108) & 0.042(0.018) & 0.045(0.006) \\ 
   & JS.Mix.Finite & 0.070(0.162) & 0.060(0.151) & 0.060(0.151) & 0.057(0.018) & 0.058(0.006) \\ 
   & DACT & 0.020(0.089) & 0.060(0.151) & 0.080(0.172) & 0.061(0.017) & 0.064(0.006) \\ 
   & MT.Comp & 0.090(0.182) & 1.000(0.000) & 0.240(0.271) & 0.274(0.023) & 0.074(0.005) \\ 
   & JM.Max & 0.000(0.000) & 0.000(0.000) & 0.000(0.000) & 0.043(0.022) & 0.052(0.007) \\ 
   & JM.Product & 0.000(0.000) & 0.000(0.000) & 0.000(0.000) & 0.046(0.024) & 0.052(0.007) \\ 
   & JM.EmptyPoset & 0.000(0.000) & 0.000(0.000) & 0.000(0.000) & 0.043(0.023) & 0.051(0.007) \\ 
   \midrule
  \multirow{7}*{$q=0.2$}  & JS.Mix.Asy & 0.200(0.254) & 0.170(0.239) & 0.150(0.227) & 0.167(0.038) & 0.185(0.014) \\ 
   & JS.Mix.Finite & 0.210(0.259) & 0.290(0.288) & 0.260(0.279) & 0.211(0.033) & 0.224(0.010) \\ 
   & DACT & 0.060(0.151) & 0.300(0.291) & 0.380(0.309) & 0.226(0.030) & 0.245(0.008) \\ 
   & MT.Comp & 0.230(0.267) & 1.000(0.000) & 0.850(0.227) & 0.425(0.022) & 0.221(0.008) \\ 
   & JM.Max & 0.000(0.000) & 0.060(0.151) & 0.030(0.108) & 0.167(0.035) & 0.185(0.010) \\ 
   & JM.Product & 0.000(0.000) & 0.040(0.125) & 0.040(0.125) & 0.191(0.042) & 0.202(0.012) \\ 
   & JM.EmptyPoset & 0.000(0.000) & 0.040(0.125) & 0.010(0.063) & 0.187(0.042) & 0.199(0.011) \\ 
 \bottomrule
\end{tabular}
}
\caption{Mean and the corresponding standard deviation (in the round bracket) of FDPs for various methods. The results are based on 100 independent replications.\label{tab:FDPApp}}
\end{table}

\begin{table}[h!]
\centering
\resizebox{\linewidth}{!}{
\begin{tabular}{cllllll}
 \toprule
  Target &   \multirow{3}*{Method} &    \multicolumn{5}{c}{Hypothesis Configuration}\\ 
   \cmidrule(lr){3-7}
 FDR&  & \multirow{2}*{GNull} &  \multirow{2}*{SNull} &  \multirow{2}*{DNull} &  \multirow{2}*{SAlter} &  \multirow{2}*{DAlter} \\ 
Level &&&&&&\\
  \hline
  \multirow{7}*{$q=0.05$} & JS.Mix.Asy & 0.140(0.324) & 0.030(0.108) & 0.030(0.108) & 0.047(0.021) & 0.046(0.006) \\ 
   & JS.Mix.Finite & 0.140(0.324) & 0.060(0.151) & 0.060(0.151) & 0.063(0.021) & 0.059(0.006) \\ 
   & DACT & 0.040(0.178) & 0.060(0.151) & 0.080(0.172) & 0.066(0.018) & 0.065(0.006) \\ 
   & MT.Comp & 0.180(0.364) & 1.000(0.000) & 0.240(0.271) & 0.274(0.023) & 0.074(0.005) \\ 
   & JM.Max & 0.000(0.000) & 0.000(0.000) & 0.000(0.000) & 0.048(0.025) & 0.054(0.007) \\ 
   & JM.Product & 0.000(0.000) & 0.000(0.000) & 0.000(0.000) & 0.048(0.025) & 0.052(0.007) \\ 
   & JM.EmptyPoset & 0.000(0.000) & 0.000(0.000) & 0.000(0.000) & 0.045(0.025) & 0.051(0.007) \\ 
   \midrule
  \multirow{7}*{$q=0.2$}   & JS.Mix.Asy & 0.400(0.509) & 0.171(0.241) & 0.150(0.227) & 0.204(0.053) & 0.203(0.017) \\ 
   & JS.Mix.Finite & 0.420(0.518) & 0.291(0.290) & 0.260(0.279) & 0.266(0.047) & 0.249(0.012) \\ 
   & DACT & 0.120(0.302) & 0.300(0.291) & 0.380(0.309) & 0.276(0.040) & 0.265(0.009) \\ 
   & MT.Comp & 0.460(0.535) & 1.000(0.001) & 0.850(0.227) & 0.426(0.022) & 0.221(0.008) \\ 
   & JM.Max & 0.000(0.000) & 0.063(0.160) & 0.030(0.110) & 0.204(0.046) & 0.202(0.011) \\ 
   & JM.Product & 0.000(0.000) & 0.040(0.125) & 0.040(0.125) & 0.197(0.043) & 0.202(0.012) \\ 
   & JM.EmptyPoset & 0.000(0.000) & 0.042(0.130) & 0.010(0.063) & 0.205(0.047) & 0.202(0.012) \\ 
 \bottomrule
\end{tabular}
}
\caption{Mean and the corresponding standard deviation (in the round bracket) of mFDPs for various methods. The results are based on 100 independent replications.\label{tab:mFDPApp}}
\end{table}

\begin{table}[h!]
\centering
\begin{tabular}{clll}
 \toprule
  \multirow{2}*{ Target FDR Level}&   \multirow{2}*{Method} &    \multicolumn{2}{c}{Hypothesis Configuration}\\ 
   \cmidrule(lr){3-4}
&   &  SAlter & DAlter \\ 
  \hline
  \multirow{7}*{$q=0.05$}
 & JS.Mix.Asy & 0.772(0.061) & 0.875(0.032) \\ 
   & JS.Mix.Finite & 0.812(0.053) & 0.895(0.028) \\ 
   & DACT & \textbf{0.825(0.051)} & 0.906(0.026) \\ 
   %& MT.Comp & 0.978(0.013) & 0.949(0.017) \\ 
   & JM.Max & 0.763(0.078) & 0.887(0.030) \\ 
   & JM.Product & 0.807(0.067) & \textbf{0.928(0.020)} \\ 
   & JM.EmptyPoset & 0.765(0.078) & 0.913(0.025) \\ 
   \midrule
  \multirow{7}*{$q=0.2$} 
   & JS.Mix.Asy & 0.901(0.036) & 0.965(0.012) \\ 
   & JS.Mix.Finite & 0.921(0.031) & 0.973(0.010) \\ 
   & DACT & \textbf{0.929(0.028)} & \textbf{0.977(0.009)} \\ 
   %& MT.Comp & 0.993(0.007) & 0.990(0.005) \\ 
   & JM.Max & 0.902(0.035) & 0.965(0.012) \\ 
   & JM.Product & 0.928(0.038) & 0.\textbf{977(0.009)} \\ 
   & JM.EmptyPoset & 0.890(0.042) & 0.969(0.009) \\  
 \bottomrule
\end{tabular}
\caption{Mean and the corresponding standard deviation (in the round bracket) of powers for various methods. The results are based on 100 independent replications.\label{tab:POWApp}}
\end{table}

\clearpage
\bibliographystyle{asa}
\bibliography{reference}

%% file: Terminology.tex
\def\var{\mathrm{var}}

\def\FDP{\operatorname{FDP}}

\def\FDR{\operatorname{FDR}}

\def\mFDR{\operatorname{mFDR}}
\def\mFDP{\operatorname{mFDP}}

\def\ttf{\tilde{f}}

\def\tp{\tilde{p}}

\def\tt{\tilde{t}}

\def\hi{\hat{i}}

\def\hq{\hat{q}}

\def\whScal{\widehat{\Scal}}

\def\pf{\mathbf{p}}

\def\sf{\mathbf{s}}
\def\tf{\mathbf{t}}

\def\xf{\mathbf{x}}

\def\1f{\mathbf{1}}

\def\tpf{\tilde{\mathbf{p}}}

\def\ttf{\tilde{\mathbf{t}}}

\def\Eb{\mathbb{E}}

\def\Pb{\mathbb{P}}
\def\Rb{\mathbb{R}}

\def\btheta{\bm{\theta}}

\def\bTheta{\bm{\Theta}}

\def\Acal{\mathcal{A}}

\def\Fcal{\mathcal{F}}

\def\Hcal{\mathcal{H}}
\def\Kcal{\mathcal{K}}
\def\Mcal{\mathcal{M}}

\def\Rcal{\mathcal{R}}
\def\Scal{\mathcal{S}}

\def\Ucal{\mathcal{U}}
\def\Pcal{\mathcal{P}}

\def\Proj{\mathop{\mbox{Proj}}}

\def\tPcal{\widetilde{\mathcal{P}}}

\def\opn1{o_{\Pb_n}(1)}

\usepackage[normalem]{ulem}